\documentclass[useAMS,usenatbib]{mn2e}
\usepackage{graphicx,amsmath,amssymb}

\newcommand{\Msol}{M_\odot}
\newcommand{\angleIso}{\theta_0}
\newcommand{\Moideg}{^\circ}
\newcommand{\Moy}[1]{\left\langle {#1} \right\rangle}
\newcommand{\ro}{r_0}

\newcommand{\V}[1]{\boldsymbol{#1}} 

\newcommand{\Module}[1]{\left|{#1} \right|}
\newcommand{\ModuleVect}[1]{\left\|{#1} \right\|}
\newcommand{\moiPar}[1]{\left({#1} \right)}
\newcommand{\Crochets}[1]{\left[{#1} \right]}

\newcommand{\grasR}{\V{r}}

\newcommand{\echExt}{\mathcal{L}_0}

\newcommand{\matInterTomoi}{\V{M^{N_t}_{\alpha_i}}}
\newcommand{\matInterTomo}{\V{M^{N_t}_{\alpha}}}
\newcommand{\matT}{\V{M^{N_t}_{0}}}
\newcommand{\matRecTomo}{\V{W_{opt}}}

\title[The FALCON concept]{The FALCON concept: multi-object adaptive  
optics and atmospheric tomography for integral field spectroscopy.  
Principles and performances on an 8 meter telescope.}
\author[F. Ass\'emat, E. Gendron and F. Hammer]{F.  
Ass\'emat$^{1,3}\footnotemark[1]$\thanks{E-mail: 
francois.assemat@durham.ac.uk(FA); eric.gendron@obspm.fr(EG);  
francois.hammer@obspm.fr(FH)}, E. Gendron$^{2}$ and F. Hammer$^{1}$\\
$^{1}$GEPI, Observatoire de Paris-Meudon, 92195 Meudon, France\\
$^{2}$LESIA, Observatoire de Paris-Meudon, 92195 Meudon, France\\
$^{3}$Physics Department, University of Durham, South Road, DH1 3LE,  
Durham, UK}
\begin{document}


\pagerange{\pageref{firstpage}--\pageref{lastpage}} \pubyear{2006}

\maketitle

\label{firstpage}

\begin{abstract}
Integral field spectrographs are major instruments to study the  
mechanisms involved in the formation and the evolution of early  
galaxies. When combined with multi-object spectroscopy, those  
spectrographs can behave as machines used to derive physical parameters  
of galaxies during their formation process. Up to now, there is only  
one available spectrograph with multiple integral field units, e.g.  
FLAMES/GIRAFFE on the VLT. However, current ground based instruments  
suffer from a degradation of their spatial resolution due to  
atmospheric turbulence. In this article we describe the performance of  
FALCON, an original concept of a new generation multi-object integral  
field spectrograph with adaptive optics for the ESO Very Large  
Telescope. The goal of FALCON is to combine high angular resolution  
($0.25~arcsec$) and high spectral resolution ($R>5000$) in J and H  
bands over a wide field of view ($10\times 10~arcmin^2$) in the VLT  
Nasmyth focal plane. However, instead of correcting the whole field,  
FALCON will use multi-object adaptive optics (MOAO) to perform locally  
on each scientific target the adaptive optics correction. This requires  
then to use atmospheric tomography in order to use suitable natural  
guide stars for wavefront sensing. We will show that merging MOAO and  
atmospheric tomography allows us to determine the internal kinematics  
of distant galaxies up to $z\approx 2$ with a sky coverage of 50\%,  
even for objects observed near the galactic pole. The application of  
such a concept to Extremely Large Telescopes seems therefore to be a  
very promising way to study galaxy evolution from $z=1$ to redshifts as  
high as $z=7$.
\end{abstract}


\section{Introduction}
Thanks to the Hubble Space Telescope, astronomers have been able to  
determine the morphology of distant galaxies located at $z>0.5$,  
showing that galaxies in the past were mostly irregular and smaller  
than those in the local universe \citep{Abraham-a-2001,  
VanDenBergh-a-2002}, and had higher merging rates  
\citep{Lefevre-a-2000,Bundy-a-2004}. The use of spatially resolved  
color maps has also allowed to highlight regions of star formation as  
well as showing that galaxies cores were bluer than today's galaxies  
bulges \citep{Abraham-a-1999,Zheng-a-2004}.

Spectroscopic studies have also taught us that star formation rates  
(SFR) were higher in distant galaxies, in particular in Luminous  
Infrared Galaxies (LIRGs) where SFR can reach 100 $\Msol /yr$, and that  
the density of star formation was much higher at $z=1$ than today  
\citep{Madau-a-1996,Flores-a-1999}. Therefore, when the re-emission  
from the dust in the IR is taken into account, a simple integration of  
the global star formation history shows that half of the present-day  
stars have been forming since $z\approx 1-1.5$ \citep{Dickinson-a-2003,  
Hammer-a-2005}.

However, the dynamical information of distant galaxies, crucial for  
their studies, still is not well known. Indeed, spatially resolved  
color maps allow us to show where the star formation occurs, but not to  
know how masses and gas are distributed in those galaxies, and neither  
to know the physical and chemical properties of the gas. Also, the  
question of the evolution of the fraction of barred galaxies is still  
controversial today \citep{Abraham-a-1999b,Sheth-a-2003,Zheng-a-2005}.  
At last, the way galaxies in merging systems were assembling their  
masses in the past is still mysterious, as well as the influence of  
dark matter into those exchanges. Kinematics and chemistry of galaxies  
up to $z \approx 2$ is therefore required in order to evaluate the  
velocity fields of their main components (bulges and disks), to  
determine how important are the respective roles of the merging  
phenomenon and the dark matter, and finally to establish the physical  
origin of the Hubble Sequence. At these redshifts, the important  
features such as the $H\alpha$ emission lines or the stellar absorption  
lines are redshifted in the near-infrared, in J and H bands. This means  
that integral field spectroscopy (IFS) is required, with a high  
spectral resolution ($3000 \leq R \leq 15000$), but also with a good  
spatial sampling (1-2 kpc) of the velocity field, implying a spatial  
resolution ($0.15-0.25~arcsec$) better than the atmospheric seeing.  
Moreover, there are others important issues such as the the field of  
view (FoV) and the multiplex capability. Indeed, it is important to  
observe galaxies on scales greater than the correlation length (4 to 9  
Mpc), leading to a minimal FoV of 100 square arcmin. Observations of  
several cosmological fields will then allow us not to be sensitive to  
cosmic variance effects, and a multiplex capability is then required so  
that several objects can be observed simultaneously, providing a gain  
in exposure time.\newline

In order to reach the required image quality, the solution is to use  
adaptive optics \citep{Roddier-b-1999}. Adaptive Optics (AO) allow us  
to compensate in real-time for the wavefront's distorsion caused by the  
turbulence in the Earth's atmosphere, providing to the telescopes their  
nominal angular resolution (diffraction-limited imaging). However, the  
FoV on which this resolution is achieved is very low. Indeed, all the  
AO systems built before now work with the \textit{classical} AO method,  
where a guide star (natural or artificial), bright enough and close  
enough to the science target, is used to properly measure the incoming  
turbulent wavefront and compensate for it. In these conditions, the  
compensated FoV around the guide star is equal to a few times the  
isoplanatic patch $\angleIso$, \textit{eg} typically 10 to $20~arcsec$  
in H band. When applied to extragalactic astronomy, this makes the use  
of natural guide stars impossible for AO, at least under its  
"classical" form, as the sky coverage would then be lower than 3\% due  
to the lack of sufficiently bright guide stars, even at a galactic  
latitude of $30\Moideg$.

Several concepts have therefore been proposed in the last years to  
increase the sky coverage and the corrected FoV of AO systems. Laser  
Guide Stars\footnote{hereafter LGS} \citep{Foy-a-1985} create an  
artificial star in the sodium layer of the atmosphere at an altitude of  
90 kms thanks to a laser. However this method suffers from the cone  
effect and the tilt determination problem  
\citep{Tallon-a-1990,Rigaut-a-1992}, meaning that a natural GS is  
always required close to the science target to measure the low orders  
of the turbulent wavefront (tip-tilts). Moreover, in the case of  
multi-objects instruments such as those required for extragalactic  
astronomy, one LGS is required per scientific target as well as a  
dedicated wavefront sensor to measure the low orders. This can  
dramatically increase the cost of such a multi-object instrument.  
Multi-Conjugated Adaptive Optics\footnote{hereafter MCAO}  
\citep{Dicke-a-1975,Johnston-a-1994,Fusco-a-2001,LeLouarn-a-2002} have  
also been proposed to improve the FoV of AO systems, using several  
deformable mirrors conjugated to the dominant turbulent layers. But the  
compensated FoV provided by such systems remains generally insufficient  
for extragalactic studies. As an example, the compensated FoV provided  
by the Gemini-South MCAO system or the MAD system for the VLT will have  
a diameter of only 2 arcmin \citep{Ellerbroek-p-2003,Marchetti-p-2003}.  
At last, Ground Layers Adaptive Optics\footnote{hereafter GLAO}  
\citep{Rigaut-p-2002} have also been proposed to widen the FoV of AO  
systems. But this approach assumes that most of the turbulence is  
located close to the telescope pupil. Moreover, the corrected FoV  
remains generally insufficient for extragalactic studies, and most of  
the GLAO systems studied until today always use at least one LGS to  
reach decent performance \citep{LeLouarn-a-2004,Morris-p-2004}.\newline

Such considerations initiated therefore the FALCON concept  
\citep{Hammer-p-2002}. FALCON (\textit{Fiber optics spectrograph with  
Adaptive optics on Large fields to Correct at Optical and  
Near-infrared}) is a concept of a new generation multi-object integral  
field unit (IFU) spectrograph, working with adaptive optics over a very  
wide field of view ($10 \times 10~arcmin^2$) at the Nasmyth focus of  
the VLT and suitable for extragalactic studies. But instead of  
correcting the whole FoV, the AO correction is only performed on the  
regions of interest, eg the spectroscopic IFUs observing galaxies,  
thanks to a Deformable Mirror (DM) conjugated to the pupil for each  
IFU. However, with such an architecture, the sky coverage problem of  
classical AO remains. We have focused on this problem, which is solved  
by using a new type of wavefront reconstruction, inspired from MCAO  
techniques \citep{Fusco-a-2001,Tokovinin-a-2001}, where the WFS  
measurements from several off-axis NGS around each galaxy are used to  
extrapolate the on-axis galaxy wavefront in the pupil. Preliminary  
performance of this concept have been given in \citet{Assemat-p-2004}  
and \citet{Hammer-p-2004}.\newline

In this paper, we will show more accurately the expected performance  
achieved by FALCON thanks to the combination of 3D spectroscopy with  
AO. Firstly, we will present the technical specifications of FALCON,  
derived from the higher level scientific requirements. We will then  
show the gain AO can bring to Integral Field Spectroscopy. Following  
that, we will accurately describe the principle of the FALCON AO  
system. The next section will show the expected performance of such a  
system in the case of median atmospheric conditions for the Cerro  
Paranal. The discussion will then compare these performance with those  
from a GLAO system working in the same conditions, and will deal with  
additional source of errors whose influence remains to be quantified.

\section[High level specifications]{High level specifications}
\subsection{Wavelength range}
As explained in the introduction, the dynamical information is required  
to improve our knowledge of the mechanisms responsible for the  
formation of galaxies. The velocity field of the galaxies has therefore  
to be probed by measuring the redshifts of emission lines, that we  
choose to peak in the $1.00-1.85~\mu m$ wavelength range (covering J  
and H bands), thus avoiding the thermal domain ($\lambda>1.95$ microns)  
where the instrumental thermal background will dominate the noise.  
Several emission lines can then be used, such as  
$[OII]~(\lambda=3727\AA)$, $H\beta~(\lambda=4861\AA)$,  
$[OIII]~(\lambda=5007\AA)$ and in particular  
$H\alpha~(\lambda=6563\AA)$. Indeed, this latter suffers less from  
extinction than the other emission lines \citep{Liang-a-2004b}, and can  
be used to map the dynamical information up to $z=1.8$. The use of  
shorter wavelength emission lines will allow to observe galaxies up to  
$z=2.5$, however extinction may then be a problem  
\citep{Liang-a-2004b}.

\subsection{Angular resolution}
Morphological studies of galaxies located into the Hubble Deep Fields  
and the GOODS fields have shown that galaxies with $z\geq 1$ had  
half-light radii $r_{1/2}$ smaller than $0.5~arcsec$  
\citep{Marleau-a-1998,Ferguson-a-2004}, with an average  
$\Moy{r_{1/2}}=0.25~arcsec$ at $z=2$ \citep{Bouwens-a-2004b}.
As we want to be study the dynamical processes occuring within the  
galaxies, we must be able to resolve their
half-light radius, up to $z=2$. Therefore, an angular resolution of at  
least $0.25~arcsec$ is required, implying a pixel sampling of  
$0.125~arcsec$. Such a spatial resolution is far beyond abilities of  
current ground-based, seeing-limited, integral fields spectrographs, as  
it is definitely better than the atmospheric seeing. This strongly  
suggests the use of adaptive optics, as we will see later in section  
\ref{sec:sec3}.

\subsection{Spectral resolution}

We have to observe in the near infra-red (NIR) to fulfill our  
scientific requirements. However, the spectral window we want to  
explore is crowded with thin, intense terrestrial OH emission lines  
\citep{Maihara-a-1993,Rousselot-a-2000,Hanuschik-a-2003}. Observations  
have therefore to be performed between these lines, implying to  
separate them. This leads therefore to a minimum spectral resolution  
$R=5000$. Considering now the impact on the determination of the galaxy  
dynamics, this is equivalent to resolve velocity dispersions $\sigma  
\geq 25.5$ km/s at $z=0$.

However, for such an instrument dedicated to the observation of distant  
galaxies, the optimal spectral resolution is a compromise between the  
scientific goals on one hand, and the drastic loss in spectral  
signal-to-noise ratio with increasing $R$ on the other. As an example,  
spectral resolutions $R \leq 15000$ will allow us to resolve velocity  
dispersions $\sigma \geq 8/(1+z)$ km/s, i.e. greater than 4 km/s for a  
galaxy located at $z=1$. A value of $15000$ seems therefore to be a  
reasonable upper limit to consider for the spectral resolution $R$.  
Recent observations of $z=0.6$ galaxies with GIRAFFE  
\citep{Flores-a-2006} have in fact shown that a slightly lower spectral  
resolution $R=10000$ is definitely required to probe precisely the  
kinematics of distant galaxies with various morphologies, including  
compact galaxies. Moreover, a spectral resolution $R=10000$ also allows  
us to resolve the $[OII]3727$ doublet, and then to retrieve the  
kinematics of galaxies with higher redshift.

\subsection{Field of view and multiplex capability}
Observations of Lyman Break Galaxies with $z\geq 2$ show a strong  
spatial clustering, with correlation lengths greater than 4 Mpc  
\citep{Giavalisco-a-2002}. Similar results are also confirmed by cosmic  
web simulation codes, based on $\Lambda CDM$ theory  
\citep{Hatton-a-2003}. The observation of distant galaxies over scales  
greater than this correlation length should therefore be achievable, in  
order to avoid statistical biases due to the cosmic variance effects.  
As an example, such effects have been observed on the WFPC2 Hubble Deep  
Fields \citep{Labbe-a-2003}, where the FoV ($\approx 6~arcmin^2$) is  
unsufficient. It is then required to encompass the correlation lengths  
for all redshifts, leading to a instrument with a minimum FoV of  
$100~arcmin^2$.

The instrument's FoV is also linked to the number of IFUs as well as  
the wished multiplex gain. As an example, \citet{Steidel-a-2004} found  
a density of 3.8 galaxies per arcmin$^2$, for objects with redshifts  
$1.5 < z < 2.0$ and magnitudes $R\leq 25.5$. If we assume now that we  
only observe objects with magnitudes $R \leq 23.5$ or even $R \leq  
23.0$ (because of SNR issues), we find a lower source density,  
comprised between 0.31 and 0.84 galaxies per arcmin$^2$. From our past  
experience on ESO large program with GIRAFFE \citep{giraffe}, we think  
that 15 to 30 IFUs should be a reasonable number while providing an  
efficient gain in terms of exposure time. However, in order to optimise  
the exposure time, it is required to observe objects in the same  
magnitude range, and the appropriate density to consider becomes the  
number of objects per magnitude and per arcmin$^2$. Moreover, a non  
negligible fraction of the objects will see their emission lines match  
with atmospheric OH lines, making them impossible to observe. As a  
consequence, this might make the source density definitely smaller than  
the values written above, and low enough to warrant a wide FoV, equal  
to $10 \times 10~arcmin^2$. The VLT Nasmyth focal plane is therefore  
ideally suited for a field of view, as the vignetting on a $10~arcmin$  
diameter FoV is less than $1\%$. Spreading several IFUs over the  
Nasmyth focus will therefore allow us to perform the simultaneous 3D  
spectroscopy of several galaxies with similar redshifts, thus providing  
a very important gain in exposure time.

\subsection{Conclusion}
In this section we have given high level specifications for the FALCON  
instrument, derived from scientific requirements. As it has been shown,  
a  multi-object spectrograpy with both high spatial ($FWHM \leq  
0.25~arcsec$) and spectral ($R \geq 5000$) resolutions is required,  
moreover over a wide field of view (FoV$\geq 100~arcmin^2$).

We would especially like to insist now on the spatial resolution we  
require ($0.25~arcsec$), which is not achievable with current seeing  
limited instruments such as GIRAFFE. This requires the image quality to  
be enhanced, and the best way to achieve this, in particular in terms  
of signal to noise ratio (SNR), is to use Adaptive Optics. We therefore  
show in the following section the gain brought by Adaptive Optics to  
Integral Field Spectroscopy.

\section[3D Spectroscopy and Adaptive Optics]{3D Spectroscopy and Adaptive Optics}\label{sec:sec3}
\subsection{Introduction: why use AO with 3D Spectroscopy}\label{whyAo3dSpec}

Adaptive optics (AO) is a technique that allows us to restore in  
real-time the flatness of a wavefront distorded by the effects of the  
atmopheric turbulence, thus improving the image quality at the focal  
plane of an instrument. For classical imaging applications, the main  
interest of AO is to allow us the partial restoration of spatial  
frequencies in the images up to the telescope's cut-off frequency  
($D/\lambda$), far beyond the turbulent cut-off frequency. This is the  
consequence of the morphology of compensated point-spread function  
(PSF), which is quite complex: it cannot
only be summarized by its FWHM, as it depends on the structure function  
of the phase residuals, and in particular on the compensation order.  
The more important this latter, the more intense is the "coherent core"  
(with a width of $\lambda/D$) on top of the PSF halo. Deconvolution  
processes can then be applied, whose performance increase together with  
signal-to-noise, i.e. with the restoration quality. Expecting this high  
frequency restoration makes it mandatory to obey the Nyquist critera  
for the image sampling.

Expectations are different when studying faint galaxies with 3D  
spectroscopy at $R\approx10000$. The intrinsic faintness of the  
dispersed signal prohibits, anyway, any attempt to spatially sample it  
at Nyquist frequency. Hopefully for us, this latter is not an issue,  
since we are not interested in the ultimate telescope resolution, but  
only in a moderate resolution of $0.25~arcsec$. Our concern is  
therefore to maximise the flux from the object within a square aperture  
of $0.25 \times 0.25~arcsec^2$ -our spatial resolution element. This  
has two consequences. Firstly, it increases the signal from the object  
within the spatial resolution element, whereas the noise level  
(especially thermal and sky background) will remain constant: the  
spectroscopic SNR will increase. Secondly, the flux spread in the  
neighbouring pixels will be minimised, thus avoiding spatial  
contamination.

These improvements of the image quality expected in spectroscopy  
(resolution, SNR improvement, reduction of confusion) are difficult to  
relate with the parameters commonly used in adaptive optics to  
characterise the quality of the restored image. Classical parameters  
are, as an example, the full width at half maximum (FWHM) or the Strehl  
ratio (SR ; this latter is defined as the central intensity of a  
restored PSF, divided by the central one of the diffraction-limited  
PSF, taking both PSFs with the same total energy). For 3D spectroscopy  
applications, the criterium traditionally used, and that defines the  
spectroscopic image quality, is in fact the fraction of the total PSF  
flux ensquared within the spatial resolution element. We will therefore  
use this criterium, the \textit{ensquared energy}, in the following  
sections of this paper.

\subsection{The need for a high-order compensation.}\label{sec:sec32}
FALCON had initially began with the wrong idea that a low-order, or a moderate compensation should be sufficient to achieve a goal that, at least at the first glance, appeared to be modest : obtain a resolution of 0.25 arcsec. We demonstrate in the following sections that getting such a resolution implies a finer spatial sampling (0.125 arcsec) compared to existing instruments, and consequently a huge loss in signal-to-noise per spatial sample. This loss has to be compensated by an increase of the ensquared energy. We point out that a high ensquared energy can be obtain only by using a high-order compensation. This effect is implicitly demonstrated at the end, in the last result section of our article, where we will see that the ensquared energy still take advantage of a high-order correction. However, we want to insist on this particular point from now, because this explains why we have especially focused a large part of our study on the number of freedom of the system shall have, and why we study all the compensation levels up to 120 Zernike modes in this paper.

We illustrate here how the ensquared energy is a parameter sensitive to the spatial frequency range of the phase residuals. We will compare two illustrative cases : they are comparable, in the sense that their wavefront flatness is the same (i.e. same amount of residual phase variance). The two cases just differ by the spatial frequency of the phase residuals.

Any publication about AO compensation focuses on phase residual variance, which is considered as the key parameter that optimizes the image quality. We exhibit here two striking examples of PSFs where the ensquared energy is fundamentally different (more than a factor of 2), while their associated phase variance, Strehl ratio, and FWHM are the same.

\begin{figure}
\centering
\includegraphics[width=1\linewidth]{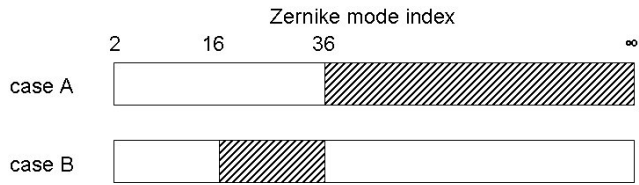}
\caption
{The hatched area of the bars indicate the spatial frequency range of the residuals, described by the index of the Zernike mode. Case A has been chosen because it mimics a case typical of a low-order AO system. Numbers of case B (residuals is the range $Z_{17-36}$) are chosen to provide the same wavefront flatness (same phase variance) as in case A. The two PSF will have the same phase variance, same SR, and same FWHM.
}
\label{fig:highlow}
\end{figure}

\begin{figure}
\centering
\begin{tabular}{cc}
\includegraphics[width=0.4\linewidth]{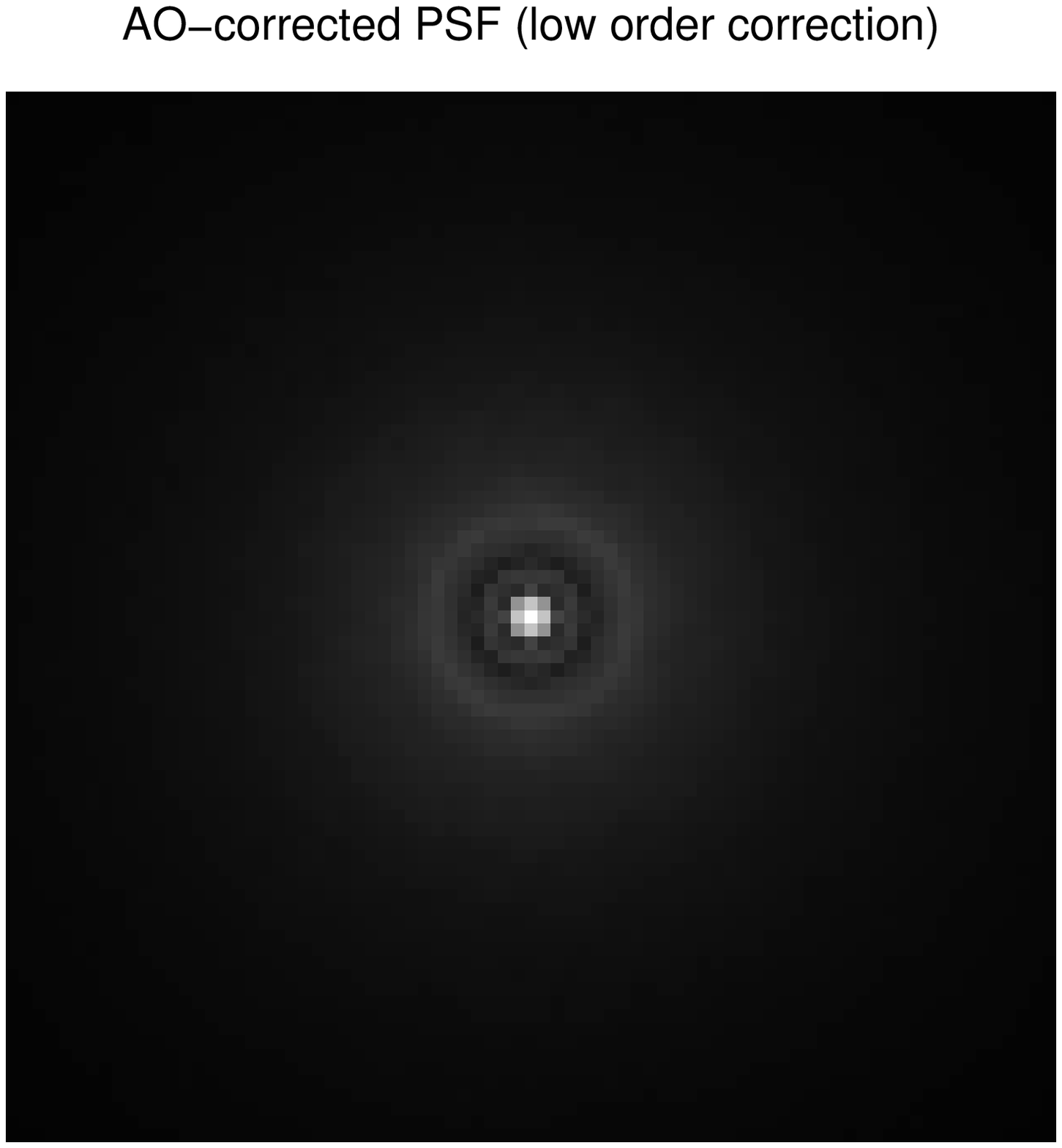}&
\includegraphics[width=0.4\linewidth]{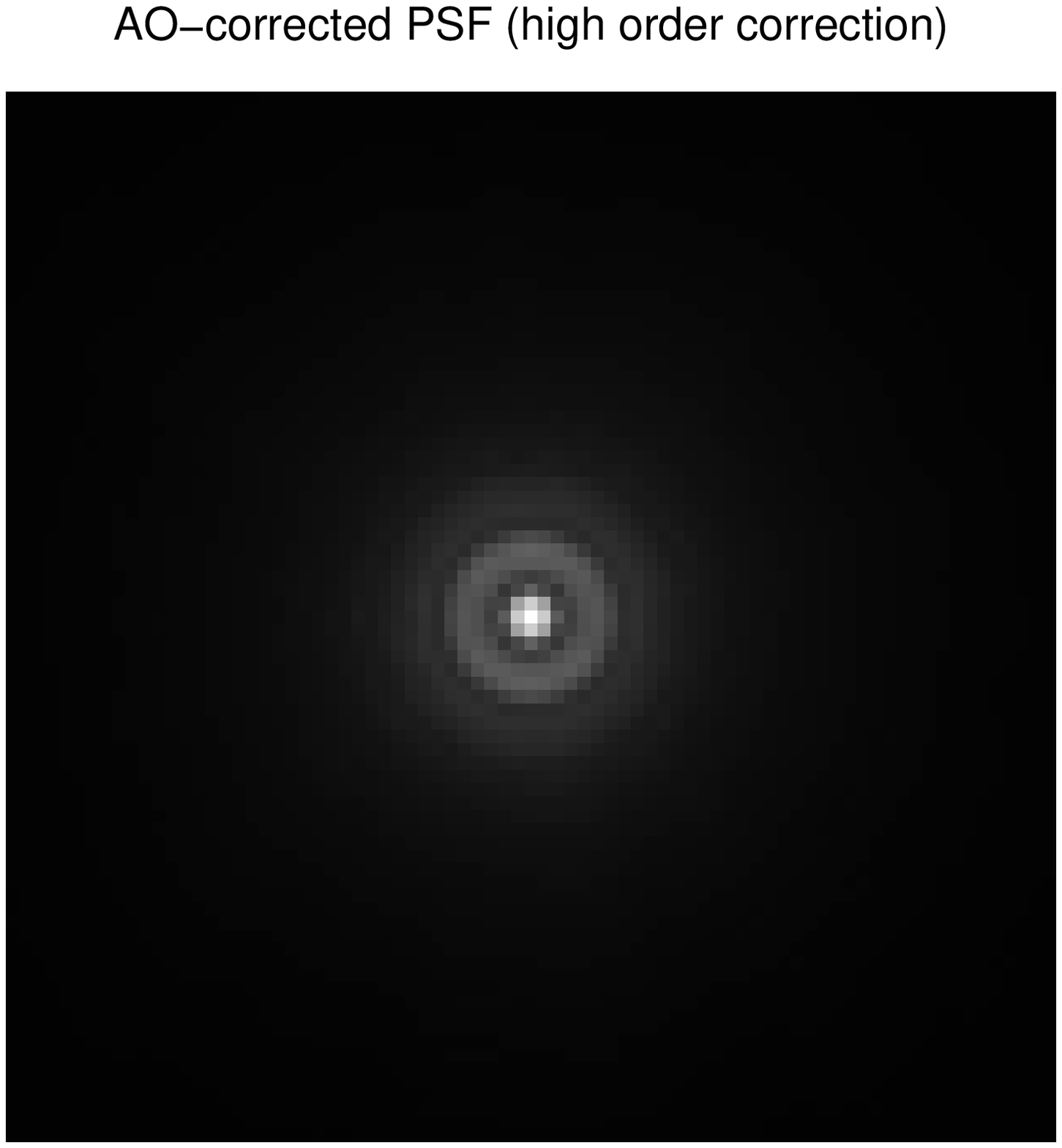}\\
\end{tabular}
\caption
{
AO-corrected PSFs in the case of low order correction (perfect  
correction of ZPs up to $j_{max}=36$, left) and high order correction  
(perfect correction of ZPs 2 to 16 and 37 to $\infty$, right). 8 meter  
diameter telescope, seeing of $0.81~arcsec$ at $\lambda=0.5~\mu m$, H  
band ($\lambda=1.65~\mu m$). The simulated field of view is equal to  
$1.70 \times 1.70~arcsec^2$.}
\label{fig:fig2}
\end{figure}

For our illustration, we performed some Monte-Carlo numerical simulations of  
an AO corrected PSF, where the wavefront (from a fully developped  
Kolmogorov spectrum) is corrected by zeroing some coefficients of its  
expansion into Zernike polynomials \citep{Noll-a-1976}. As  
we just study the influence of the compensation order, here we focus  
only on the spatial aspect of the correction, ignoring errors due to  
time delay, anisoplanatism or noise. The conditions of the simulation  
are a 8-meter telescope (VLT case), a seeing of $0.81~arcsec$ at  
$0.5~\mu m$ (median seeing of VLT site, \citet{Martin-a-2000}), leading  
to a $\ro$ equal to $52~cm$ at the imaging wavelength of $1.65~\mu m$  
(H band).
We compared two clear-cut cases of correction, similar since they both lead to  
the same amount of residual phase variance  
$\sigma^2_{res}=0.0132(D/\ro)^{5/3}~rad^2$, but different because the  
compensation concerns different domains of spatial frequencies, as illustrated in figure \ref{fig:highlow}:
\begin{itemize}
\item for the case A, we mimic a typical low-order AO system, and we compensate for the polynomials $Z_2$ to  $Z_{36}$. Phase residuals are  
high-order perturbations $Z_{>37}$.
\item for the case B, we consider the phase residuals are now  
low-order perturbations in the range $Z_{17-36}$. Those last numbers are chosen so that the spatial frequency contents of the phase does not overlap with case A, but has exactly the same variance.  
\end{itemize}

\begin{figure}
\centering
\begin{tabular}{cc}
\includegraphics[width=0.4\linewidth]{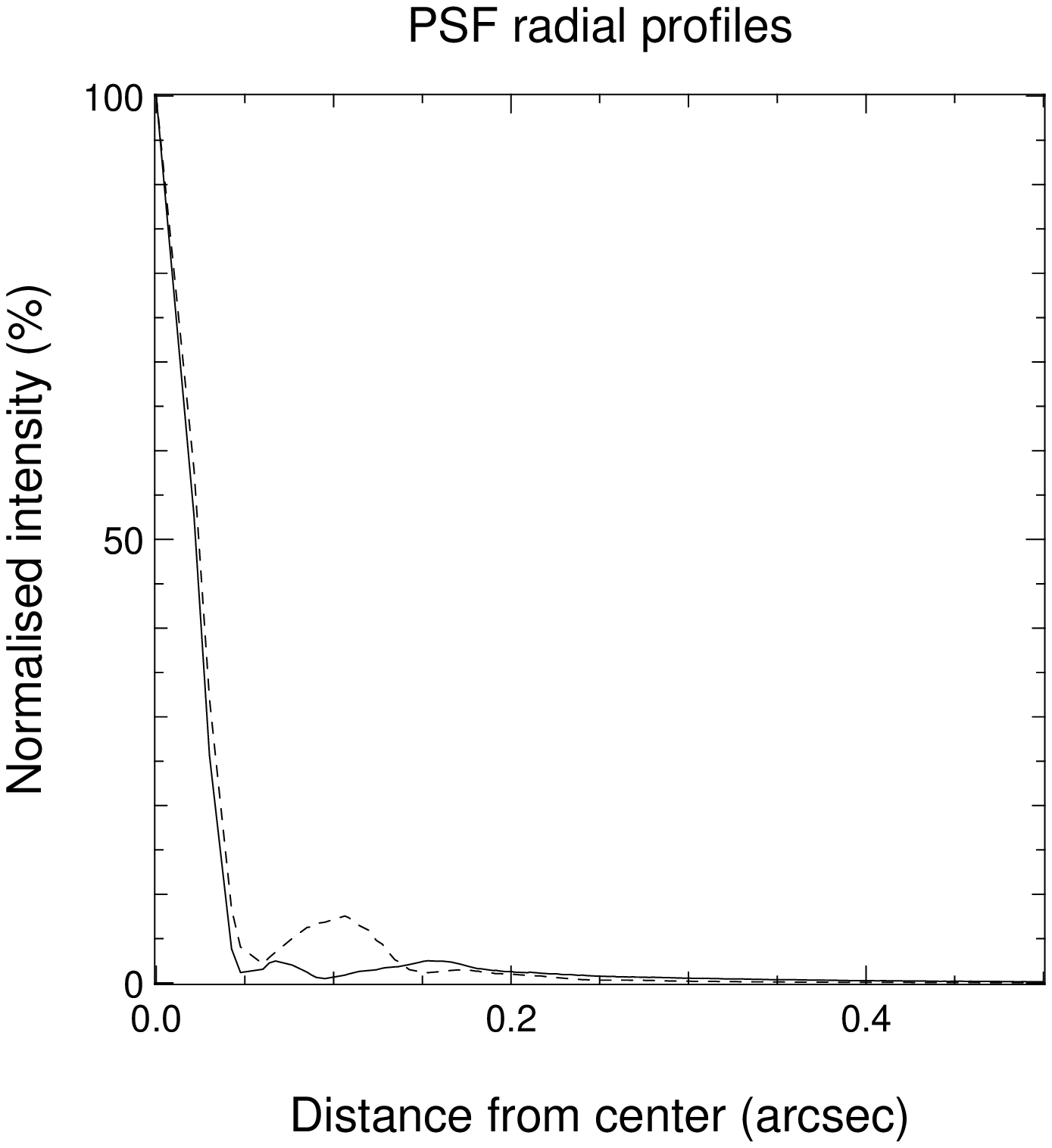}&
\includegraphics[width=0.4\linewidth]{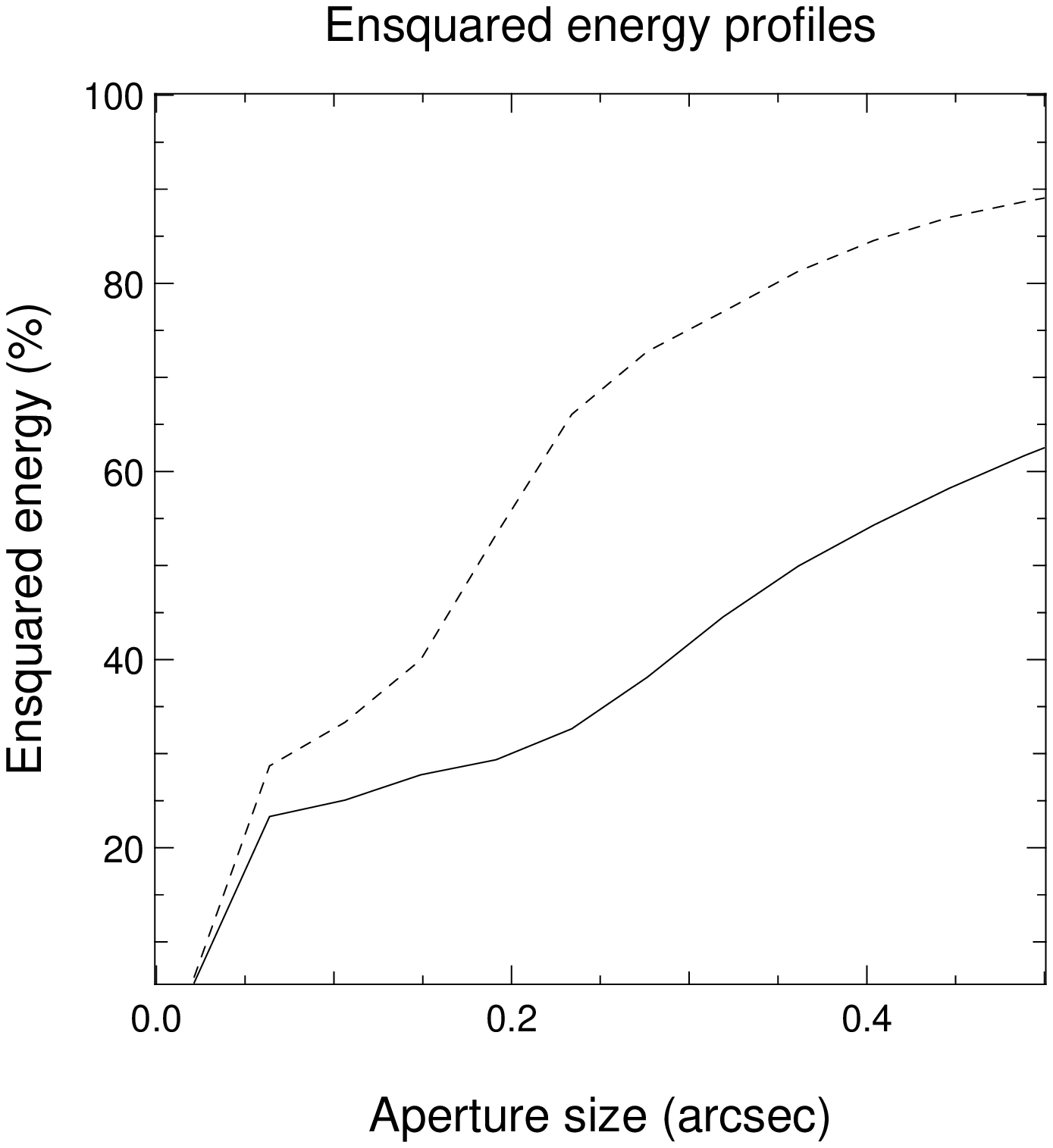}\\
\end{tabular}
\caption
{
Radial profiles (left) and ensquared energy profiles (right) of the  
AO-corrected PSFs. The solid line corresponds to low order correction,  
the dashed line to high order correction.}
\label{fig:fig3}
\end{figure}

The AO corrected PSFs in H band are shown on the figure \ref{fig:fig2},  
whereas figure \ref{fig:fig3} shows the radial (left) and ensquared  
energy profiles (right). Those figures aim to demonstrate how different  
the shapes of the PSFs can be, although they simultaneously exhibit  
identical residual phase variances, identical central intensities  
(predictable from the phase variance: both Strehl ratios are close to  
30\%), and identical FWHMs: the shapes of the coherent cores are  
identical, only the energy distribution of the halo is different,  
because it is strongly determined by the spatial contents of the phase  
residuals.

In terms of astrophysical implications, a PSF such as in case B  
(low-order residuals, right of figure \ref{fig:fig2}) would be  
catastrophic for exoplanet detection, where the intensity in the first  
rings has to be minimised because of contrast issues. On the opposite,  
such a PSF is interesting for 3D spectroscopy as the ensquared energy  
is increased: this is particularly noticeable on the right of the  
figure \ref{fig:fig3}.

One has therefore to keep in mind that for a fixed amount of phase  
variance removed, the ensquared energy will strongly be enhanced in the  
case of a high order compensation, which is particularly effective at  
reducing the halo compared to low-order modes correction.
The conclusion is that the common rule of thumb, according to which the  
phase variance error budget should be balanced between undermodeling  
error and other sources, becomes obsolete when it is a matter of  
designing an AO system for spectroscopy. Despite the moderate  
resolution ($\approx 0.25~arcsec$) required, a rather large number of  
actuators is required to bring the light from the further parts of the  
halo into the inner parts, so as to increase the spectroscopic SNR, and  
only a high-order modes compensation can allow us to achieve  
this.\newline

However, compensating for high order modes is not going without some  
difficulties from the AO point of view, the major limitation being  
anisoplanatism: high order modes unfortunately decorrelate faster with  
the separation angle \citep{Chassat-a-1989, Roddier-a-1993}.  
Unfortunately,  the usual observing conditions of distant galaxies are  
at high galactic latitudes, where natural guide stars become scarce and  
are unlikely to lie within the isoplanatic patch $\angleIso$. This  
requires to work most of the time with large separation angles, that  
precisely forbid the compensation of high-order modes. This drawback  
makes classical AO unappropriate for the statistical studies of distant  
galaxies.

In the section \ref{sec:4} we will describe  a new approach of adaptive  
optics, which solves anisoplanatism problems, which allows to use  
natural guide stars at distances greater than $\angleIso$, and which is  
still efficient to correct high order modes.\newline

Before this, we want to be able to establish requirements about the  
ensquared energy needed for a proper statistical study on high redshift  
galaxies, in order to derive later some specifications for the AO  
system we propose.

\subsection{Case study: 3D spectroscopy of UGC 6778 observed at high  
redshift}
Thanks to numerical simulations, we show in the following subsections  
some quantitative results about the gain in spectroscopic SNR that can  
be achieved when AO and 3D spectroscopy are used together. We  
would like to emphasize here that we do not attempt to perform a  
complete simulation of a 3D spectrograph feeded by an AO system,  
because of the lack of 3D spectroscopic data currently available in the  
litterature for high redshift galaxies, and especially in the  
near-infrared wavelength range, where the $H\alpha$ emission line is  
observed for redshifts $z \geq 0.5$. We therefore do very simple  
assumptions about the instrument as well as the observed objects, in  
order to compute first estimates about the SNR achieved thanks to AO  
correction. Such results can be used to derive very preliminary  
required ensquared energy values, that can then be translated into  
instrumental requirements for the AO system. A $\Lambda CDM$ cosmology  
with $H_0=70$ km/s/Mpc, $\Omega_m=0.3$ and $\Omega_\Lambda=0.7$ is  
assumed in the rest of this paper.\newline

\begin{table*}
\centering
\begin{minipage}{140mm}
\begin{tabular}{ccccccccccc}
\hline
\hline
UGC&
Type$^{\mathrm{a}}$&
$V_{VIR}^{\mathrm{b}}$
&D$^{\mathrm{b}}$&
$M_B^{\mathrm{a}}$&
$R_D^{\mathrm{c}}$&
$\mu_D^{\mathrm{c}}$&
$i^{\mathrm{a}}$&
PA$^{\mathrm{a}}$&
$V_{max}^{\mathrm{a}}$\\

ID&
&
(km s$^{-1}$)&
(Mpc)&
(mag)&
(kpc)&
V/arcsec$^2$
&($\Moideg$)
&($\Moideg$)&
(km s$^{-1}$)\\
\hline
3893&SBc&1193&17.04&-20.6&1.84&20.17&30&165&305\\
\hline
\end{tabular}
\begin{list}{}{}
\item[$^{\mathrm{a}}$] from \cite{Garrido-a-2002}
\item[$^{\mathrm{b}}$] from LEDA  
(http://www-obs.univ-lyon1.fr/hypercat/)
\item[$^{\mathrm{c}}$] from \cite{Baggett-a-1998}
\end{list}
\caption{Photometric and dynamic characteristics of UGC 6778. From left  
to right: the identifier of the galaxy, its morphological type, the  
recession velocity corrected from the local's group infall toward Virgo  
$V_{VIR}$, the distance $D$ in Mpc, the absolute magnitude in B band  
$M_B$, the exponential disk scale lenght $R_D$, the disk surface  
brightness $\mu_D$, the inclination $i$ and the position angle $PA$,  
and the maximum velocity observed on the edge of the disk}
\end{minipage}
\end{table*}\label{tab:table1}

We used for this study the $H\alpha$ image of the UGC 6778 galaxy  
provided by members of the GHASP survey  
\citep{Garrido-a-2002,Garrido-a-2003,Garrido-a-2004,Garrido-a-2005},  
with a spatial sampling $s_0=0.960~arcsec/pixel$. The table 1 lists the  
photometric and dynamic properties of this galaxy. These data were then  
used to simulate two science cases, where we assumed that the galaxy  
would be respectively located at $z=0.9$ or $z=1.5$, and that a  
microlens+fibers integral field spectrograph similar to GIRAFFE would  
be used to perform 3D spectroscopy on this galaxy.\newline

The first steps required to perform these simulations are the following:
\begin{enumerate}
	\item Use of the original $H\alpha$ image to simulate the distant  
galaxy image, taking into account the decrease of the apparent diameter  
due to redshift as well as the change of spatial sampling (see  
\citealt{Giavalisco-a-1996} for more information). We assumed a spatial  
sampling in the simulated galaxy image equal to half of the telescope's  
diffraction limit (Nyquist sampling), i.e. $\lambda/2D$.
	\item Computation of the apparent magnitude $m$ in the observing band  
(J or H) from nearby galaxy's absolute magnitude $M$, using distance  
modulus, and taking into account the \textit{morphological  
K-correction} $K(z)$ due to redshift into the distance modulus  
equation: $m-M=5\log (D_L/10)+K(z)$, where $D_L$ is the luminosity  
distance. We used values from \citet{Mannucci-a-2001} for the  
K-correction. The flux in the continuum $f_{cont}$ (in  
$ergs~s^{-1}cm^{-2}\AA^{-1}$) can then be computed from the apparent  
magnitude $m$.
	\item Computation of the total flux (in $ergs~s^{-1}cm^{-2}$) of the  
galaxy into the redshifted $H\alpha$ line, and normalisation of the  
simulated $H\alpha$ image.
	\item Simulation of the distant galaxy continuum image assuming pure  
Freeman exponential profile, taking into account the disk scale length  
$R_D$ and the angular distance $D_A$. The flux normalisation is done  
using $f_{cont}$.
\end{enumerate}
\begin{figure}
\centering
\begin{tabular}{ccc}
\includegraphics[width=0.2\linewidth]{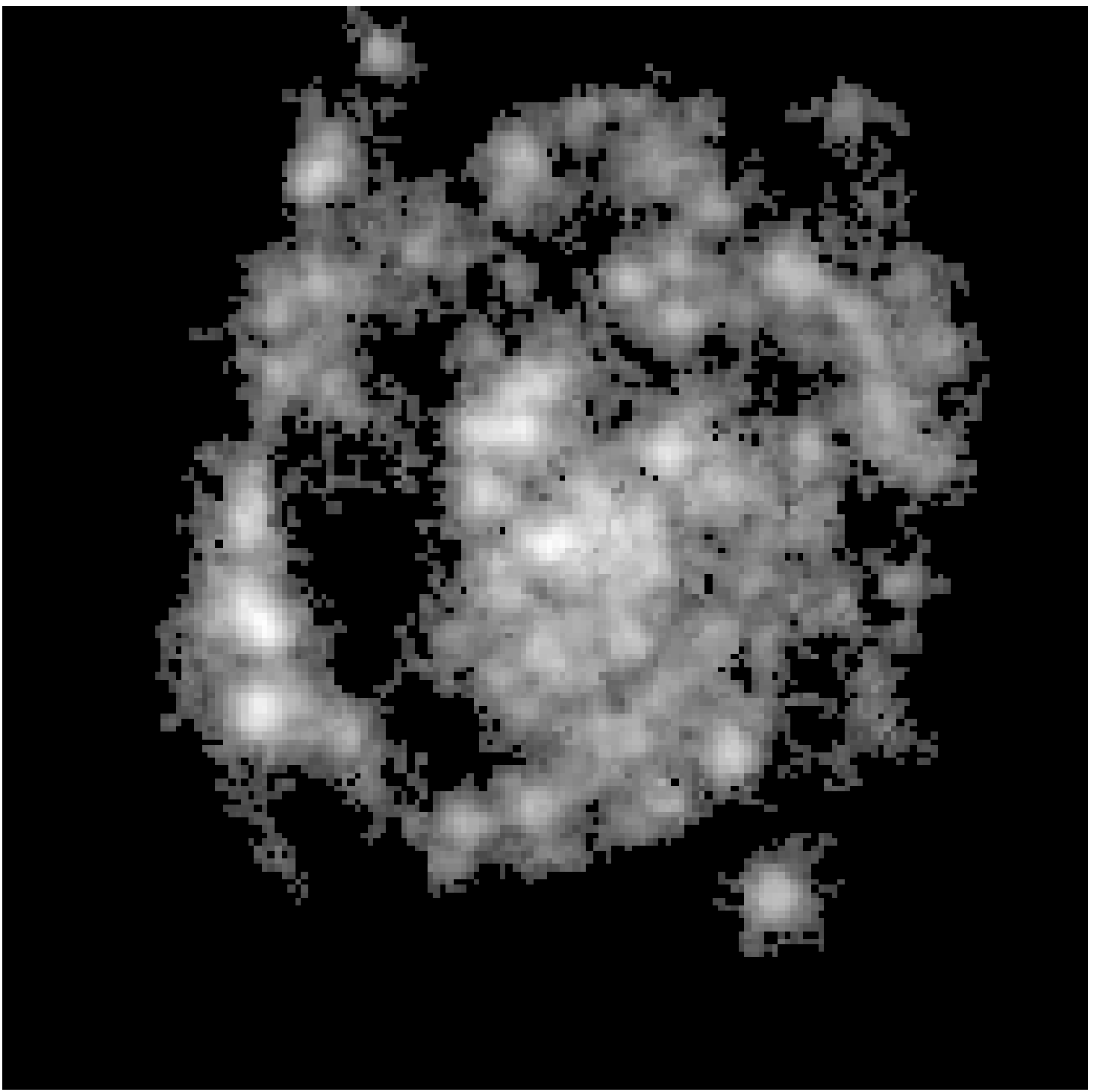}&\includegraphics[width=0.2\linewidth]{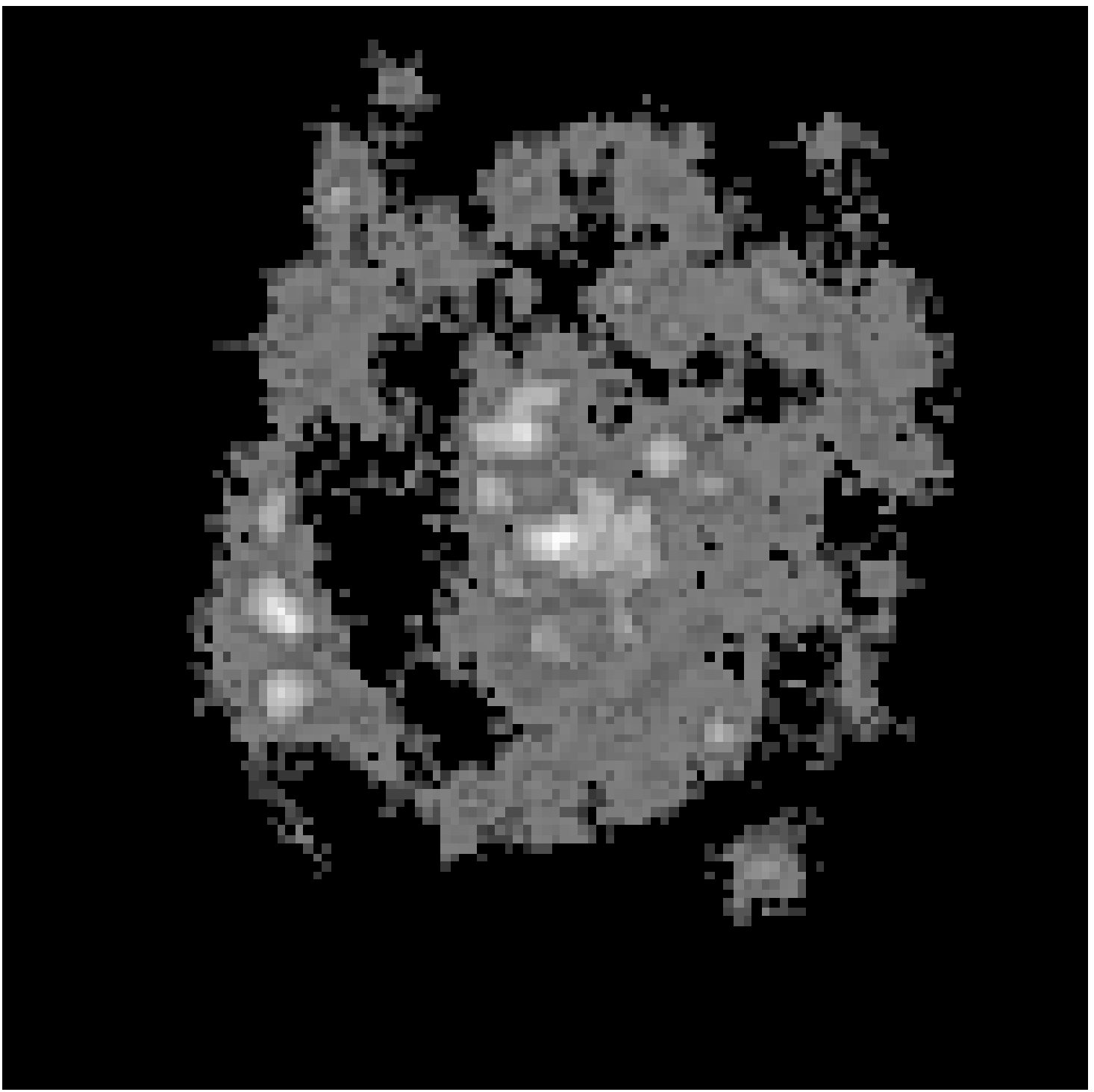}&\includegraphics[width=0.2\linewidth]{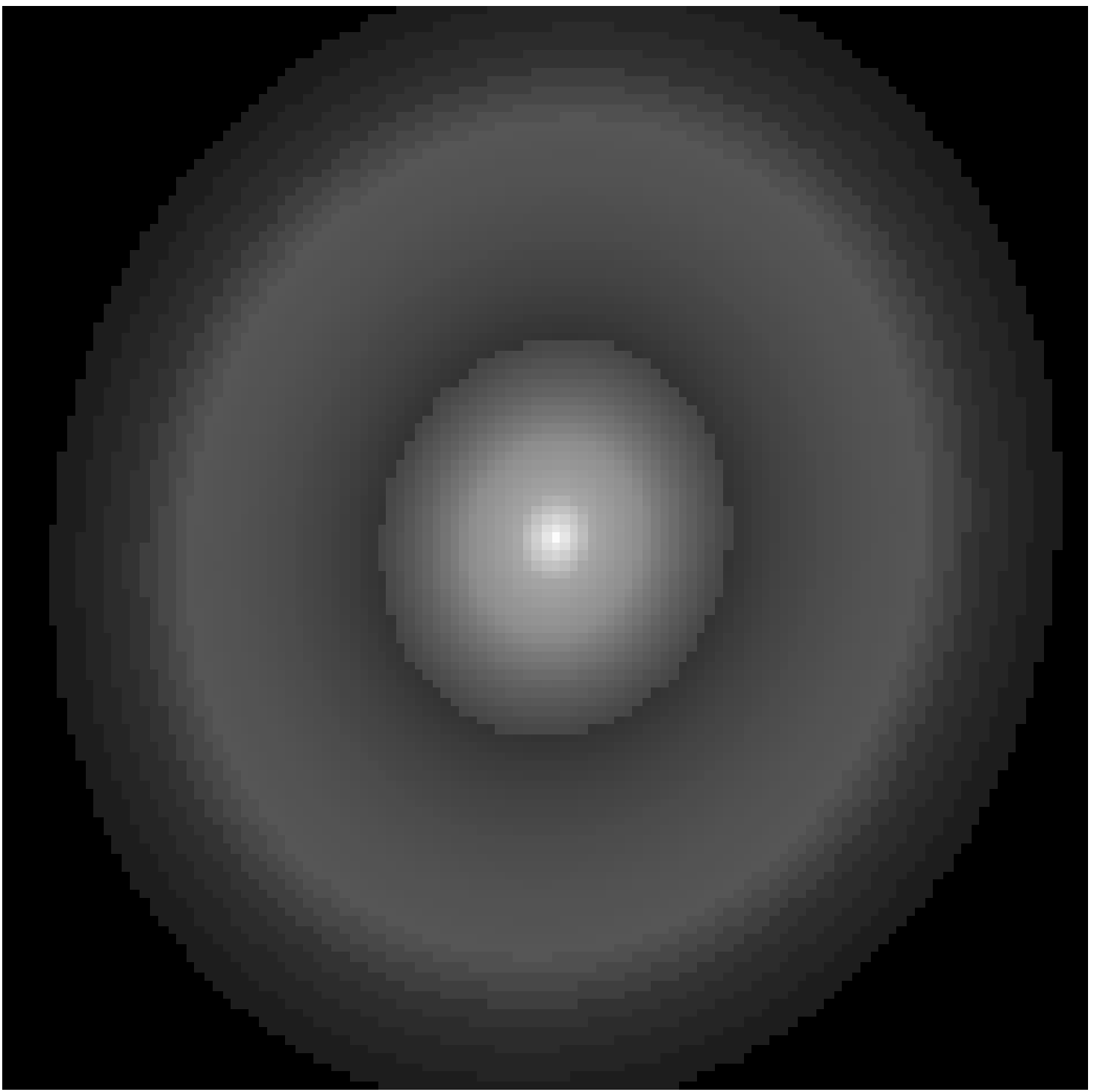}\\
Original $H\alpha$ image&Simulated $H\alpha$ image&Continuum map\\
$z\approx 0.0$, $\lambda=0.656~\mu m$&$z=0.9$, $\lambda=1.250~\mu m$&\\
$2.62 \times 2.62~arcmin^2$&$1.80 \times 1.80~arcsec^2$&\\
$s_0=0.96~arcsec$&$s_z=0.016~arcsec$&\\
\end{tabular}
\caption
{
Original $H\alpha$ image (left), simulated $H\alpha$ (middle) and  
continuum map (right) of the UGC 6778 galaxy as if it was observed at  
$z=0.9$ (in J band), with the same physical size than today. The  
simulated field is $1.80 \times 1.80~arcsec^2$.}
\label{fig:fig5}
\end{figure}

The figure \ref{fig:fig5} shows therefore the original $H\alpha$ image  
coming from the GHASP survey ($z\approx~0.0$, pixel size  
$s_0=0.96~arcsec$), as well as the simulated $H\alpha$ and continuum  
map corresponding to the first science case ($z=0.9$, pixel size  
$s_z=0.016~arcsec$).\newline

We are going now to focus on the two $HII$ regions on the bottom left  
of the $H\alpha$ image. \citet{Garrido-a-2002} have indeed shown that  
they have very similar velocities ($\approx 890$ km/s). We therefore do  
the assumption (sufficient for the rest of this study) that the  
integral field spectrograph we use has a sufficient spectral resolution  
$R$ so that the flux sampled around the wavelength $\lambda_0(1+v/c)$  
(where $\lambda_0$ is the observing wavelength, i.e. 1.25 or 1.65 $\mu  
m$) is the flux only emitted in the $H\alpha$ line by these two  
regions, $v$ being their average velocity. We therefore isolated these  
two regions in the simulated images we created.

However the simulated $H\alpha$ and continuum images suppose infinite  
spatial resolution, and do not take into account atmospheric seeing or  
AO partial correction effects.

\begin{figure}
\centering
\begin{tabular}{cccc}
\includegraphics[width=0.2\linewidth]{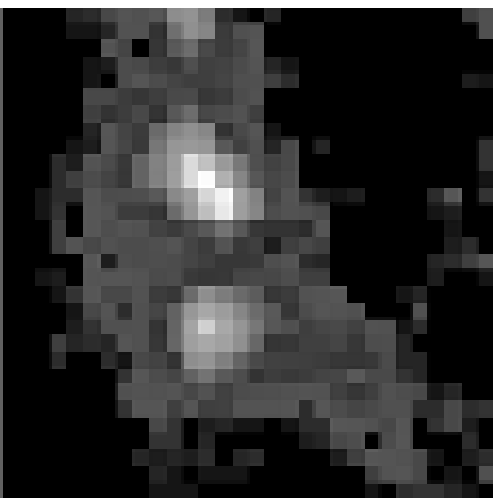}&\includegraphics[width=0.2\linewidth]{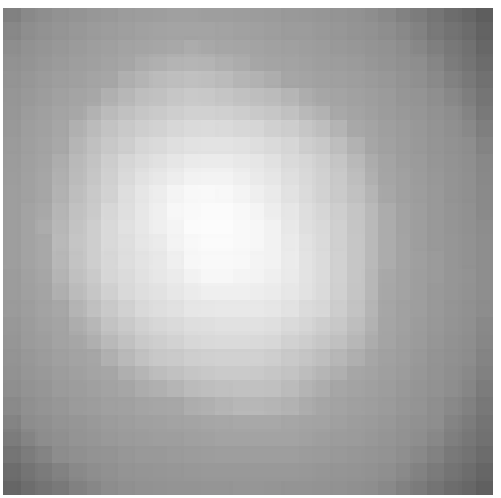}
\includegraphics[width=0.2\linewidth]{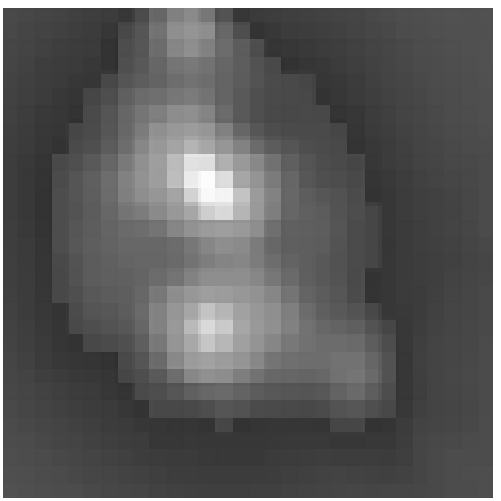}&\includegraphics[width=0.2\linewidth]{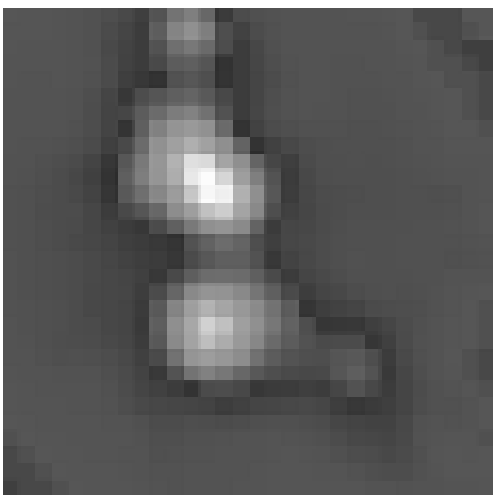}
\end{tabular}
\caption
{
Illustration of the effect of adaptive optics compensation on the images. The simulated images cover a field of $0.50\times  
0.50~arcsec^2$ ($4\times 4$ microlenses of $0.125\times  
0.125~arcsec^2$). A analysis has been done, that covers compensation orders from 0 to 120 Zernike modes. A few cases have been picked up and are presented here for illustration purpose.
From left to right:  
diffraction-limited image, seeing-limited image, AO-corrected image  
for 46 and 100 ZPs.}
\label{fig:fig6}
\end{figure}

We therefore convolved those images with  
different point spread functions (PSF) corresponding to
all the possible orders of AO correction, from 0 to 120 Zernike, at $1.25~\mu m$ and $1.65~\mu m$.
The principle of the simulation is the same
than in section \ref{sec:sec32}, and we assumed  
the same conditions (8m telescope, seeing of $0.81~arcsec$ at  
$\lambda=0.5\mu m$, no anisoplanatism, noise or temporal error).
As a consequence of this AO compensation, a better contrast in the images as
well as a better ensquared energy is observed, as already explained in section
\ref{whyAo3dSpec}. The result of the AO impact is illustrated, for a few cases, on figure \ref{fig:fig6}.\newline

Those images could then be used to compute the flux entering into the  
microlens sampling each $HII$ region, and then compute the final number  
of electrons on each pixel of the detector located at the output of the  
spectrograph, requiring therefore to know the spectral profile of the  
line. We assumed a gaussian line profile for the line, being the  
consequence of some internal velocity dispersion $\sigma_V$ in those  
$HII$ regions, as well as the finite spectral resolution $R$ of the  
spectrograph. As a result, the line profile on the spectrograph  
detector has a gaussian shape, with a standard deviation  
$\sigma_{tot}=\sqrt{\sigma_{V,\lambda}^2+\sigma_S^2}$,  
$\sigma_{V,\lambda}=\lambda_0 \times \sigma_V/c$ being the standard  
dispersion in wavelength due to the internal velocity dispersion of the  
$HII$ regions, and $\sigma_S=\lambda_0/2.354R$ being the standard  
deviation due to the spectral resolution, with $\lambda_0$ being the  
observing wavelength and $c$ the light celerity. Therefore, if we call  
$F_{ML}$ the total flux in HII region sampled by the spectrograph's  
microlens, the line's profile $f(\lambda)$ as a function of wavelength  
will be written:
\begin{equation}\label{eq:eq1}
f(\lambda)=\frac{F_{ML}}{\sigma_{tot}\sqrt{2\pi}}\exp{- 
\frac{\left(\lambda-\lambda_0\right)^2}{2\sigma_{tot}^2}}
\end{equation}

This line profile is then going to spread over two directions on the  
spectrograph's detector: the horizontal direction will correspond to  
the spectral information, and the perpendicular direction will  
correspond to the spatial information. This allows us to compute the  
number of electrons due to the $H\alpha$ line per pixel on the  
spectrograph detector, and then the spectroscopic SNR. If $\Delta  
\lambda$ is the spectral sampling of the detector, $S_{tot}$ the  
telescope area, $t_{exp}$ the exposure time, and $T_{tot}$ the total  
transmission of the system (which takes into account the detector  
efficiency), the number of electrons $N_{H\alpha}$ coming from the  
redshifted $H\alpha$ line on the pixel sampling the line at the  
$\lambda_0$ wavelength is then equal to:
\begin{equation}\label{eq:eq2}
N_{H\alpha}=2\int_0^{\Delta\lambda/2}f(\lambda)d\lambda \times S_{tot}  
\times t_{exp} \times T_{tot} \times \lambda_0/hc
\end{equation}

where $h$ and $c$ are respectively the Planck constant and the speed of  
light.\newline

We then have the flux from the line going onto the detector pixel.  
However, as we are interested in computing the spectroscopic SNR, we  
therefore need to take into account all the sources of noise, which can  
be divided in three categories: the line and continuum photon noise,  
the background photon noise, and the detector noise. As both line and  
continuum signals have poissonnian statistics, therefore their  
associated photon noise has a variance equal to their respective value,  
i.e. $N_S$ and $N_{cont}$, where  the number of electrons $N_{cont}$ due  
to the continuum is equal to:
\begin{equation}\label{eq:eq3}
N_{cont}=f_{cont} \times S_{tot} \times t_{exp} \times \Delta \lambda  
\times T_{tot} \times \lambda_0/hc
\end{equation}

$f_{cont}$ being the flux from the continuum entering the microlens  
after convolution of the continuum image by the AO corrected PSF.
Concerning the background noise, this latest is the sum of the sky  
background as well as the instrumental thermal background. Then, if we  
call $N_B$ the total number of electrons coming from the backgroud, as  
it has also Poisson statistics, its associated photon noise will then  
have a variance equal to $N_B$. At last, detector noises, i.e. dark  
current noise and readout noise, have to be considered.\newline

Knowing all these values, we can then compute the SNR per spatial  
resolution element and per spectral pixel, which will be written:
\begin{equation}\label{eq:eq4}
SNR=\frac{N_{H\alpha}}{\sqrt{N_{H\alpha}+N_{cont}+N_B+N_{pix,spat}(N_D+N 
_{exp}\sigma_R^2)}}
\end{equation}

with:
\begin{itemize}
	\item $N_B$ the number of electrons coming from the instrumental and  
sky background
	\item $N_D=DC \times T_{exp}$ the number of electrons dues to the dark  
current DC (in $e^-/s/pix$)
	\item $N_{exp}$ the number of elementary exposures
	\item $\sigma_R$ the readout noise (in $e^-/pix$)
	\item $N_{pix,spat}$ the number of pixels on which the flux is spread  
in the spatial direction on the detector\newline
\end{itemize}

The equation \ref{eq:eq4} therefore gives the expression SNR on the  
measured quantities. However, the figure \ref{fig:fig6} shows that as  
the AO correction improves, the image contrast becomes better, with an  
increase of the effective light coming from the $HII$ region. As a  
result, the flux going into the microlens sampling each HII region is  
therefore the sum of two terms: the effective flux emitted by the $HII$  
region, and a \textit{pollution} term coming from adjacent regions,  
which is going to degrade the spectrum of the area sampled by the  
microlens, provided as an output of the 3D spectrograph. This has  
leaded us to define another signal-to-noise ratio, which we will call  
\textit{Effective signal-to-noise ratio} (ESNR), and which gives a  
measure of the effective gain provided by the coupling of AO with 3D  
spectroscopy. Its expression is:
\begin{equation}\label{eq:eq5}
ESNR=\frac{N_{H\alpha,eff}}{\sqrt{N_{H\alpha}+N_{cont}+N_F+N_{pix,spat}( 
N_D+N_{exp}\sigma_L^2)}}
\end{equation}

\begin{figure}
\centering
\begin{tabular}{cccc}
\includegraphics[width=0.2\linewidth]{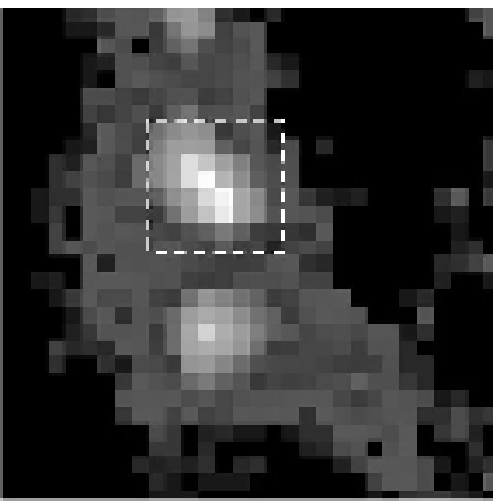}&\includegraphics[width=0.2\linewidth]{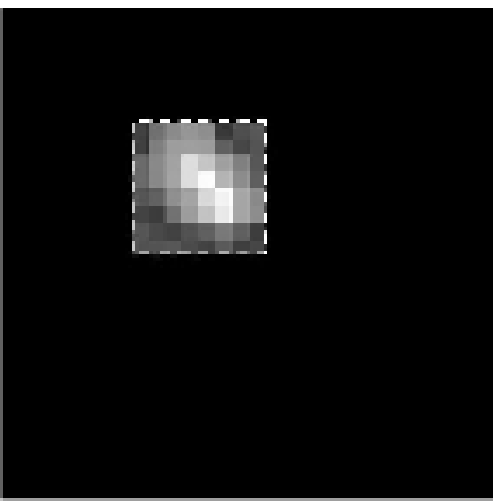}
\includegraphics[width=0.2\linewidth]{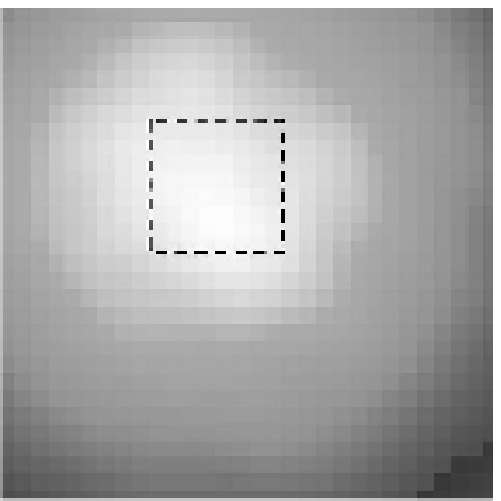}&\includegraphics[width=0.2\linewidth]{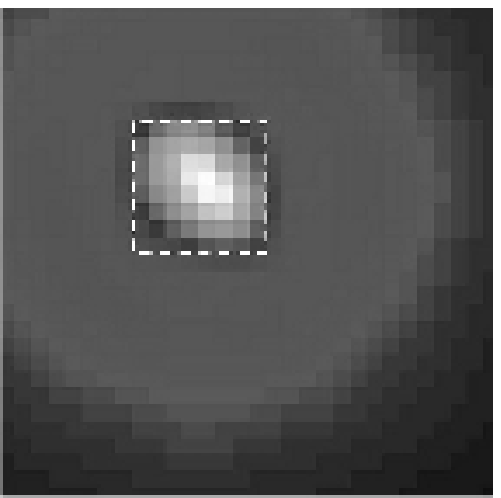}
\end{tabular}
\caption
{
Explanation of the \textit{effective flux} notion. The first image on  
the left shows the original image, with the two $HII$ regions and the  
superposed microlens (in dashed line). The second image shows only the  
area sampled by the microlens. The third and fourth show this latter  
convolved by the uncorrected PSF (third) and by an AO-corrected PSF  
(fourth image, correction of 60 ZPs). The effective flux from the  
$H\alpha$ line is then equal to the flux of the region after  
convolution by the PSF going into the microlens.}
\label{fig:fig7}
\end{figure}

where $N_{H\alpha,eff}$ is the number of electrons due to the effective  
$H\alpha$ flux emitted by the area sampled by the microlens after  
convolution by the AO corrected PSF, with no contribution from the  
adjacent areas around the microlens. This notion of \textit{effective  
flux} is explained on the figure \ref{fig:fig7}.\newline

One will notice that denominators in the equations \ref{eq:eq4} and  
\ref{eq:eq5} are the same, and therefore that the ESNR is a lower bound  
of the SNR. This is explained by the fact that in the case of the ESNR,  
all the noise sources must appear in the denominator. Concerning the  
line, this latter is perturbed by a noise, which is the sum of two  
terms: the photon noise of the effective flux, and a \textit{pollution}  
term due to all the light coming from the area around the microlens and  
entering the microlens because of the convolution. As both quantities  
have poissonnian statistics, and as the convolution operator is  
distributive, the sum of these two terms is equal to the photon noise  
of the measured light entering the microlens.\newline

We would however like to emphasize here that the ESNR defined  
at the equation \ref{eq:eq5} and the classical SNR definition shown in  
equation \ref{eq:eq4} cannot be directly compared, in particular when  
the ensquared energy changes. As it is shown in figure \ref{fig:fig7},  
the ESNR numerator is the flux coming \textbf{only} from the region  
sampled by the microlens after convolution by the AO corrected PSF,  
whereas the SNR numerator is the flux not only coming the microlens,  
but also the flux from adjacent areas and polluting the microlens  
because of the convolution. As different images are convolved for ESNR  
and SNR computations, there is no direct relation (i.e. multiplication)  
between these numbers. Moreover, as explained in the section  
\ref{whyAo3dSpec}, an increase of the ensquared energy is due to a  
better compensation order provided by the AO system. Thus different  
ensquared energy values correspond to different AO corrected PSFs, and  
the results of the convolution of the same high-resolution galaxy image  
by these different PSFs also cannot directly be comparable.\newline

Once these two different quantities (SNR and ESNR) defined, we  
performed some numerical simulations in order to quantify the gain in  
ESNR provided by the combination of AO with 3D spectroscopy. The  
results of these simulations are shown in the next section.

\subsection{Results}\label{sec:sec34}
Thanks to simulations, we are now able to show some preliminary results  
about the improvement in ESNR thanks to AO. Two science cases,  
corresponding to the observation of UGC 6778 at two different  
redshifts, are considered in the two next subsections.\newline

For the first one, we assumed that the observed galaxy would be located  
at $z=0.9$. In that case, the $H\alpha$ emission line is redshifted at  
a wavelength of $1.25~\mu m$, corresponding to the central wavelength  
of J band. Moreover we assumed that the galaxy had the same properties  
than the distant large spirals observed in the CFRS  
\citep{Lilly-a-1998,Zheng-a-2004}. In particular the physical size of  
the simulated galaxy is the same than the one observed in the local  
universe. As there are no direct measurements of $H\alpha$ fluxes  
available in the litterature for CFRS galaxies at these redshift, we  
therefore computed the $H\alpha$ luminosity from $[OII]3727$ fluxes  
\citep{Hammer-a-1997}, in order to normalise the simulated  $H\alpha$  
image.\newline

For the second one, we assumed that the observed galaxy would be  
located at $z=1.5$. In that case, the $H\alpha$ emission line is  
redshifted at a wavelength of $1.65~\mu m$, corresponding to the  
central wavelength of H band. Moreover we assumed that the galaxy had  
$H\alpha$ fluxes similar to the ones observed on the population of "`BM  
galaxies"' introduced by \citet{Steidel-a-2004}. The apparent size of  
the galaxy was then chosen to cover a field of $1~arcsec^2$, consistent  
with apparent sizes observed on distant galaxies at these redshifts in  
deep surveys \citep{Bouwens-a-2004b}.\newline

We focus now on the two $HII$ regions we used to compute the SNR and  
ESNR. Together they contribute at $10\%$ of the total $H\alpha$ flux in  
the simulated $H\alpha$ image. We also assumed that they had an  
internal velocity dispersion with $FWHM_V=30~\mathrm{km.s^{-1}}$ (eg  
$\sigma_V \approx 13~\mathrm{km.s^{-1}}$), leading as explained  
previously to a dispersion in wavelength on the spectrograph's detector  
equal to $\sigma_{\lambda,V}=\lambda_0 \times \sigma_V / c$,  
$\lambda_0$ being equal to 1.25 or 1.65 $\mu m$. The spectral  
resolution was then fixed to $R=10000$, allowing to resolve the  
velocity dispersions of the $HII$ regions. This then allowed us to  
define two cases of spectral sampling, corresponding to the horizontal  
direction on the spectrograph's detector:  $\Delta\lambda=0.3\AA / pix$  
(for J band) or $\Delta\lambda=0.6\AA / pix$ (for H band). Such  
spectral samplings allow us to sample the spectral resolution with more  
than two pixels, thus providing a better sampling than Nyquist  
sampling, and then do a precise estimation of the barycenter of the  
$H\alpha$ emission line in order to recover the galaxy's velocity  
field. For the vertical direction on the spectrograph's detector,  
corresponding to the spatial information, we considered a number of  
pixels $N_{pix,spat}=3$, similar to the number of pixels used on the  
GIRAFFE spectrograph's detector. Such a number of pixels, here again  
better than Nyquist sampling, is quite convenient to resolve spectrum  
deblending problems on the spectrograph's detector.\newline

For the sky background, we assumed that the spectrograph is observing  
between atmospheric OH lines, so the sky noise is equal to the sky  
continuum. \citet{Cuby-a-2000} has measured the sky continuum at the  
Cerro Paranal observatory, and found respective values of 1200  
$ph/s/m^2/\mu m/arcsec^2$ at a wavelength of 1.25~$\mu m$, and 2300  
$ph/s/m^2/\mu m/arcsec^2$ at a wavelength of 1.65~$\mu m$. We used the  
characteristics of the VLT's ISAAC spectrograph to simulate the  
instrument: the instrumental background was assimilated to a blackbody  
of temperature $T=273K$ and an emissivity equal to $25\%$, a total  
transmission (atmosphere+telescope+instrument+detector) of $13\%$ at  
$1.25~\mu m$  (central wavelength of J band) and $19\%$ at $1.65 \mu m$  
(central wavelength of H band) was assumed, and the detector had a dark  
current $DC=30e-/h/pix$ and a readout noise $RON=10~e-/pix$.\newline

\subsubsection{First science case: 3D spectroscopy of UGC 6778 at $z=0.9$}
Here we give the results of simulations of the 3D spectroscopy of UGC  
6778 as if it was observed at $z=0.9$. We assumed that the galaxy had  
the same physical size as it has now. However the flux in the $H\alpha$  
line is consistent with the one observed of distant galaxies. Such  
physical conditions correspond to the ones observed on large spirals in  
the distant universe \citep{Lilly-a-1998,Zheng-a-2004}.\newline

The original $H\alpha$ image covers a field of $2.62 \times  
2.62~arcmin^2$, and the galaxy has a distance $D=17.04$ Mpc, which  
leads to a physical field of $d_1 \times d_1~kpc^2$, with  
$d_1=13.40~kpc$. Now, if we assume that the same galaxy has a redshift  
$z$, we can then compute its luminosity-distance $D_L$ and its angular  
distance $D_A$, and we found that the observed field has therefore an  
apparent size equal to $\theta_z \times \theta_z~arcsec^2$, with  
$\theta_z=d_1/D_A=1.80~arcsec$. Moreover, the spatial sampling is not  
the same as we simulate the galaxy as if it was observed by the VLT (8  
meter diameter telescope) at its diffraction limit, meaning the pixel  
sampling $s_z$ is equal to Nyquist sampling ($\lambda/2D$), leading to  
a pixel size $s_z=0.016~arcsec$. Therefore the number of pixels in the  
original and simulated images are not the same: as shown in figure  
\ref{fig:fig5}, the original image has $164\times 164$ pixels, whereas  
the simulated image has $111\times 111$ pixels.

This simulated image must be then normalized so that its flux is equal  
to the flux in the $H\alpha$ line. As explained before, there are not  
direct measurements of $H\alpha$ fluxes on the CFRS galaxies, therefore  
we used $[OII]3727$ luminosities to compute $H\alpha$ fluxes.  
\cite{Hammer-a-1997} found that the average rest-frame equivalent width  
of the $[OII]3727$ line in CFRS galaxies located at $z=0.9$ was equal  
to $EW([OII]3727)=40\AA$. Moreover, \citet{Kennicutt-a-1992b} found  
that there is a relation between the equivalent width in the  
$[OII]3727$ and $H\alpha$ line, in the form $EW(H\alpha)=2.5 \times  
EW([OII]3727)$. We therefore assumed that the $H\alpha$ line had a rest  
frame equivalent width $EW(H\alpha)=100\AA$.

Once $EW(H\alpha)$ is known, the flux in the continuum $f_{cont}$ must  
be computed to know the flux $f_{H\alpha}$ in the $H\alpha$ line, as  
both are linked to the relation: $f_{H\alpha}=f_{cont} \times  
EW(H\alpha) \times (1+z)$. $f_{cont}$ is directly linked to the  
apparent magnitude of the galaxy. We see on the table \ref{tab:table1}  
that UGC 6778 has an absolute magnitude
in B band $M_B=-20.6$. However, as the $H\alpha$ line is redshifted in  
J band, we are interested by the galaxy's absolute magnitude in J band. For a  
SBc galaxy like UGC 6778, \cite{Mannucci-a-2001} give the following  
colours: $(B-V)=0.70$, $(V-K)=3.03$, $(J-H)=0.66$ et $(H-K)=0.25$. We  
then find an absolute rest-frame magnitude $M_J=-23.42$.  
\cite{Mannucci-a-2001} give also the following \textit{K-correction}  
for a Sc galaxy: $K(z=0.9)=0.155$. The redshift $z$ can then be used to  
compute the luminosity distance $D_L$ and the distance modulus. We  
therefore find an apparent magnitude $J=20.41$, corresponding to a flux  
in the continuum $f_{cont}=8.09\times 10^{-19}~ergs/cm^2/s/\AA$,  
leading to a $H\alpha$ flux of $4.17 \times 10^{-17}~erg/s/cm^2$. This  
flux was then used to normalise the simulated $H\alpha$ image.\newline

As explained before, the continuum emission has also to be taken into  
account. This latter is not equally spread over the galaxy, as the  
intensity of disk in spiral galaxies follows a decreasing exponential  
law \citep{Freeman-a-1970}, in the form $I(r) \propto \exp^{-r/R_d}$,  
$R_d$ being the disk scale length, whose value for UGC 6778 is given in  
table \ref{tab:table1}. It is therefore possible to compute $R_{D,z}$,  
the angular size of of the disk scale length when the galaxy is  
observed at a redshift $z$, by the relation $R_{D,z}=R_D/D_A$, $D_A$  
being the angular diameter distance, and we found a value of  
$R_{D,z}=0.25~arcsec$. This value, as well as the position angle $PA$  
of the galaxy and the inclination $i$ (which are given in the table 1),  
were used to compute the continuum map of the galaxy. This latter was  
then normalised so that its total flux was equal to $f_{cont}$. The  
simulated $H\alpha$ image as well as the continuum map are shown on the  
figure \ref{fig:fig5}.\newline

\begin{figure}
	\centering
	\begin{tabular}{c}
		\includegraphics[height=0.3\textheight]{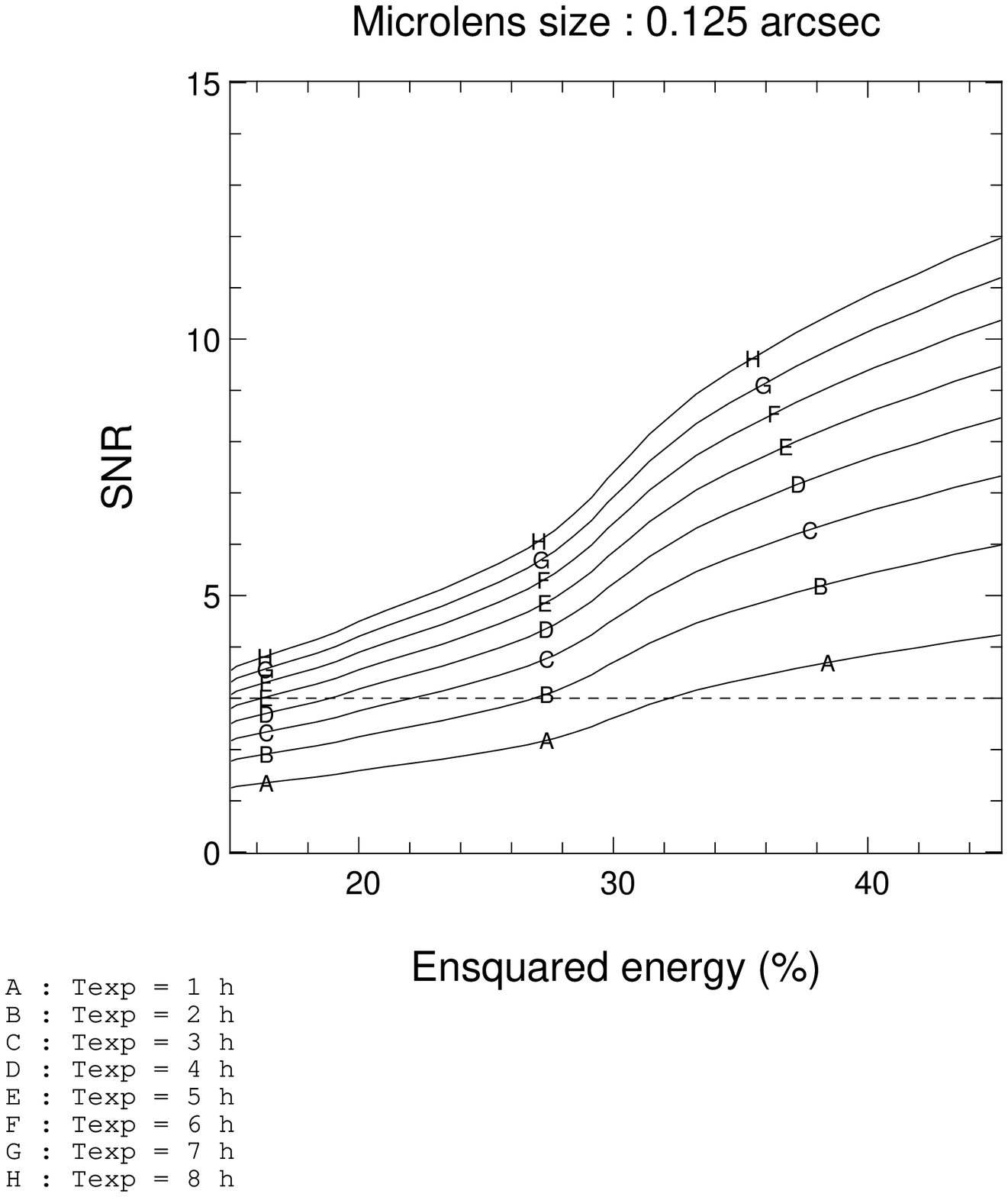}\\
		\includegraphics[height=0.3\textheight]{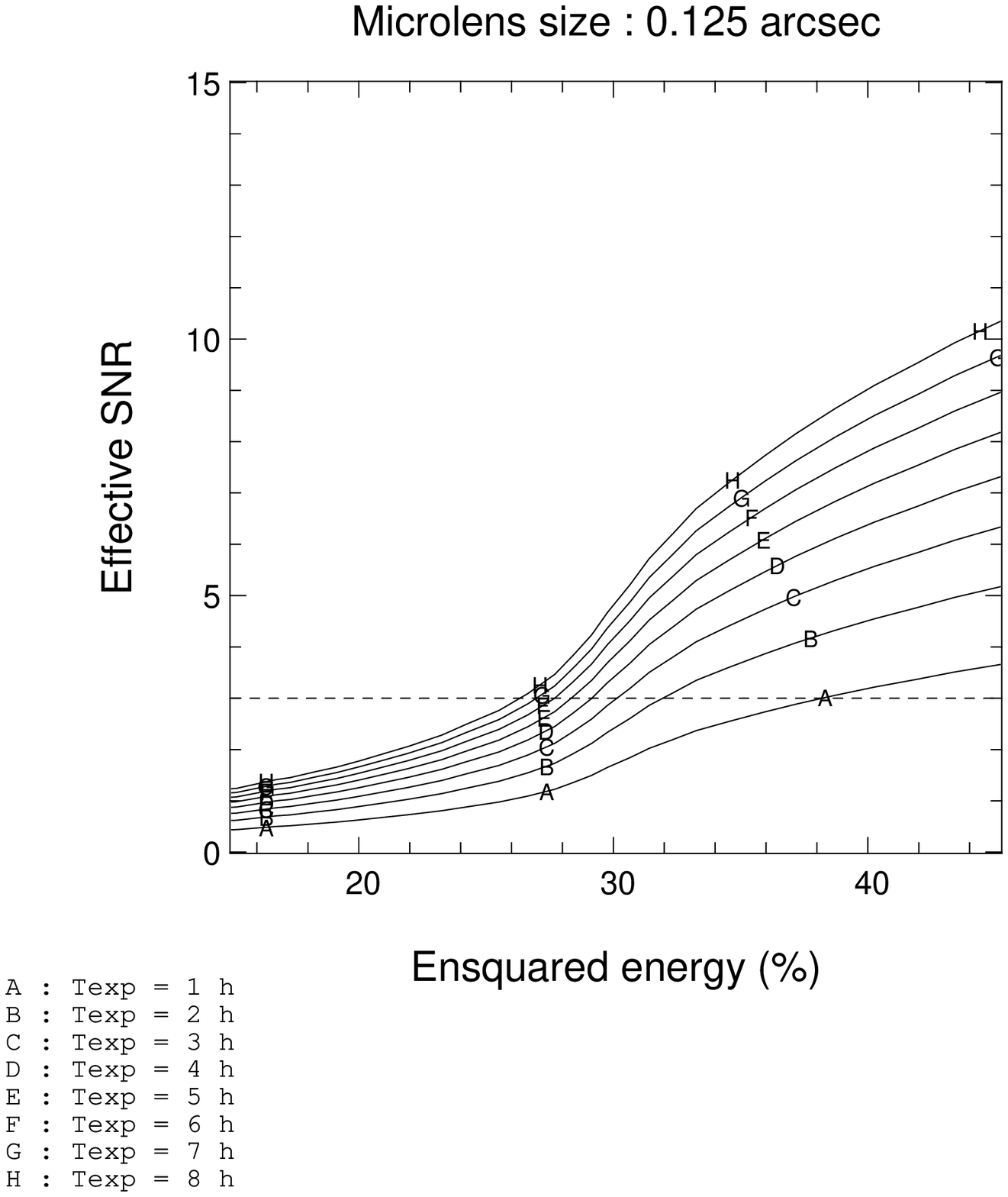}\\
	\end{tabular}
\caption{Measured spectroscopic SNR (top) and effective spectroscopic  
SNR (ESNR, bottom) on the $H\alpha$ line for a microlens of  
$0.125~arcsec$ as a function of the ensquared energy in a $0.25\times  
0.25~arcsec^2$ square aperture. The considered galaxy is located at  
$z=0.9$ (observation in J band) and has the same characteristic of  
distant large spirals. Each curve corresponds to the average of the SNR  
of each microlens sampling the HII regions, microlens, with exposure  
times going from 1 hour to 8 hours. The dashed line corresponds to  
$SNR=3$.}
\label{fig:fig8}
\end{figure}

Once those two images created, it is possible to compute the measured  
spectroscopic and effective SNR per spectral pixel. However, as our  
goal is to show the gain given by the combination of AO with 3D  
spectroscopy, we convolved these latter images by simulated AO  
corrected PSFs, with a degree of correction going from 0 ZPs  
(seeing-limited PSF) to 120 ZPs, leading to an increase of the  
ensquared energy as the number of corrected modes increases. We also  
considered exposure times between 1 and 8 hours, with individual  
exposure times of 1 hour. The figure \ref{fig:fig8} shows the evolution  
of these SNRs as a function of the ensquared energy into a $0.25 \times  
0.25~arcsec^2$ square aperture. More precisely, if we look at the  
evolution of the ESNR which \textit{in fine} says the real improvment  
provided by AO correction, we see that a minimum ensquared energy of  
$26\%$ allows to reach an ESNR of 3 after an exposure time of 8 hours.  
However, a slightly increase to an ensquared energy of $30\%$ allows to  
reach the same ESNR value, but only after 3 hours of exposure  
time.\newline

\begin{figure*}
	\centering
	\begin{minipage}{200mm}	
	\begin{tabular}{ccc}
		\includegraphics[width=0.25\textwidth]{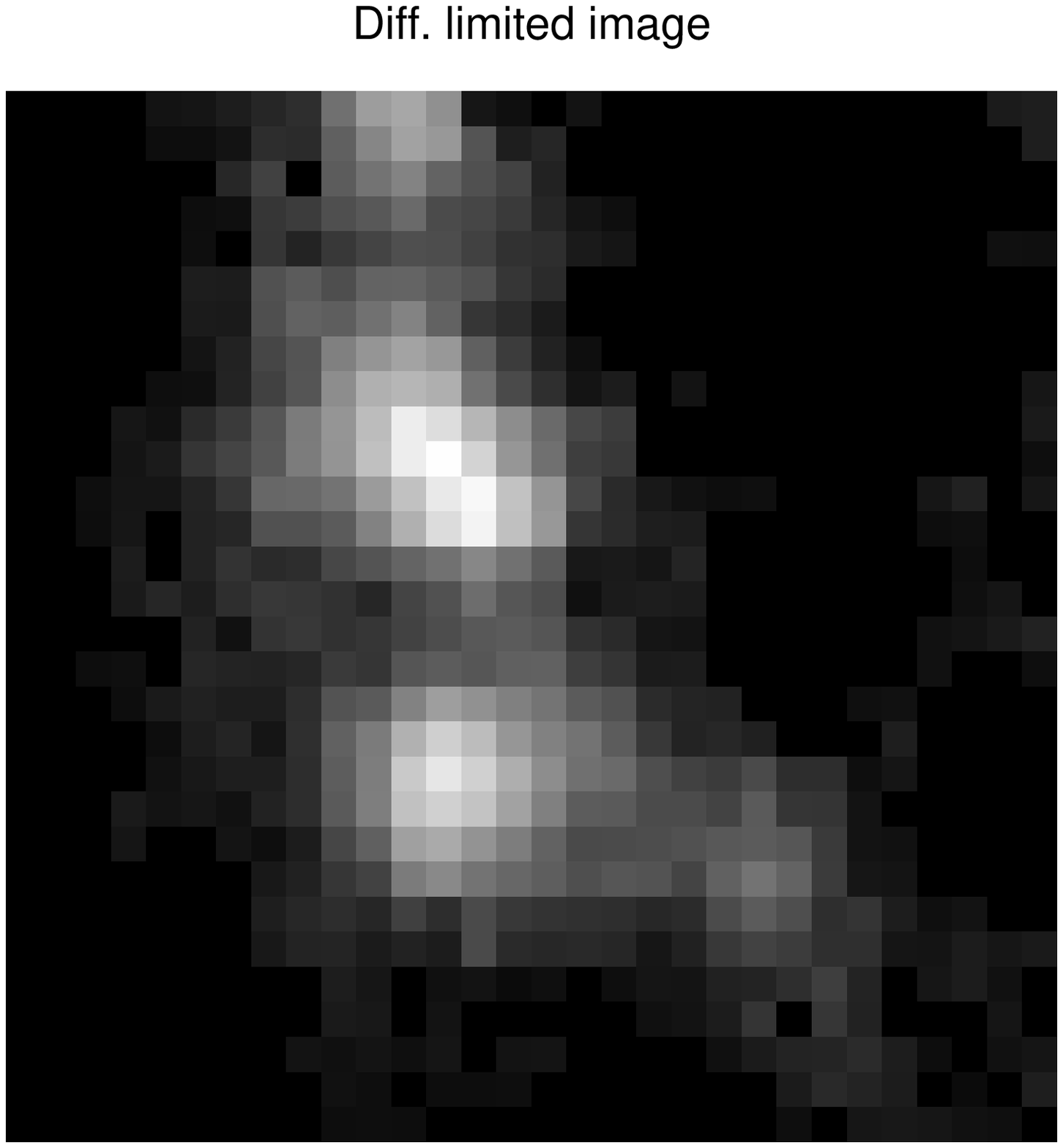}&
		\includegraphics[width=0.25\textwidth]{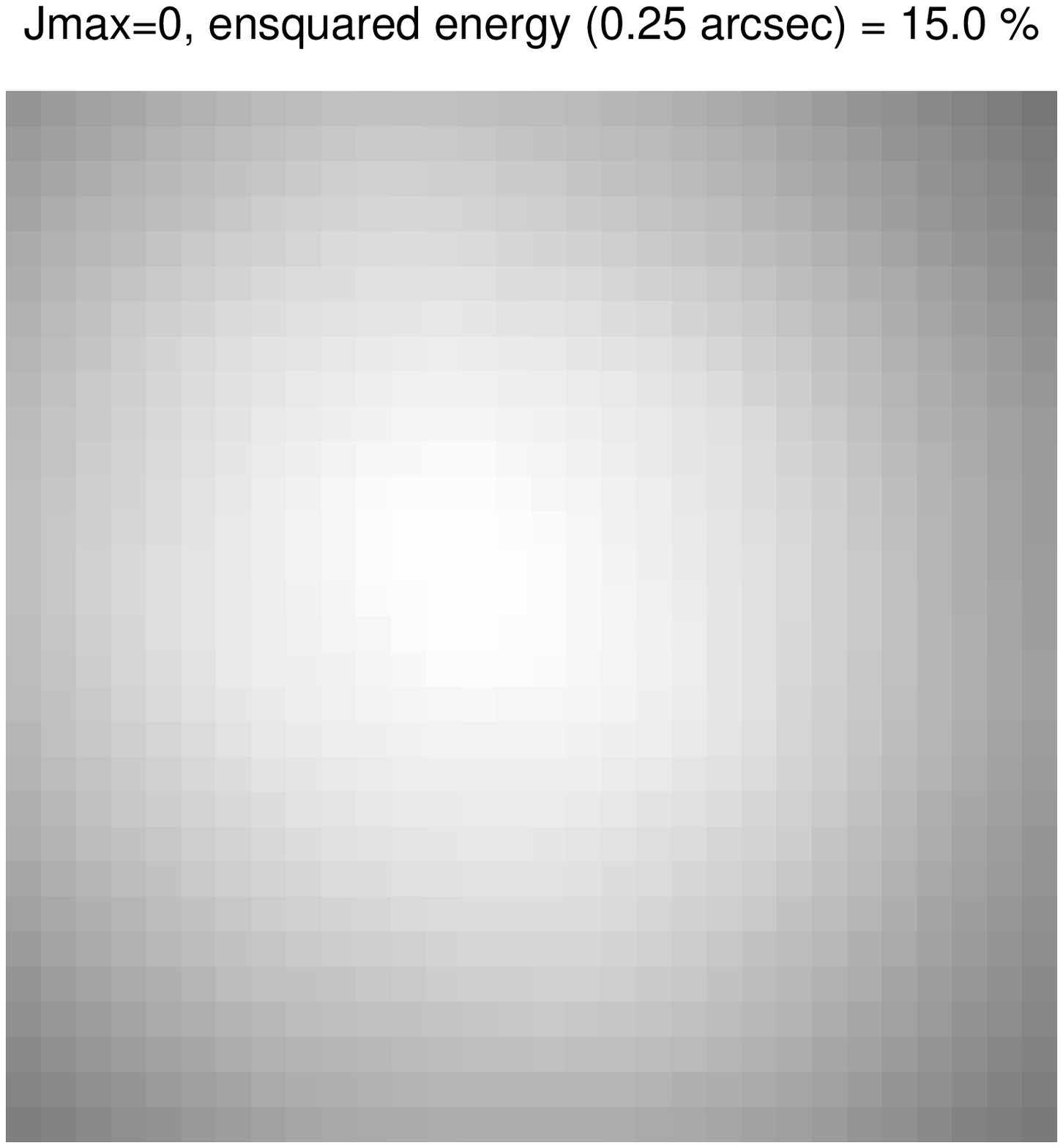}&
		\includegraphics[width=0.25\textwidth]{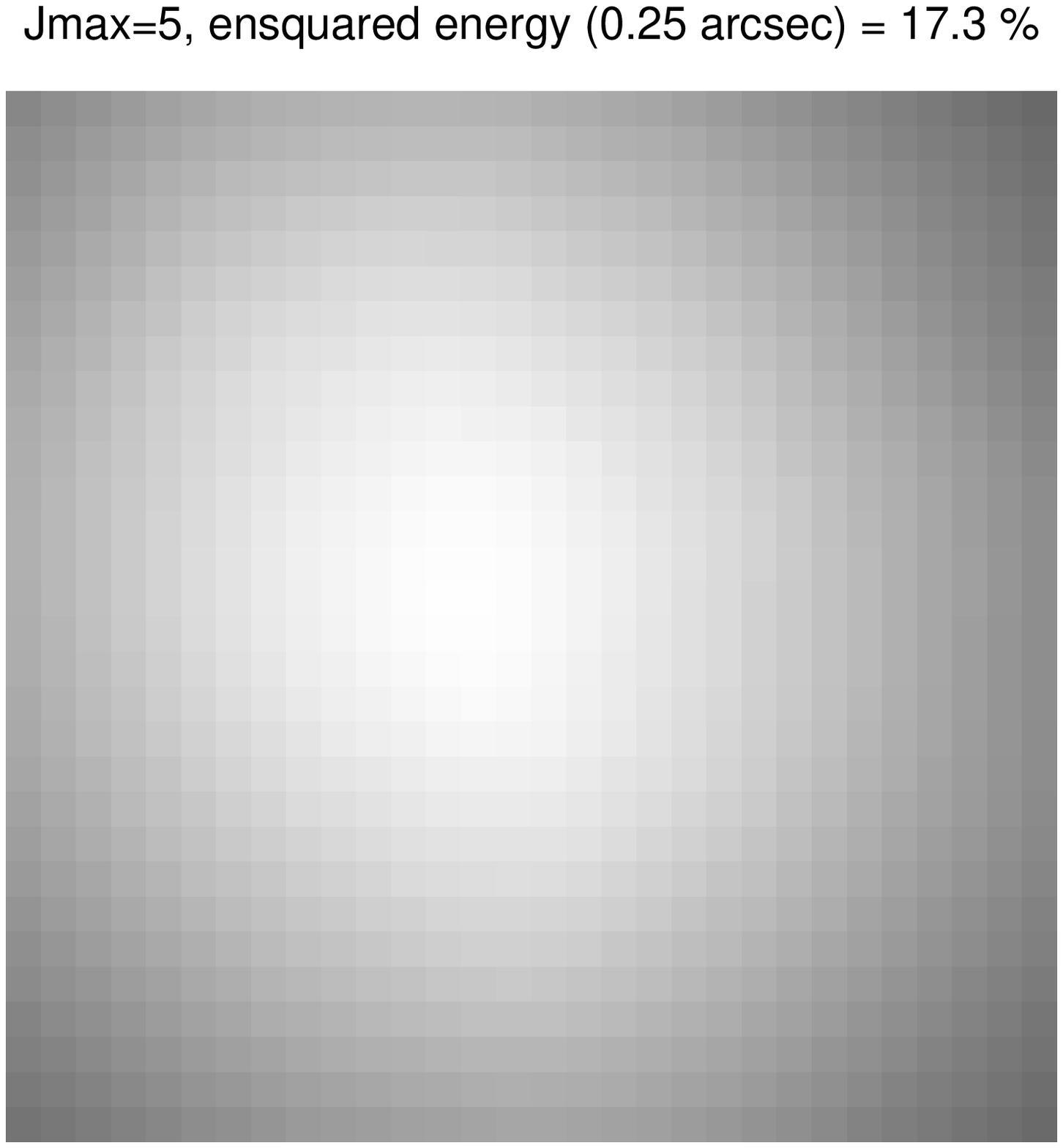}\\
		\includegraphics[width=0.25\textwidth]{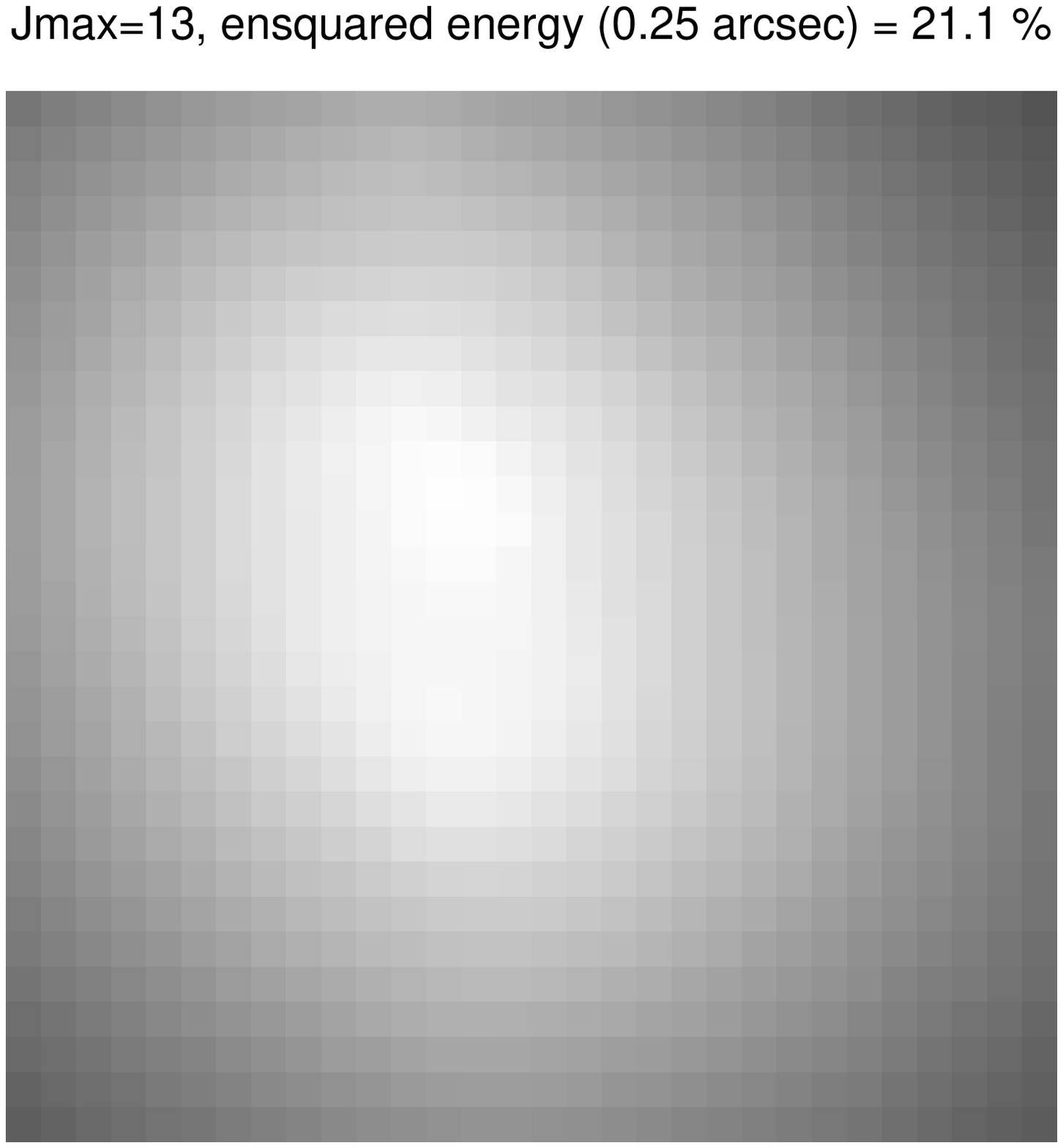}&
		\includegraphics[width=0.25\textwidth]{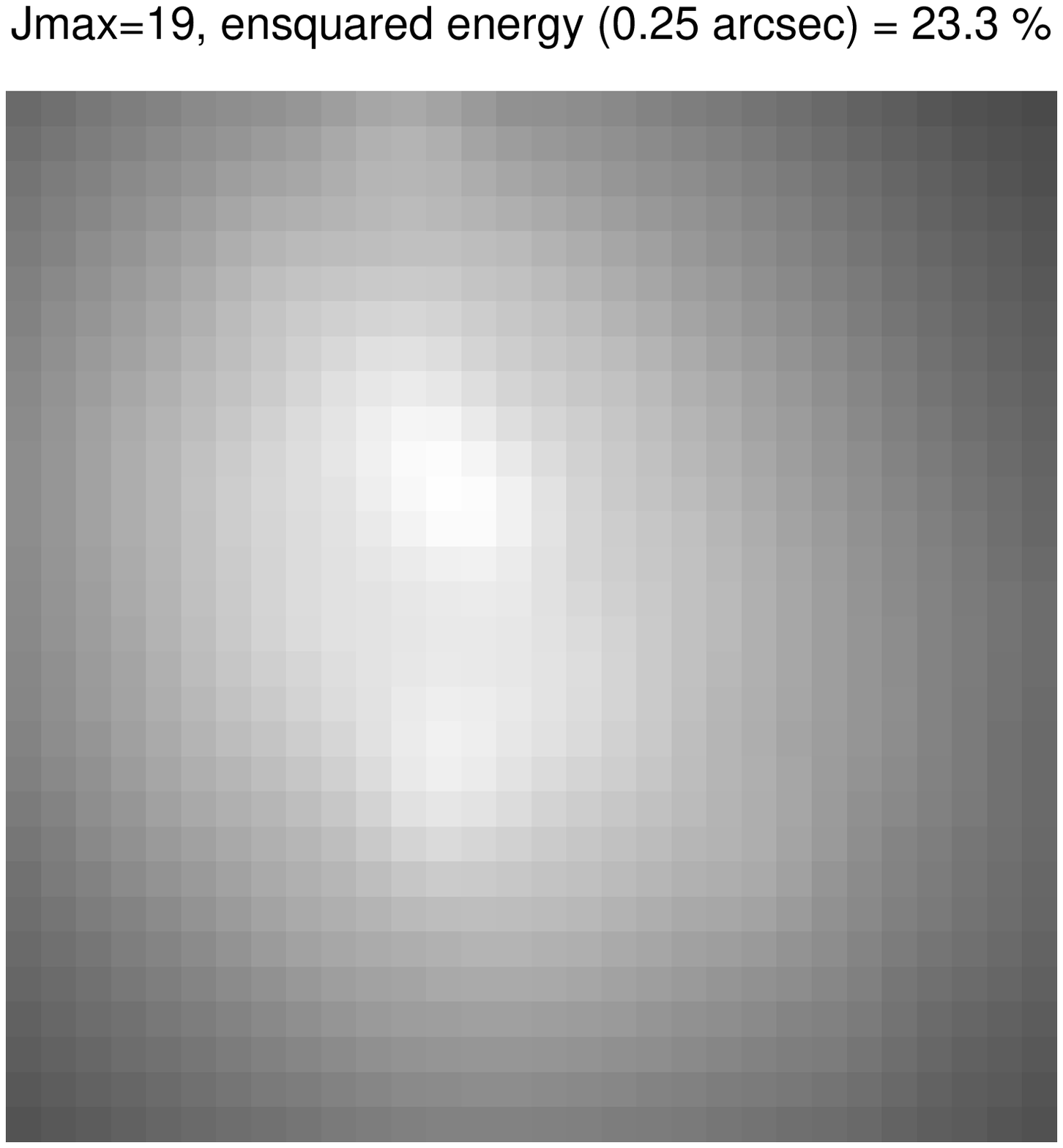}&
		\includegraphics[width=0.25\textwidth]{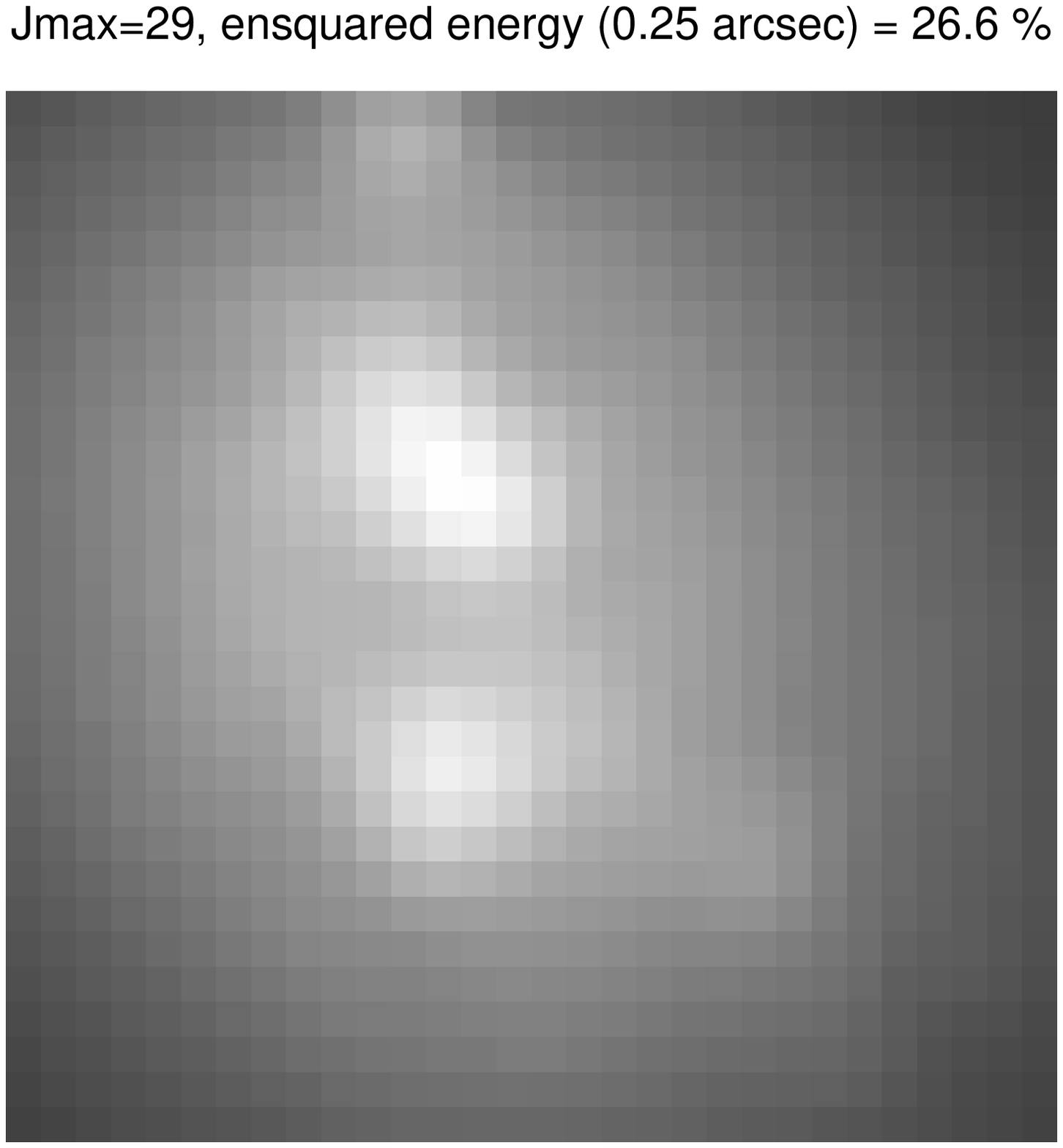}\\
		\includegraphics[width=0.25\textwidth]{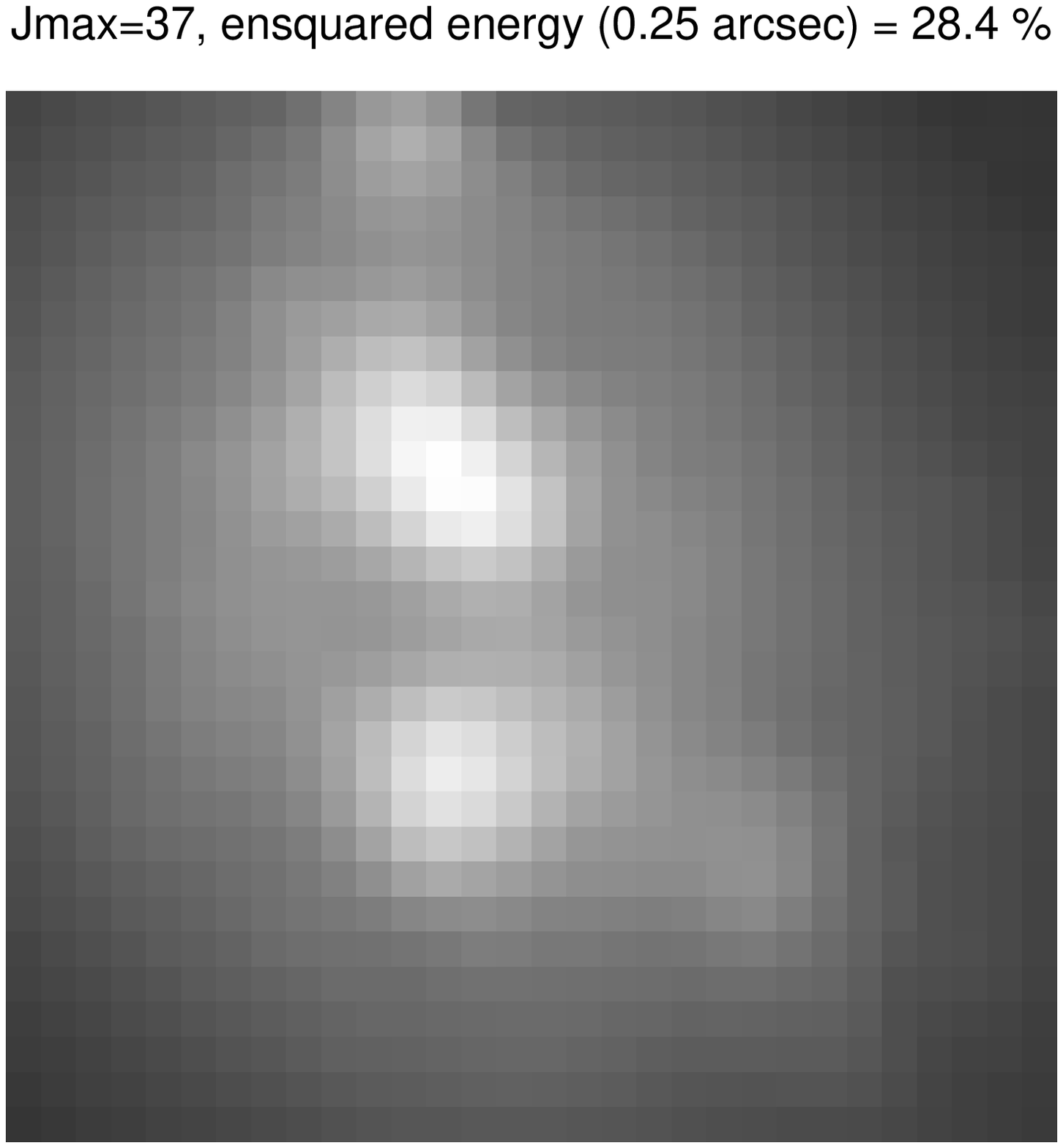}&
		\includegraphics[width=0.25\textwidth]{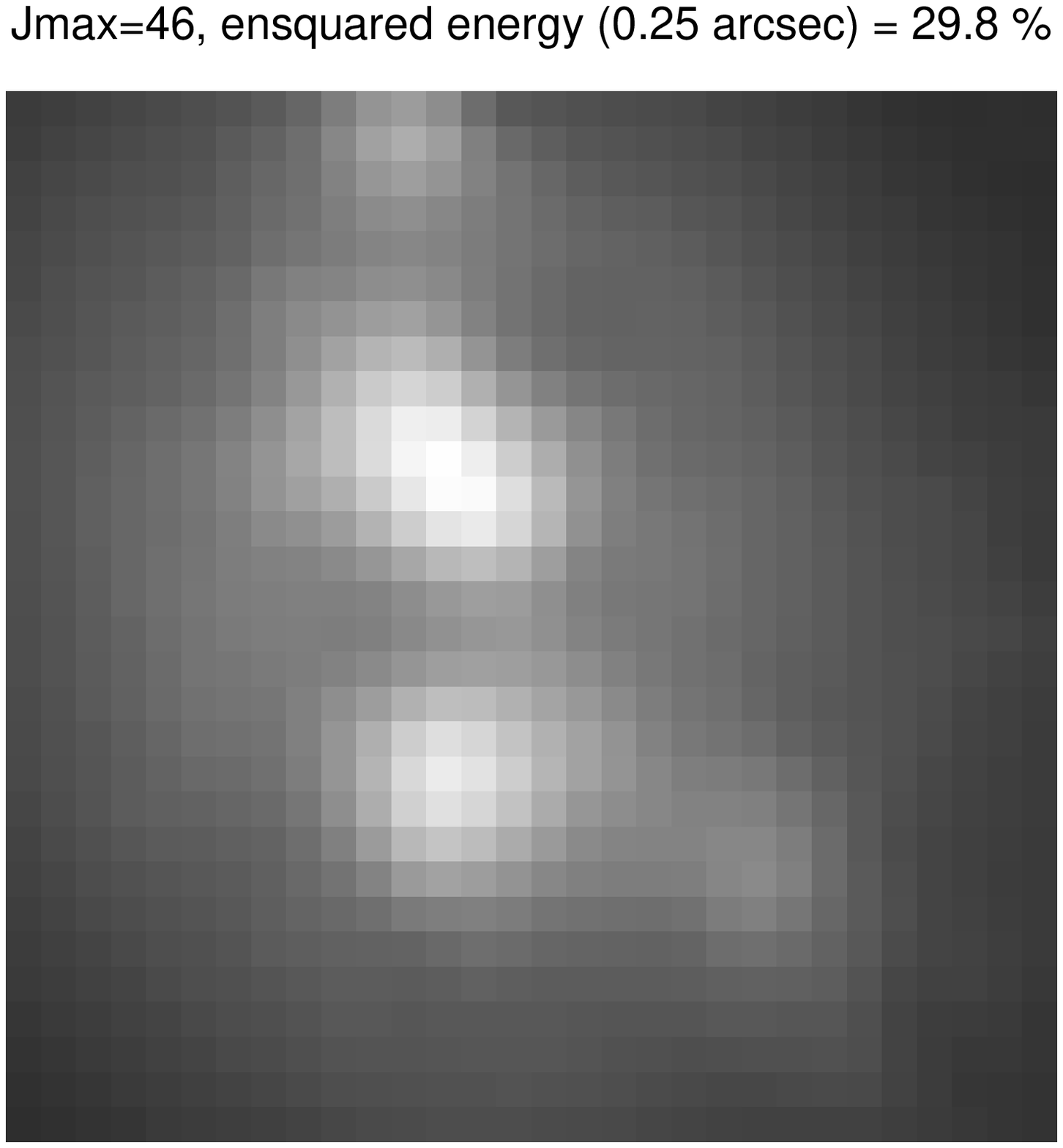}&
		\includegraphics[width=0.25\textwidth]{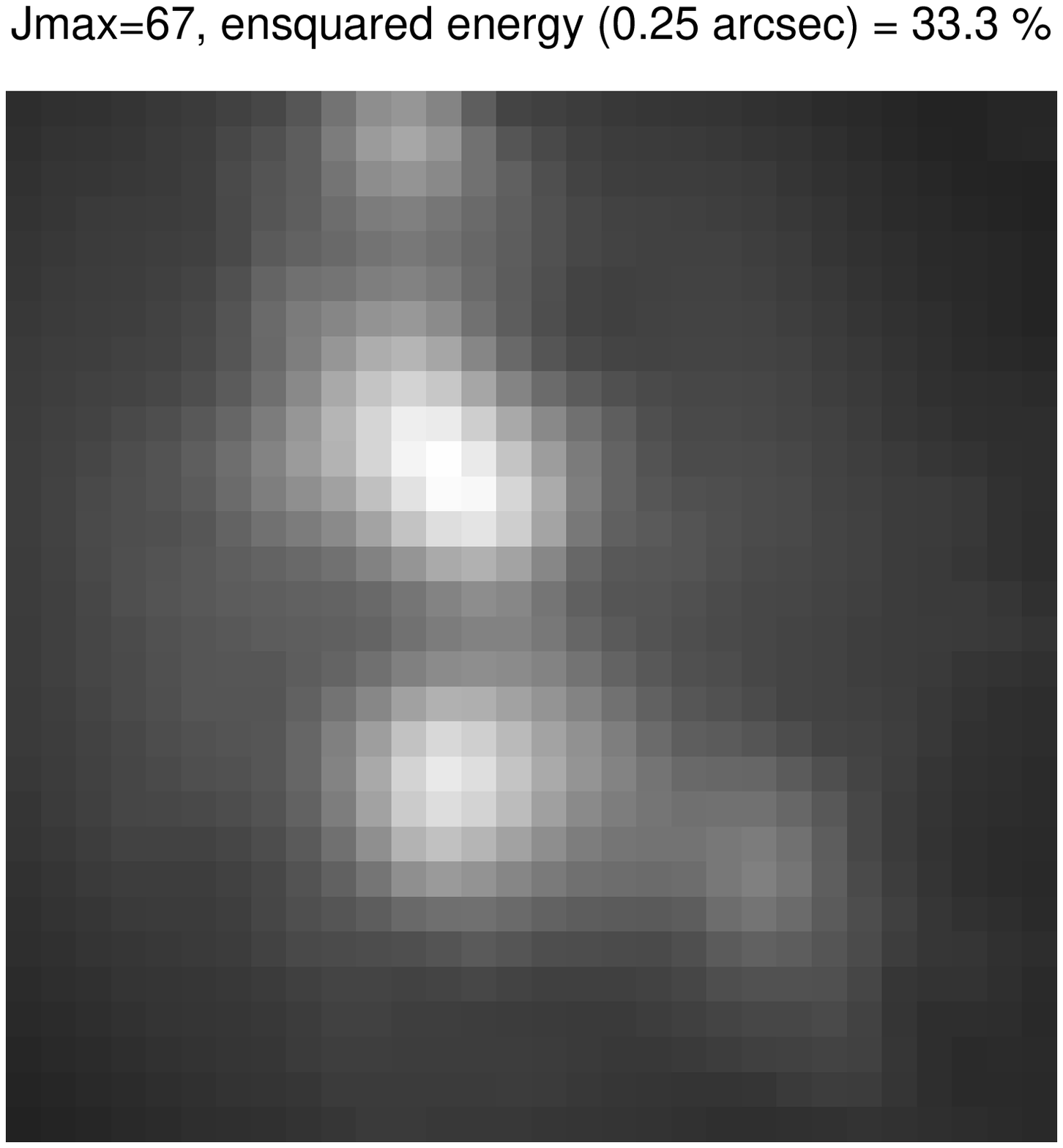}\\
		\includegraphics[width=0.25\textwidth]{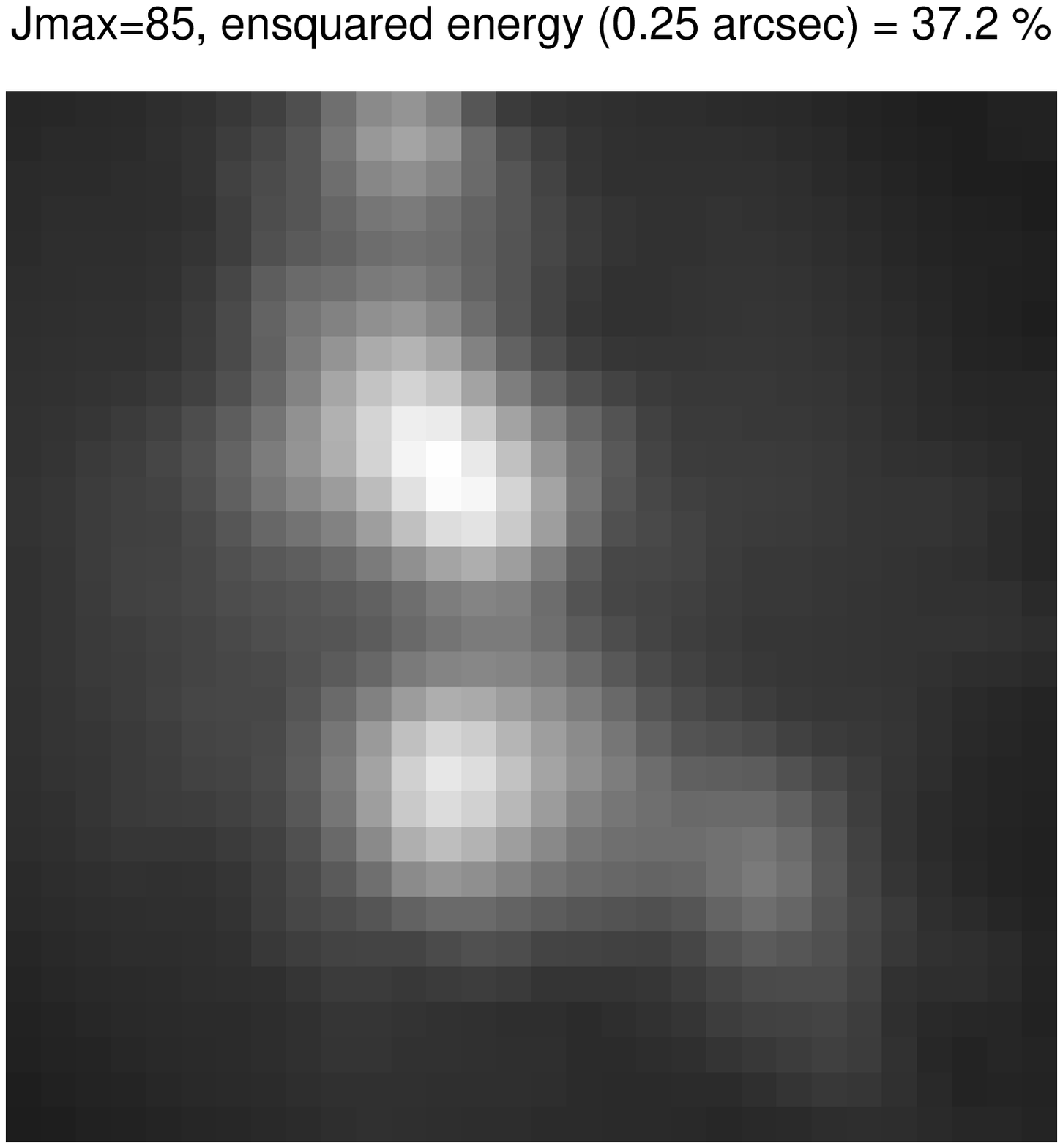}&
		\includegraphics[width=0.25\textwidth]{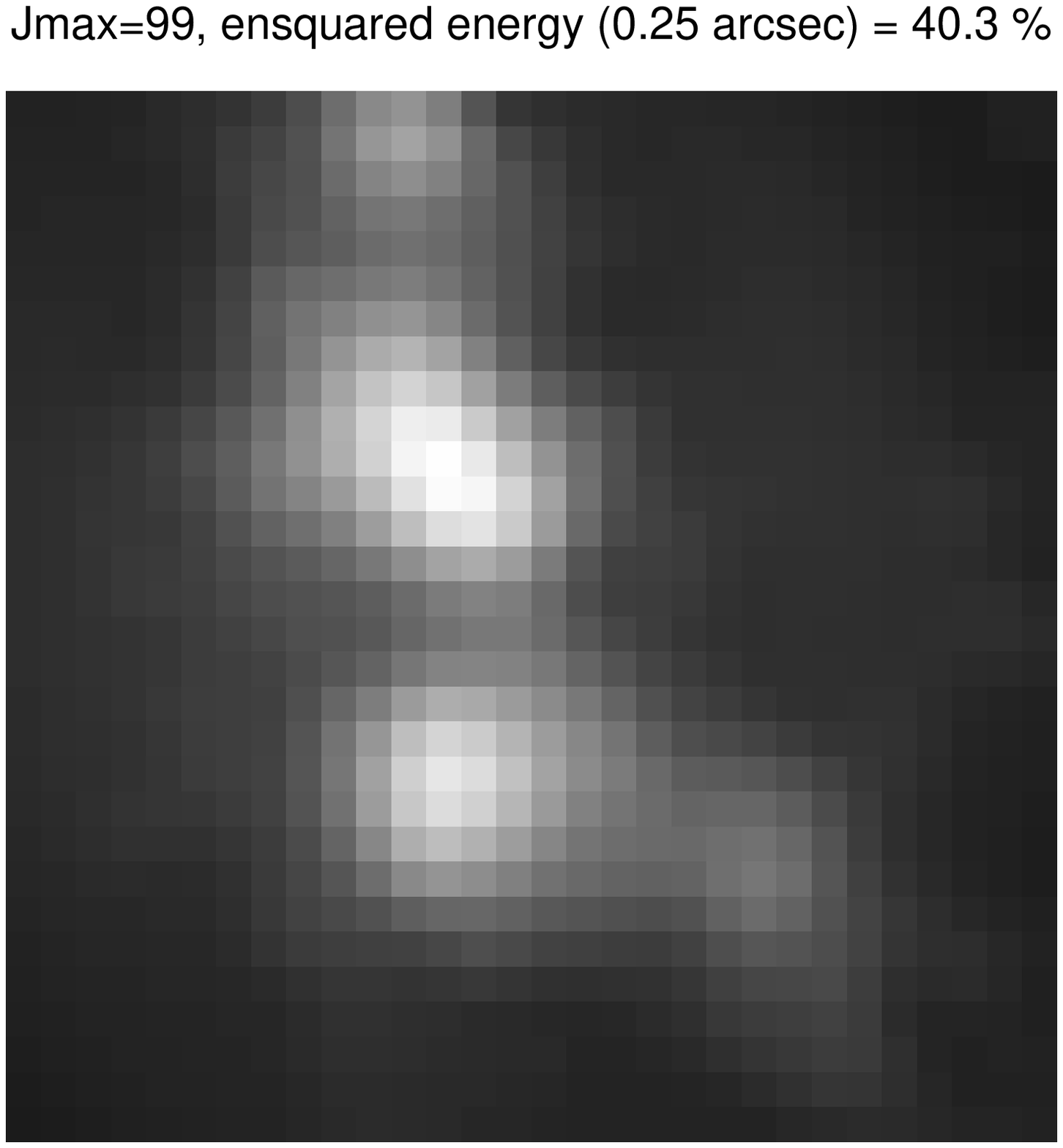}&
		\includegraphics[width=0.25\textwidth]{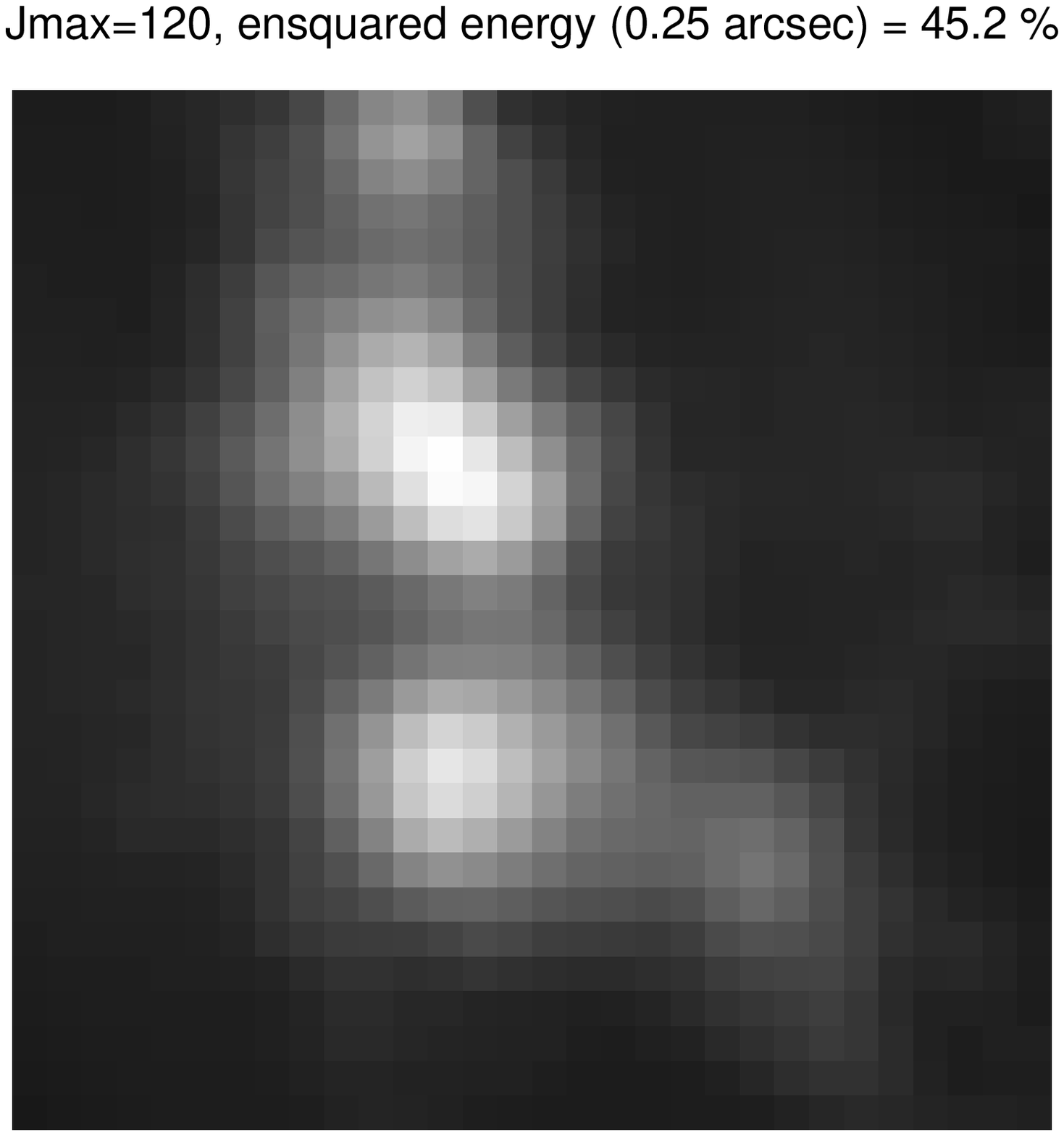}\\
	\end{tabular}
	\end{minipage}
\caption{Images of the HII regions convolved with the AO corrected PSF,  
for different levels of AO correction in J band. The title of each  
image shows the number of corrected modes and the corresponding  
ensquared energy into a $0.25 \times 0.25~arcsec^2$. The field of view  
is equal to $0.50 \times 0.50~arcsec^2$ ($4 \times 4$ 0.125 arcsec  
microlenses).}
\label{fig:fig9}
\end{figure*}

As it was shown previously, the AO correction not only increases the  
ensquared energy into the spatial resolution element, but it also  
increases the sharpness of the AO corrected PSF, therefore providing a  
better image contrast. This is indeed shown on the figure  
\ref{fig:fig9}, where it can be seen that the two HII regions start to  
be well separated after the perfect correction of the first 46 Zernike  
Polynomials. This level of AO correction corresponds to an ensquared  
energy equal to $28\%$ into a $0.25\times 0.25~arcsec^2$ square  
aperture.\newline

This study has therefore shown that the combination of 3D spectroscopy  
with AO is definitely effective in terms of angular resolution and SNR  
improvement. This means that the 3D spectroscopy of $z\approx 1$  
galaxies with an angular resolution of $0.25~arcsec$ (definitely better  
than atmospheric seeing) should be achievable on a 8 meter telescope  
located on the ground such as the ESO VLT. This requires the AO  
correction to provide a minimum ensquared energy of $28\%$ in a  
$0.25\times 0.25~arcsec^2$ square aperture. In that case, a spectral  
resolution $R=10000$, allowing to resolve velocity dispersions with  
$FWHM_V \geq 30~km/s$, can be used, as an effective spectroscopic SNR  
of 3 can be reached after an exposure time of only three hours.

\subsubsection{Second science case: 3D spectroscopy of UGC 6778 at $z=1.5$}\label{sec:sec342}
We give here the results of simulations of the 3D spectroscopy of UGC  
6778 as if it was observed at $z=1.5$. We assumed here that the galaxy  
covers an apparent surface of $1~arcsec^2$, which is consistent with  
apparent sizes measured on deep surveys. This apparent surface was then  
used to simulate the apparent $H\alpha$ and continuum images. We are  
then going to focus on the continuum and $H\alpha$ flux  
normalisation.\newline

We used the data from \cite{Erb-a-2003} and \cite{Steidel-a-2004}, who  
observed with the LRIS spectrograph on the Keck telescope a sample of  
BM/BX galaxies located between $z=1.4$ and $z=2.5$. \cite{Erb-a-2003}  
found on their sample ($\left\langle z \right\rangle =2.28$) an average  
apparent magnitude $\left\langle R \right\rangle=24.37$, and an average  
apparent $H\alpha$ flux $f(H\alpha)=4.6 \times 10^{-17}~ergs~s^{-1}~cm^{-2}$.
One will notice that the apparent $H\alpha$ flux is comparable to the one
computed in the previous science case $(z=0.9)$ where we simulated the observation
of a distant large spiral galaxy belonging to the CFRS. In fact the two science cases
have been chosen using previous well-known studies of distant galaxies. At $z=0.9$, the
CFRS has studied average $L^*$ galaxies, while at $z=1.5$, \cite{Erb-a-2003} have focused
on a sample of specifically $H\alpha$ luminous galaxies.

Moreover \cite{Steidel-a-2004} found that  
galaxies with a redshift $\left\langle z \right\rangle=2.23 \pm 0.31$  
had an average color $\left\langle R-K \right\rangle = 3.25 \pm 0.53$.
We therefore found that those galaxies had an apparent magnitude $\left  
\langle K=21.12 \right \rangle$. It is therefore possible to compute  
their K absolute magnitude $M_K$ using again the distance modulus,  
assuming a \textit{k-correction} $K(z)=-0.57$ at $z=2.3$ (extrapolation  
to $z=2.3$ of the \textit{k-correction} in K band given by  
\citeauthor{Mannucci-a-2001} for a Sc galaxy), leading to $M_K=-24.38$.  
Moreover, as Sc galaxies have a color $(H-K)=0.25$  
\citep{Mannucci-a-2001}, we found that such a galaxy has a rest-frame H  
absolute magnitude equal to $M_H=-24.13$. Assuming a  
\textit{k-correction} $K(z)=0.078$ at $z=1.5$ for a Sc galaxy observed  
in H band \citep{Mannucci-a-2001}, we therefore found an apparent  
magnitude $H=20.99$ for the case where such galaxies would be located  
at $z=1.5$, corresponding to a flux in the continuum  
$f_{cont}=1.67\times 10^{-19}~erg/s/cm^2/\AA$.\newline

\begin{figure*}
	\centering
	\begin{minipage}{200mm}	
	\begin{tabular}{ccc}
		\includegraphics[width=0.25\textwidth]{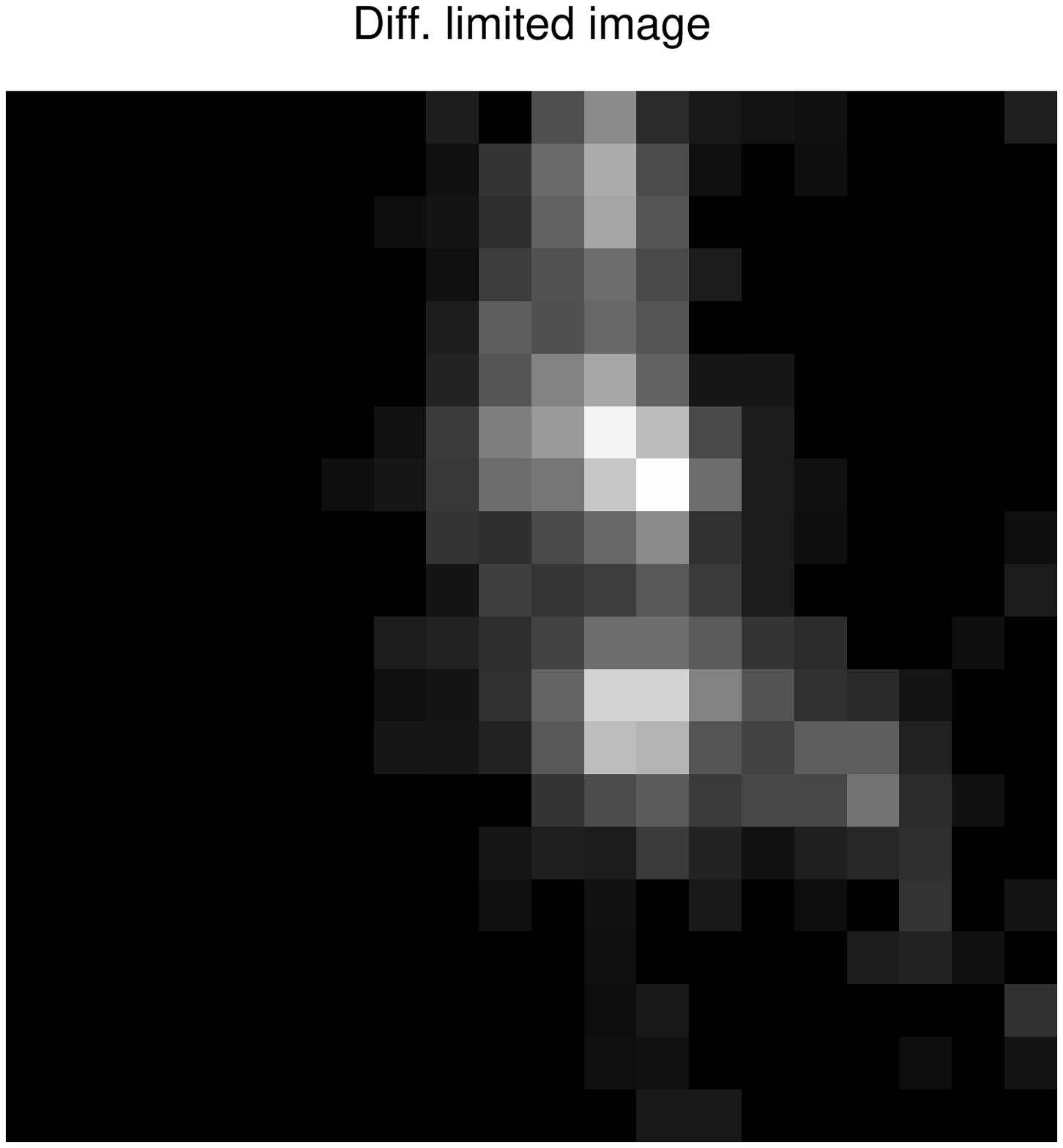}&
		\includegraphics[width=0.25\textwidth]{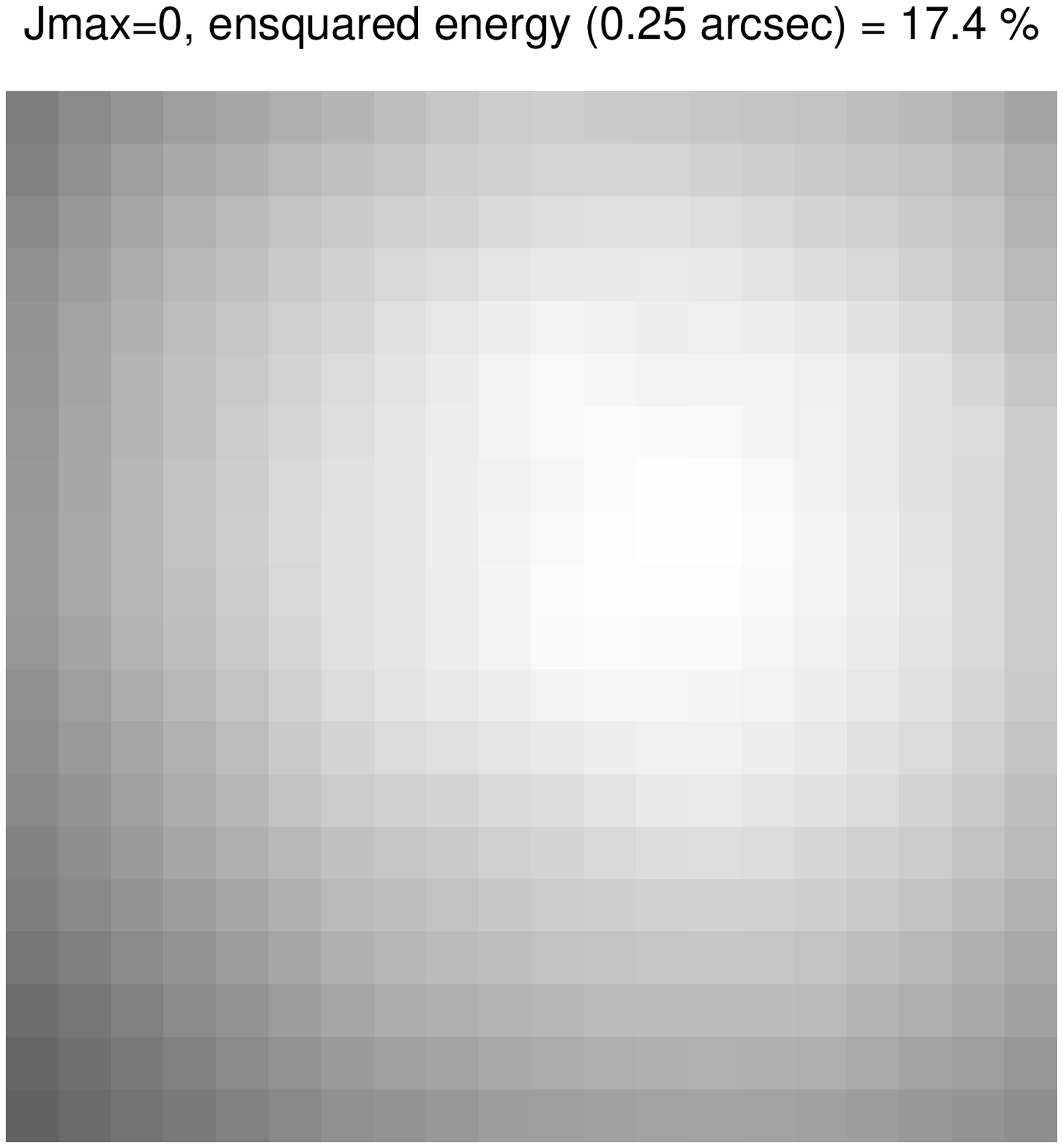}&
		\includegraphics[width=0.25\textwidth]{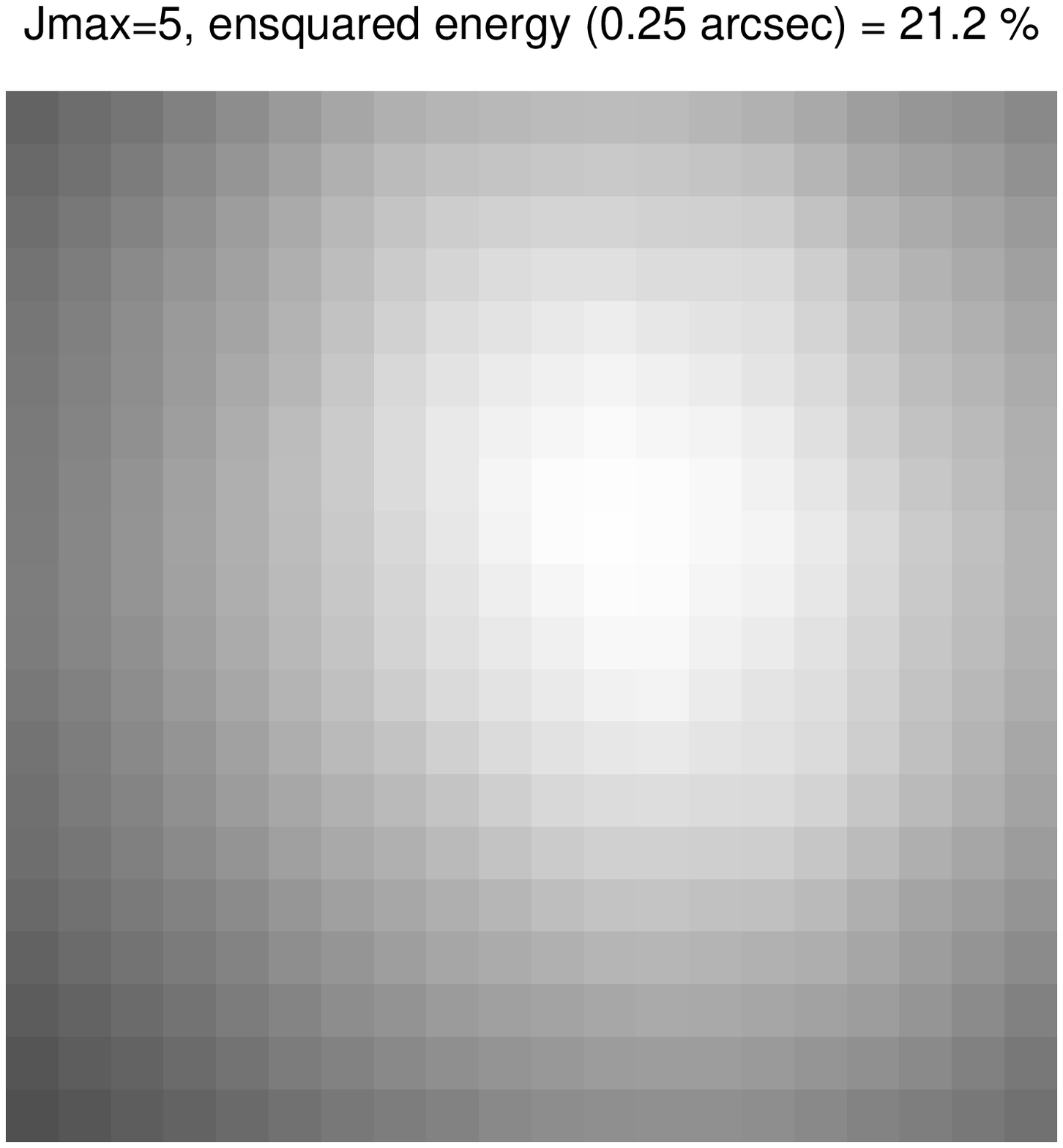}\\
		\includegraphics[width=0.25\textwidth]{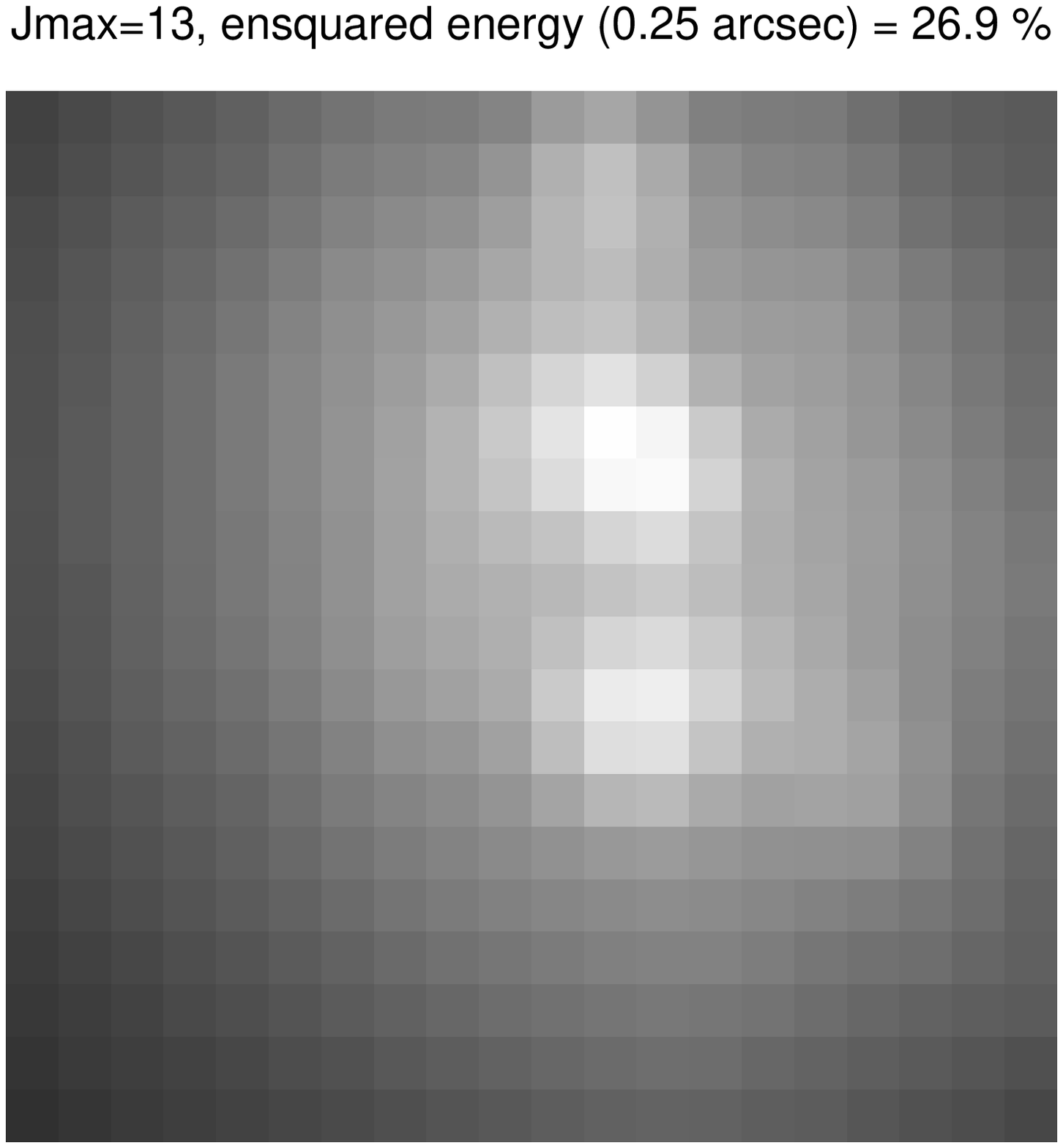}&
		\includegraphics[width=0.25\textwidth]{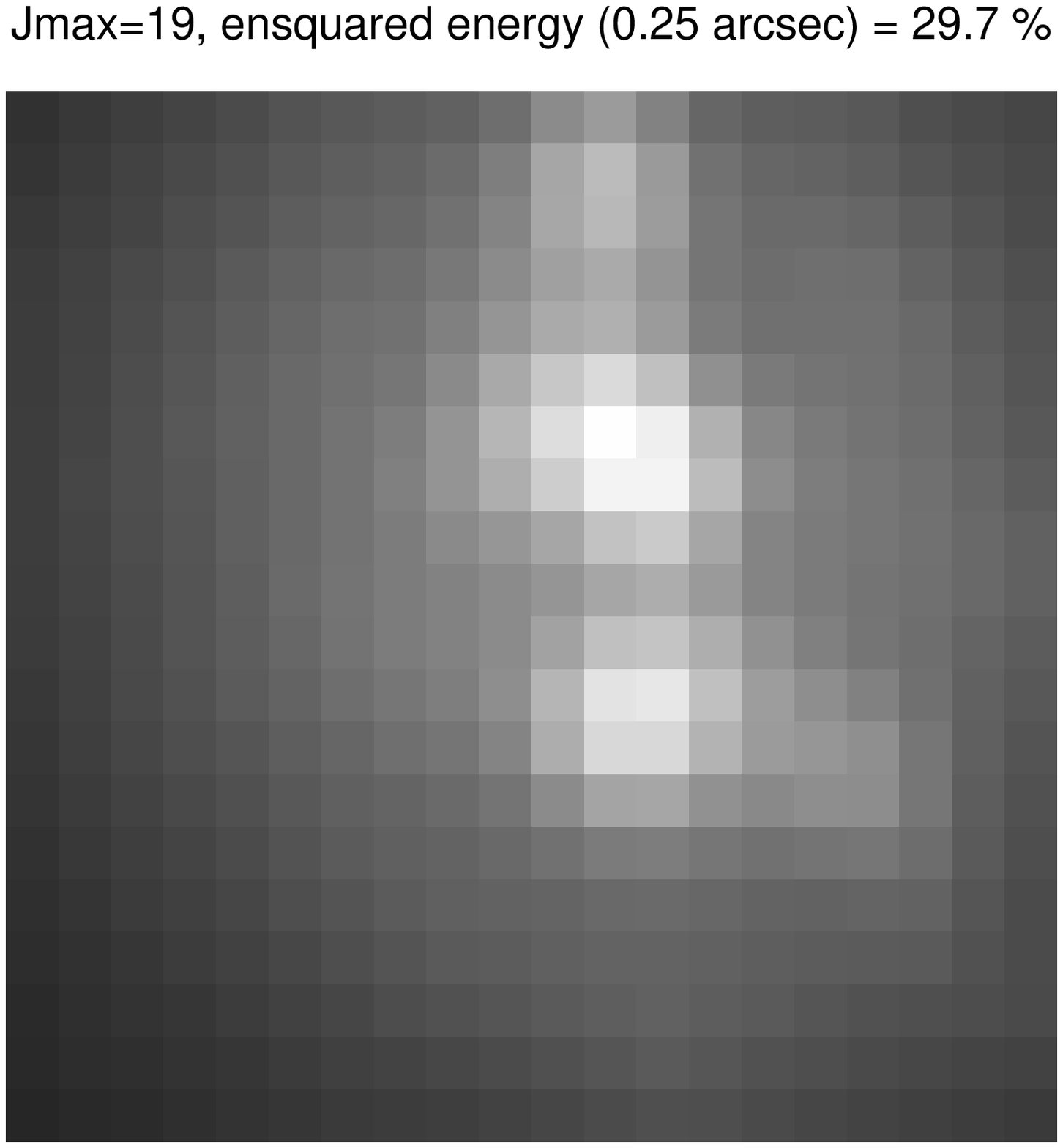}&
		\includegraphics[width=0.25\textwidth]{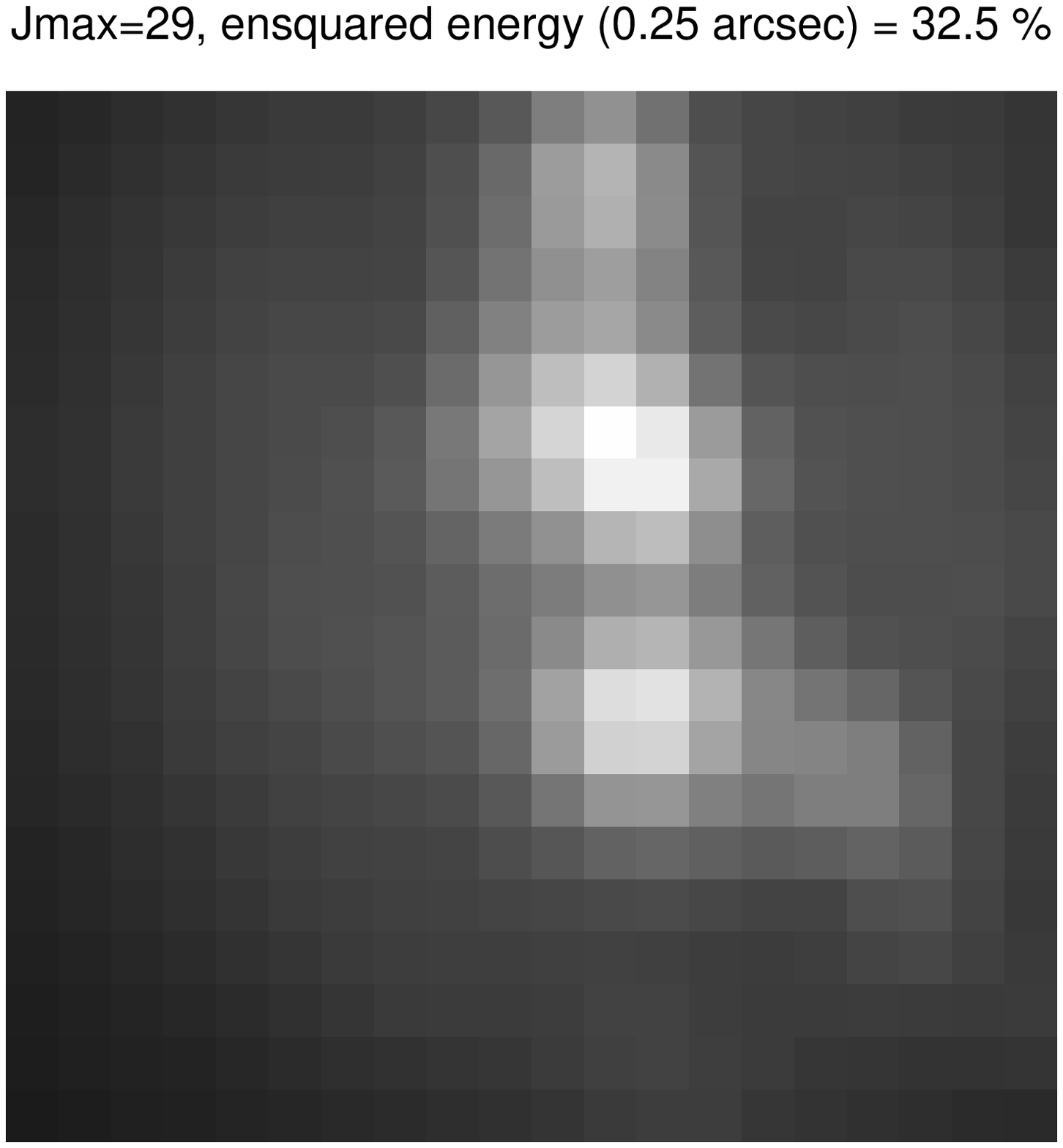}\\
		\includegraphics[width=0.25\textwidth]{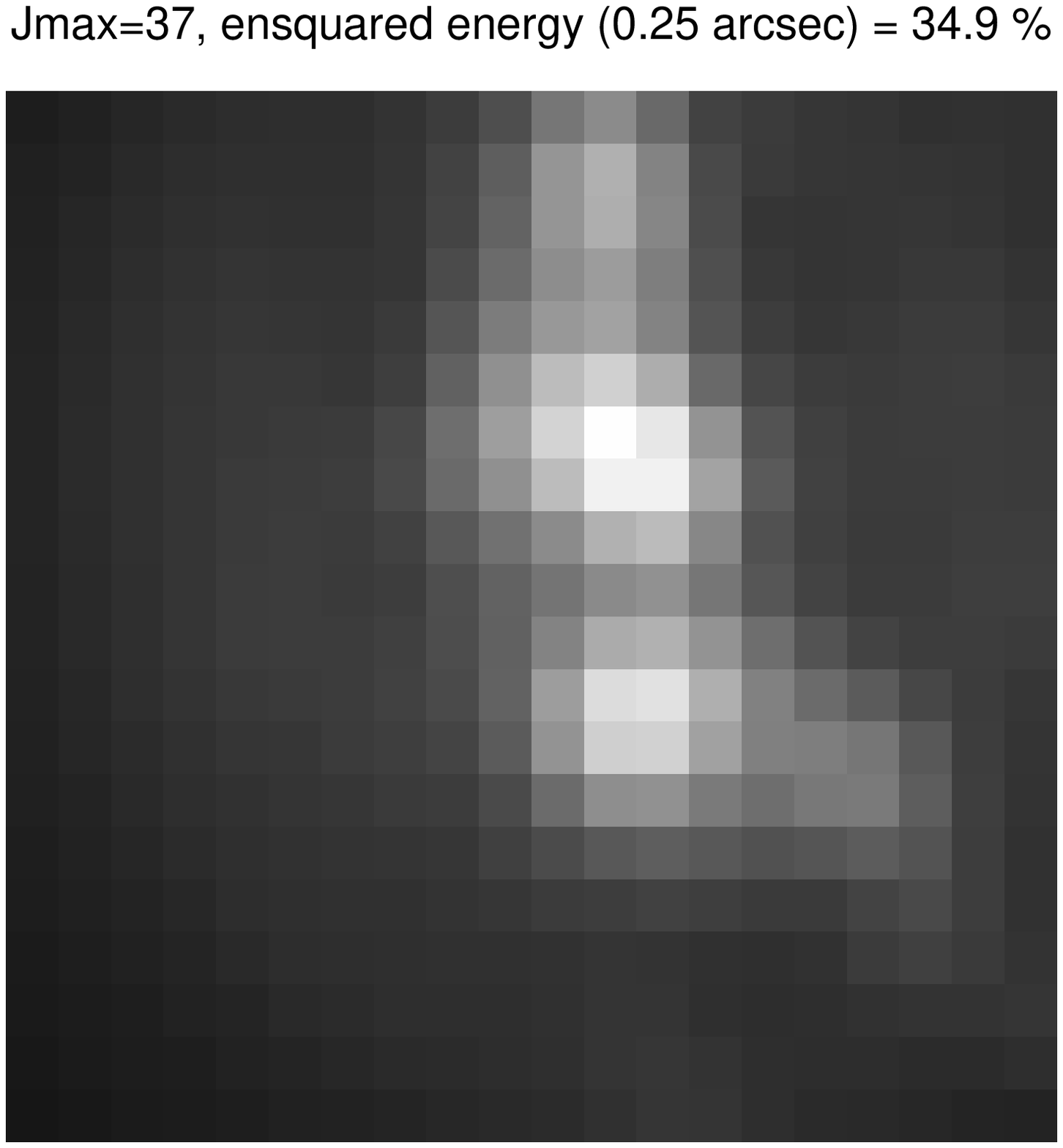}&
		\includegraphics[width=0.25\textwidth]{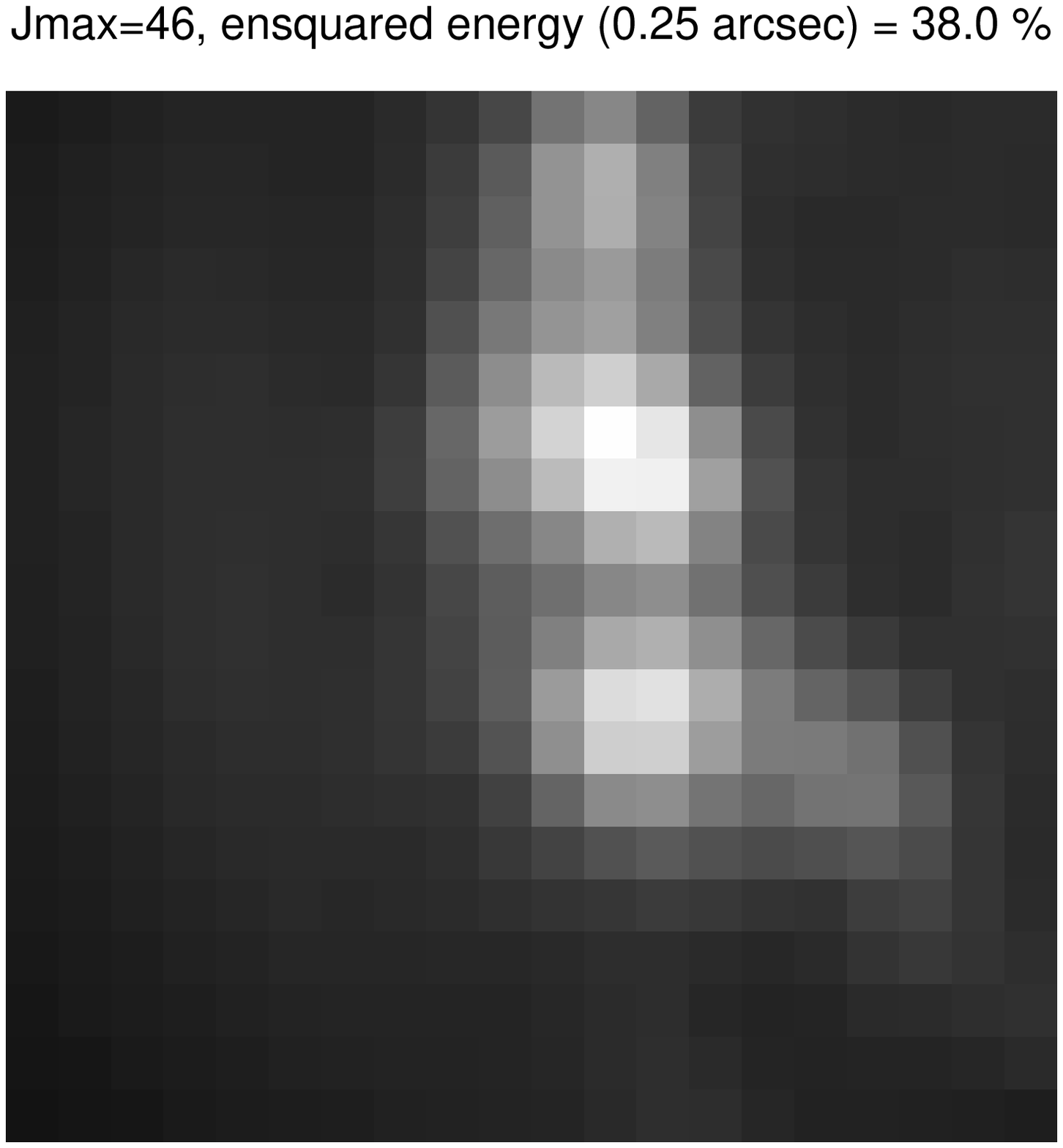}&
		\includegraphics[width=0.25\textwidth]{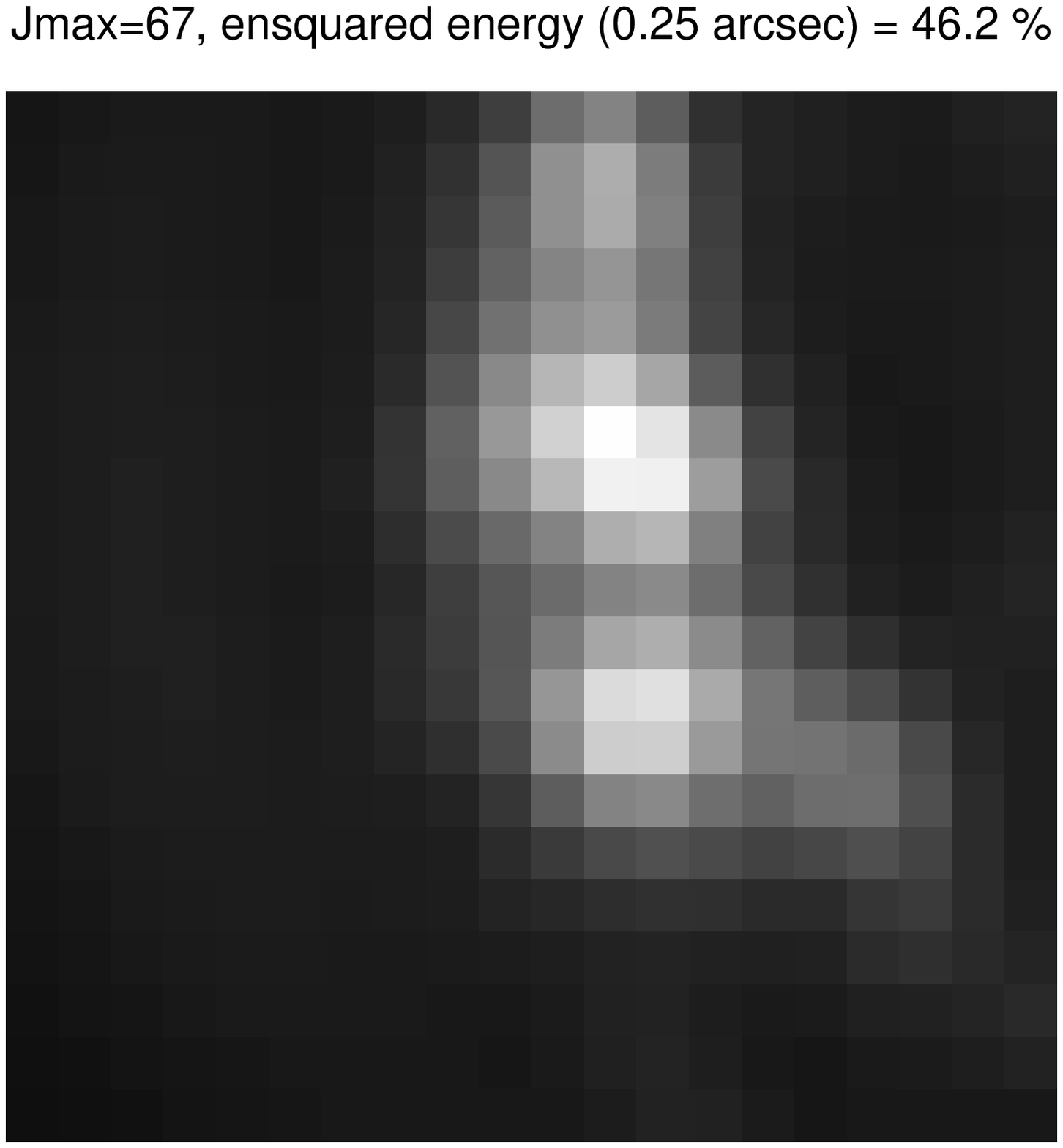}\\
		\includegraphics[width=0.25\textwidth]{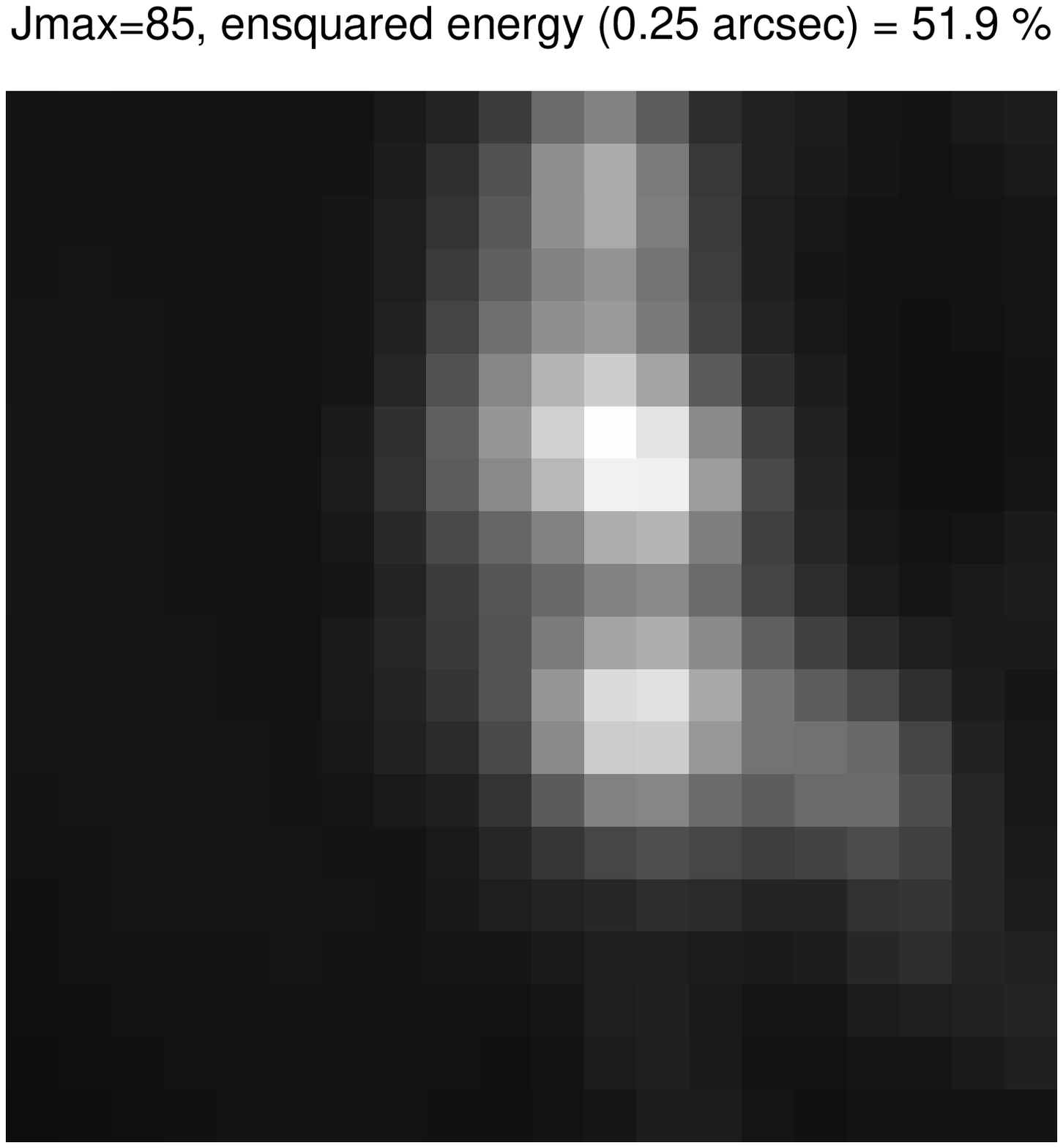}&
		\includegraphics[width=0.25\textwidth]{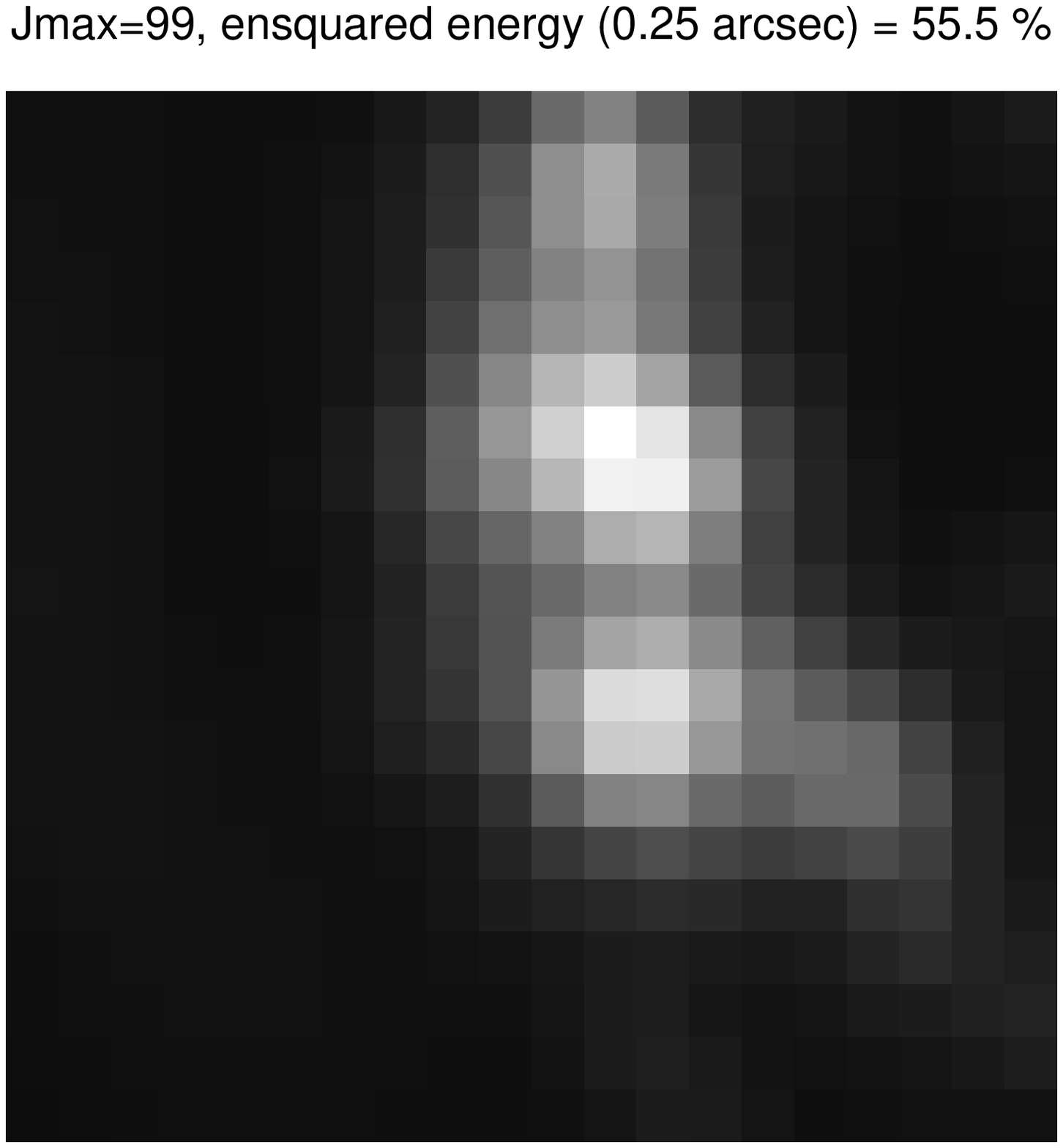}&
		\includegraphics[width=0.25\textwidth]{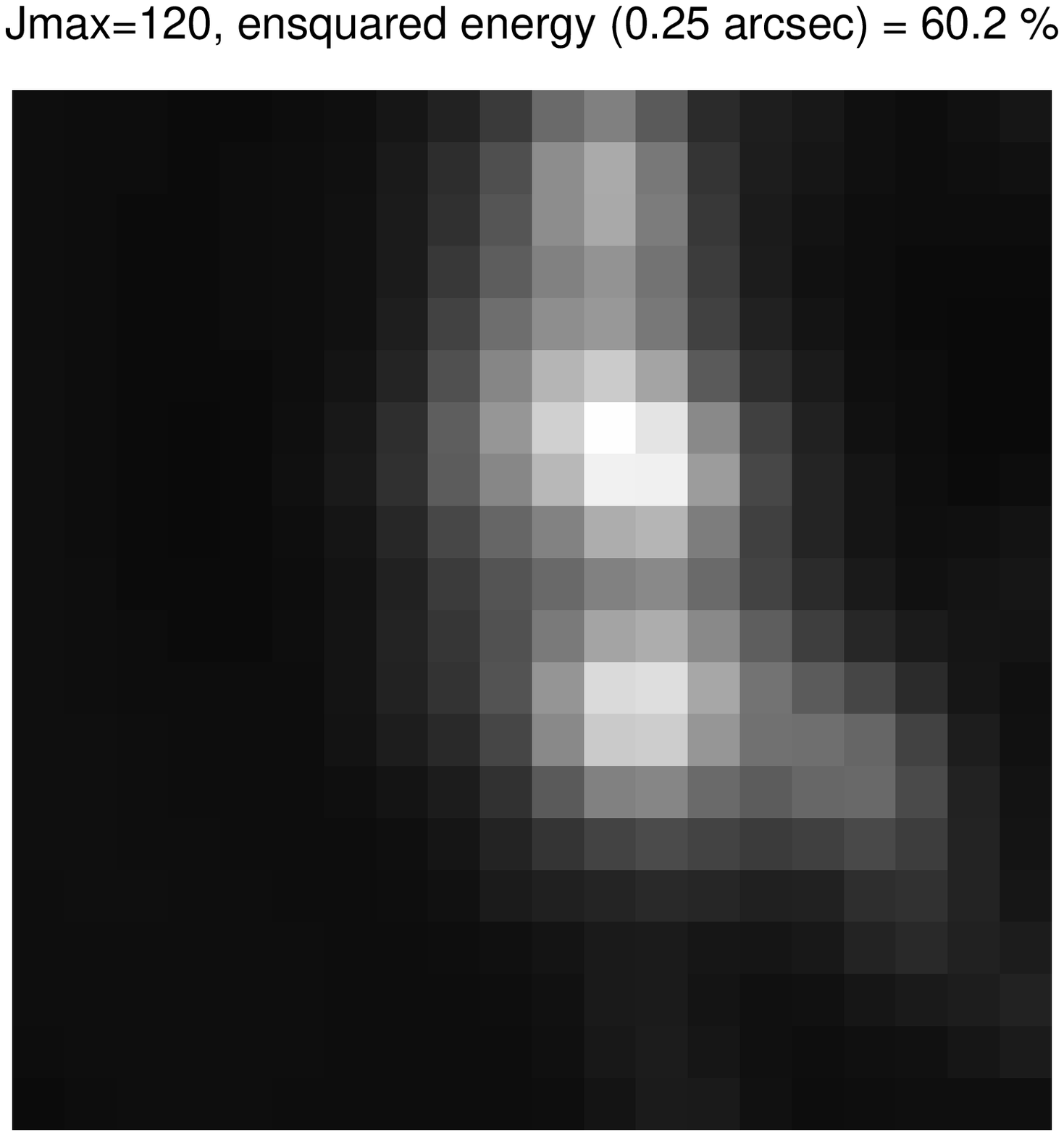}\\
	\end{tabular}
	\end{minipage}
\caption{Images of the HII regions convolved with the AO corrected PSF,  
for different levels of AO correction in H band. The title of each  
image shows the number of corrected modes and the corresponding  
ensquared energy into a $0.25 \times 0.25~arcsec^2$ square aperture.  
The field of view is equal to $0.50 \times 0.50~arcsec^2$ ($4 \times 4$  
0.125 arcsec microlenses).}
\label{fig:fig11}
\end{figure*}

\begin{figure}
	\centering
	\begin{tabular}{c}
		\includegraphics[height=0.3\textheight]{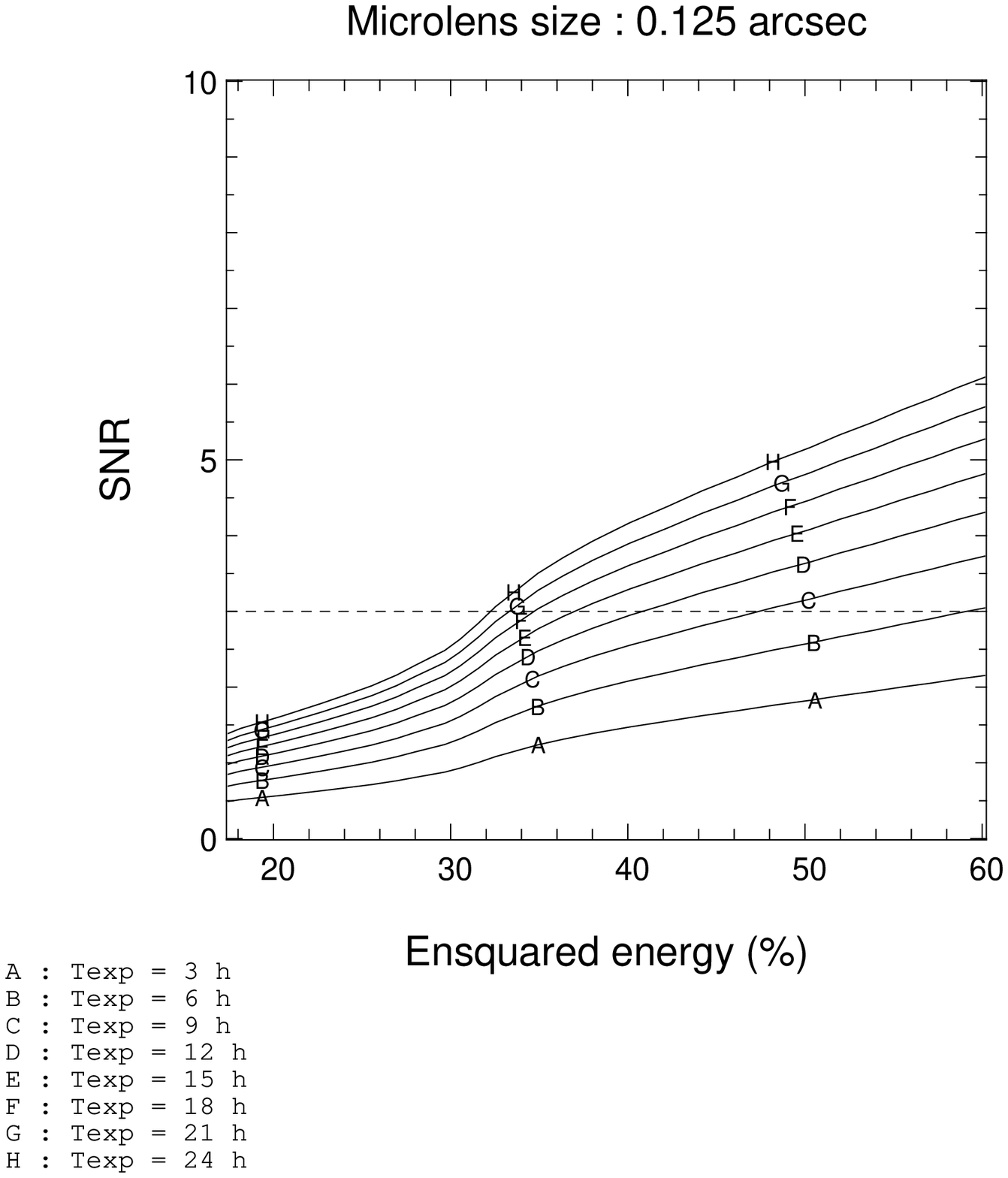}\\
		\includegraphics[height=0.3\textheight]{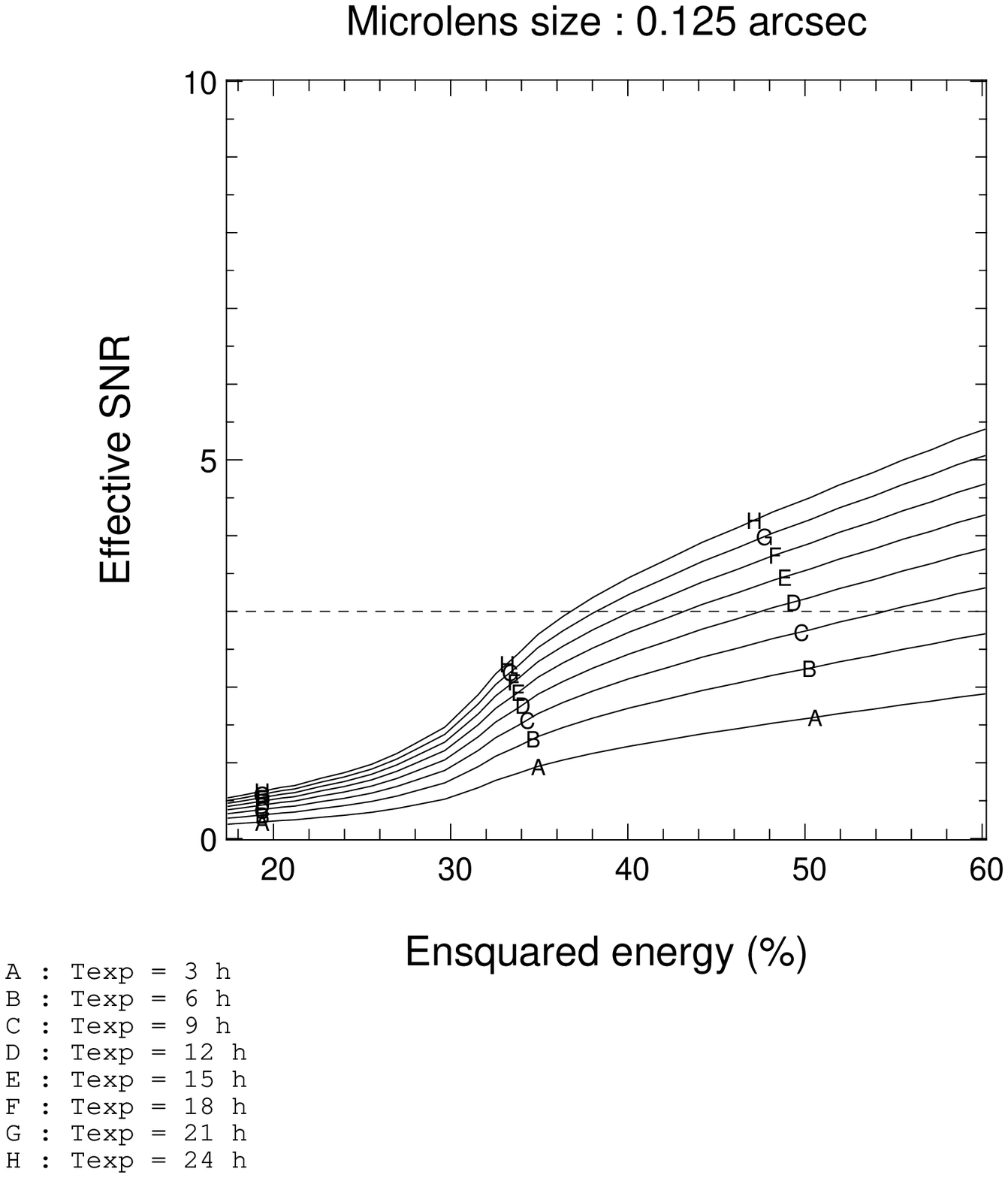}\\
	\end{tabular}
\caption{Measured spectroscopic SNR (top) and effective spectroscopic  
SNR (ESNR, bottom) on the $H\alpha$ line for a microlens of  
$0.125~arcsec$ as a function of the ensquared energy in a $0.25\times  
0.25~arcsec^2$ square aperture. The considered galaxy is located at  
$z=1.5$ (observation in H band) and covers an apparent field of  
$1~arcsec^2$. Each curve corresponds to the average of the SNR of each  
microlens sampling the HII regions, with exposure times going from 3  
hour to 24 hours. The dashed line corresponds to $SNR=3$.}
\label{fig:fig10}
\end{figure}

These flux values were used to normalise the simulated images of the  
distant galaxy. Then we did the same computations as previously  
explained, in order to compute the measured spectroscopic and effective  
SNR per spectral pixel on the spectrograph's detector, and especially  
show the improvements provided by AO correction. This time we considered
exposure times from 3 to 24 hours. The result of these  
simulations is shown on figure \ref{fig:fig10}, where we can see that  
an effective SNR of 3 can be reached after a minimum exposure time of 9 
hours, requiring an ensquared energy after AO correction equal to  
$60\%$ into a $0.25\times 0.25~arcsec^2$ square aperture. The same ESNR  
value can be reached with an ensquared energy after AO correction equal  
to $40\%$, requiring then an exposure time of 18 hours, or also with an  
ensquared energy after AO correction equal to $33\%$, requiring then an  
exposure time of 24 hours. However, in the previous section we saw that  
a minimum ensquared energy must also be reached so that the two HII  
regions can be separated. This is explained in the figure  
\ref{fig:fig11}, where we can see that the two HII regions start to be  
separated for an ensquared energy after AO correction better than  
$30\%$.\newline

This study has therefore shown that the combination of 3D spectroscopy  
with AO should allow also to perform the 3D spectroscopy of galaxies  
located at $z=1.5$, also with an angular resolution of $0.25~arcsec$  
(definitely better than atmospheric seeing) and with a spectral  
resolution $R=10000$, allowing to resolve velocity dispersions $FWHM_V  
\geq 30~km/s$, as well as reaching effective spectroscopic SNR of 3.  
This can be achieved with a minimum ensquared energy after AO  
correction of $35\%$ and an exposure time of 24 hours, knowing that the  
same ESNR value can be reached with an ensquared energy of $40\%$ and a  
shorter exposure time (18 hours).

\subsection{Conclusion}
We studied in this section the improvement provided by the combination  
of 3D spectroscopy with Adaptive Optics (AO). Firstly we have shown  
that thanks to AO, it is possible to reach angular resolution better  
than seeing, but also to increase the spectroscopic signal-to-noise  
ratio (SNR), allowing to perform 3D spectroscopy of distant galaxies,  
thus studying their kinematics. We have then performed some simulations  
allowing to give some preliminary performance of an integral field  
spectrograph providing an angular resolution of $0.25~arcsec$ and a  
spectral resolution $R=10000$, and which would observe distant galaxies  
located at $z=0.9$ and $z=1.5$. In that case, the $H\alpha$ emission  
line is redshifted in the central wavelengths of J and H bands  
(respectively $1.25~\mu m$ and $1.65~\mu m$). To quantify the  
performance of such an instrument, we have defined a new criterium  
called "`\textit{Effective signal to noise ratio}"' (ESNR). We also  
assumed for those simulations galaxies with properties consistent with  
the ones observed in distant surveys. We have therefore shown that an  
ESNR of 3 can be reached for a galaxy located at $z=0.9$, requiring a  
minimum ensquared energy  into a $0.25\times 0.25~arcsec^2$ after AO  
correction equal to $30\%$ and an exposure time of 3 hours. The same  
ESNR value can be reached for a galaxy located at $z=1.5$, requiring  
then a minimum ensquared energy  after AO correction of $35\%$ and an  
exposure time of 24 hours, however a slight increase of the ensquared  
energy to $40\%$ should allow to reach the same ESNR value after an  
exposure time equal to 18 hours.\newline

A study of the AO corrected PSF has shown that high order modes need to  
be corrected to improve the ensquared energy. This has been confirmed  
by our simulations of the 3D spectroscopy of distant galaxies, where we  
found that at least 46 Zernike polynomials need to be corrected to  
reach an ensquared energy of $30\%$ in J band and $40\%$ in H band  
(such results are not definitive, as we used the dynamical properties  
of one galaxy observed in the local universe as well as those of a few  
distant galaxies, but they can be used as a first basis, the data  
required to perform such studies being still rare in the community).  
Such a number of modes corresponds in fact to the minimum order of  
modes to correct, as we assumed a perfect AO system in our study. In a  
real AO system, other sources of error like time delay, measurement  
noise and anisoplanatism are going to degrade the performance of AO,  
meaning that a larger number of modes will be required to reach the  
same ensquared energy values and image quality. We must especially  
insist on the fact that anisoplanatism is going to become the most  
important limitation in the case of a real instrument using both AO and  
3D spectroscopy. Indeed, as extragalactic studies require to work at  
high galactic latitudes, the probability of finding a suitable star to  
perform wavefront sensing is equal to a few percents, meaning that AO,  
at least in its \textit{classical} form, cannot be used because of its  
very low sky coverage.\newline

We are therefore going to describe in the next sections the principle  
and the performance of \textbf{FALCON}, a project of a multi-object 3D  
spectrograph with AO, which solves the sky coverage's problem imposed  
by classical AO. FALCON uses the principle of Multi-Object Adaptive  
Optics (MOAO). Such a technique allows to widen the useful field of  
view of AO systems, thus to perform the simultaneous 3D spectroscopy of  
several distant galaxies in a very wide field of view, providing  
therefore a huge gain in terms of observing time efficiency.

\section[The FALCON's AO system: principle]{The FALCON's AO system: principle}\label{sec:4}
\subsection{Increase of the AO corrected FoV: MCAO and tomography}

As stated in the introduction, Adaptive Optics, in its  
\textit{classical} form where one guide star (GS) and one deformable  
mirror (DM) conjugated to the pupil are respectively used to measure  
and correct the turbulent wavefront, suffers from a very low sky  
coverage (less than $5\%$), even at galactic latitudes of $30\Moideg$.  
The reason of this limitation is that the probability to find a  
suitable GS in the isoplanatic patch is generally very low, due to the  
distribution of the stars in our galaxy.\newline

Such a limitation has leaded to the development of Multi-Conjugate  
Adaptive Optics (MCAO). In MCAO, several deformable mirrors conjugated  
to the turbulent layers are used to correct the phase in the turbulent  
volume. Indeed, this is the repartition of the turbulence in the  
altitude which is responsible for anisoplanatism, as the wavefronts  
coming from different directions do not cross the same volume of the  
atmosphere (especially for high turbulent layers), and therefore do not  
suffer from the same degradations. By correcting the phase in the  
turbulent volume, and especially in the strongest layers, it is  
possible to correct the wavefront for any direction, and to have an  
uniform compensation in a field of view larger than the isoplanatic  
patch.\newline

Before correcting the phase with the different DMs, it is required to  
know the perturbation in each layer to apply to each DM the adequate  
commands. This can be achieved thanks to \textit{tomography}  
\citep{Ragazzoni-a-1999b,Ragazzoni-a-2000,Tokovinin-a-2001b}, where the  
light coming from several off-axis GSs far outside of the isoplanatic  
patch is used to probe the 3-dimensional phase perturbations in the  
atmosphere, generally by solving an inverse problem as in medical  
imaging. However, thanks to tomography, it is also possible to know the  
integrated phase perturbation \textbf{in the pupil} for any direction  
in the field of view. This means the field where GS can be found is  
widened, and this allows a higher sky coverage for AO  
\citep{Tokovinin-a-2001}.\newline

However, the potential extension of the field of view with MCAO  
encounters some technical limitations, in particular on the optical  
system.
As an example, the ESO-MAD system for the VLT \cite{Marchetti-p-2003}  
or the Gemini-South MCAO system \cite{Ellerbroek-p-2003} will deliver a  
corrected field of view with a diameter of 2 arcmin, definitely smaller  
than the $10 \times 10~arcmin^2$ field of view required for  
extragalactic studies. We are therefore going to show in the next  
paragraph a new approach for AO, where the goal is not to provide an  
uniform correction in a very wide field of view, but only for specific  
areas into it, i.e. the galaxies on which we want to perform 3D  
spectroscopy. This new approach is called \textit{Multi-Object Adaptive  
Optics} (MOAO).

\subsection{Multi-Object Adaptive Optics}
First proposed by \citet{Hammer-p-2002}, \textit{Multi-Object Adaptive  
Optics} (MOAO) is a new method of AO correction which has essentially  
been since this date the object of a few conference proceedings papers  
\citep{Assemat-p-2004, Dekany-p-2004, Hammer-p-2004}. We summarise its  
principles in this section.\newline

Let us recall here that our goal is to measure the internal kinematics  
of distant galaxies spread over the VLT Nasmyth field. In  
other words, this means that a high angular resolution is required  
\textit{only for the scientific targets}. As a result, we propose in  
this section a totally new approach for AO: instead of correcting the  
whole $10 \times 10~arcmin^2$ field (which anyway is impossible to do),  
we propose to correct locally only the regions of interest, $i.e.$ the  
integral field units positioned on the scientific targets, and we have  
therefore \textbf{one AO system per integral field unit}. As a result,  
if we suppose that we want to perform the simultaneous 3D spectroscopy  
of 20 galaxies, we arrive to the concept of an instrument using 20  
multiple AO systems, spread in the VLT Nasmyth focal plane, and working  
in parallel independently of each other.\newline

Although the complexity of such an instrument may appear insuperable at  
first sight, it can be greatly relaxed thanks to the tininess of the  
compensated field required for each independent galaxy. Indeed, as  
already demonstrated by the instrument GIRAFFE \citep{giraffe}, a field  
of view of only $3 \times 2~arcsec^2$ is sufficient to measure the  
velocity fields of large spirals at $z \leq 1$, as well as galaxies  
with smaller apparent sizes observed at greater redshifts. This is one  
order of magnitude smaller than the isoplanatic patch in the  
near-infrared, so only one unique deformable mirror conjugated to the  
pupil is required to correct the whole galaxy.

Moreover, we propose a concept where the AO system is miniaturized.  
Each IFU, laying in the telescope focal plane, will include its proper  
micro deformable mirror -with a proper tiny pupil imaging optics. The  
wavefront sensors also lay a few centimeters apart, in the focal plane  
; they are located on the guide stars surrounding the galaxy.  
Translated into length units in the VLT focal plane, an apparent size  
of $3 \times 2~arcsec^2$ covers a physical size of $1.8 \times  
1.2~mm^2$, and a median distance of 1 arcmin between galaxy and guide  
stars translates into $\approx 35~mm$. Those numbers suggest the size  
of the system we propose: the adaptive IFU, and the WFSs have to be  
integrated into suitable optomechanical devices, which should not be  
too wide (typically less than $20~mm$) in order not to obstruct the  
focal plane and allow to use of the closest suitable guide stars for  
wavefront sensing. Such an integration of the subsystems, together with  
a separation of their functions (the DM and the WFS become separate,  
physically independent items), allow to move those devices into the VLT  
focal plane just as the positioner \textit{OzPoz} does with the IFUs  
used on GIRAFFE \citep{giraffe}.

In addition to offer an obvious multiplex advantage, the proposed  
structure combined with a tomographic wavefront sensing approach allows  
us to overcome the sky coverage problem. Several guide stars, far out  
the isoplanatic patch, are used around each IFU to sense the wavefront.  
For a 8 meter diameter telescope, it has been shown  
\citep{Fusco-a-1999} that 3 WFSs are sufficient to reconstruct the  
phase in the pupil for any direction in a field of view of $\approx  
2~arcmin$ diameter and for any turbulence profile; using more WFSs only  
brings a marginal improvement.

We chose to focus our study on natural guide stars. We believe that  
using Laser Guide Stars (LGS), although very attractive and apparently  
drastic solution at first sight, could be technically exponentially  
difficult to implement when dealing with an increasing number of  
beacons (minimum of 20 beacons here), while bringing its set of  
inherent problems such as cone effect, beacon elongation,  
superimposition of Rayleigh scattering and beacons, etc. Our approach  
was then to start the study on NGS, at the risk of doing the study  
again with LGS in the case where it would be demonstrated that NGS do  
not allow to fill the requirements.

We emphasize that the architecture described above corresponds to an  
open-loop system, since the WFS does not get any optical feedback from  
the deformable mirror of the adaptive IFU: this drastically differs  
from classical closed-loop AO systems. This is of course extremely  
challenging in terms of AO components. Firstly, the drifts, the  
linearity, the hysteresis and the calibration of the DM must be kept to  
a very low error level in order to be able to control it accurately  
enough in open-loop. Secondly the WFS must have a high dynamic range in  
order to provide reliable absolute open-loop measurements, together  
with a high sensitivity versus flux level. Such a WFS does not  
currently exist and needs to be developped, although some similar  
attempts have been done in the field of \textit{a posteriori} speckle  
imaging \citep{Lane-p-2003}. Thirdly, the miniaturisation of the  
components will undoubtly demand a large technical effort.

On the other hand, an architecture where the optical system is  
distributed all over the field has the considerable advantage to  
simplify the optics. In particular, the field selection functionnality  
becomes here a built-in function, when the amount of technical  
difficulties and the sharp specifications are often underestimated in  
classical systems. Not only this critical part does not exist any more,  
but the overall optical throuput of the instrument is boosted, as, at  
any moment, no dichroic plate is needed to split the light, and because  
the number of optical surfaces is reduced to its strict minimum. It is  
important to keep in mind that the efforts put in this instrument are  
undertook to gain not only in spatial resolution, but also in  
signal-to-noise ratio. On this latter particular point, the instrument  
optical throughput does matter even more than the AO performance,  
which, in return, will be increased when the WFS optical transmission  
is better.\newline

In order to evaluate the system performance in terms of its  
dimensionning, we ran some numerical simulations. Their presentation  
and results are the scope of the following section. Those simulations  
assume that we have 3 natural guide stars per IFU, an a tomographic  
reconstructor. They are independent of any detailled system  
configuration, such as  the miniaturisation of components or the error  
inherent to open-loop aspects. In this way, they could still apply  
whatever the technical solution. As an example, a system where the  
Nasmyth field would be optically sliced into sub-fields, themselves  
sent to independent closed-loop mono-mirror tomographic multi-analysis  
systems, would lead to the same results.

\begin{figure*}
	\centering
		\includegraphics[width=0.8\textwidth]{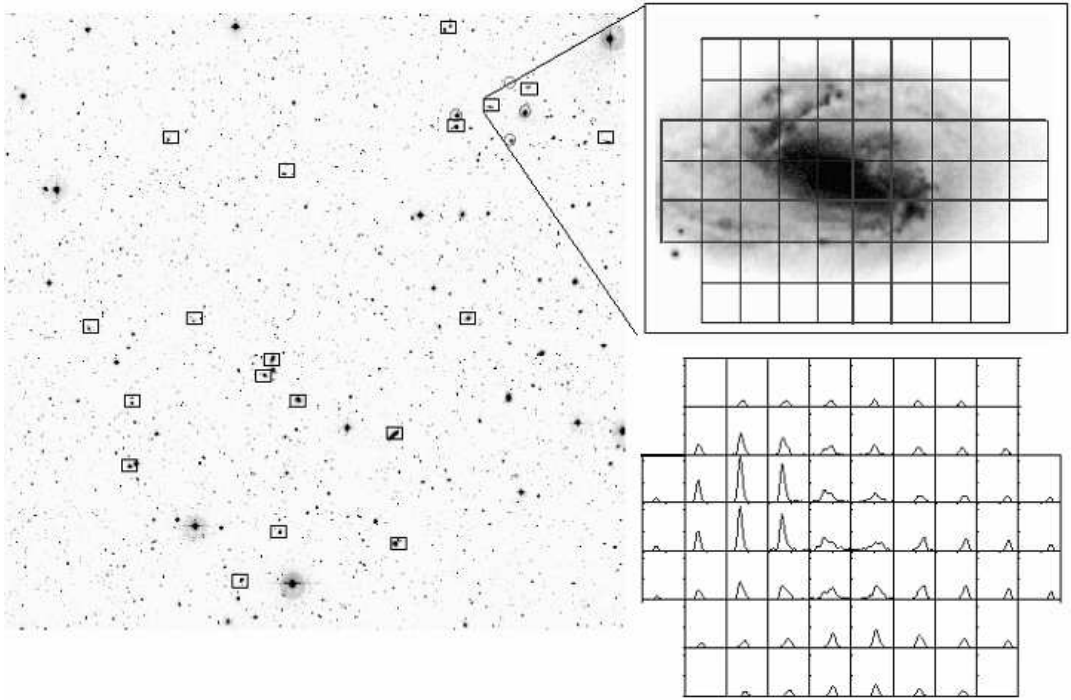}
	\caption{The Falcon concept combines MOAO and tomography: several IFUs  
(rectangles) are spread over the VLT focal plane. Each IFU is used to  
perform the 3D spectroscopy of one galaxy. Three miniaturised WFSs  
(circles) measure the wavefront coming from 3 off-axis NGS located  
around the galaxy. Their combined signals are used, through a  
tomographic reconstructor, to command the miniaturised DM into the IFU  
and correct the wavefront. It is therefore possible to recover the  
velocity field of the galaxy with a better spatial sampling.}
	\label{fig:fig12}
\end{figure*}

\section[The FALCON's AO system: performance]{The FALCON's AO system: performance}\label{sec:sec5}
\subsection{Introduction}
We show in this section the expected performance of a MOAO system like  
FALCON using atmospheric tomography methods on natural guide stars to  
reconstruct the on-axis wavefront from any geometry of off-axis  
measurements. We wrote a Monte-Carlo simulation code able to compute  
the long exposure PSF corrected by FALCON, for any atmospheric  
conditions (seeing, outer scale, profile) and for any wavelength. This  
PSF was then used to compute the Strehl Ratio, the full width at half  
maximum (FWHM) and the ensquared energy into a square aperture of any  
size.

For all the reasons explained in \ref{whyAo3dSpec}, we have focused our  
studies on the evolution of the ensquared energy in an aperture of  
$0.25 \times 0.25~arcsec^2$ (corresponding to 1 kpc for $1 \leq z \leq  
3$).
This evolution was studied
\begin{itemize}
\item as a function of the size (number of degrees of freedom) and  
sensitivity (limiting magnitude) of the AO system
\item for two wavelengths: $\lambda=1.25~\mu m$ and $\lambda=1.65~\mu  
m$ (central wavelengths of J and H bands). As we saw previously, such  
wavelengths correspond to the observation of the $H\alpha$ emission  
line redshifted at $z=0.9$ and $z=1.5$.
\item for three galactic latitudes $b\approx -90 \Moideg$, $b\approx  
-60 \Moideg$ and $b\approx -30 \Moideg$, in order to study the gain  
brought by atmospheric tomography methods as a function of guide star  
density, and estimate the corresponding sky coverage,
\item for the median atmospheric conditions of the VLT site (Cerro  
Paranal, Chile).
\end{itemize}

A particularly important aspect of this code is the simulation of the  
tomographic reconstruction, this latter being detailed in the section  
\ref{sec:43}.

We did not want to make technical assumptions about the type of the  
deformable mirror. For the sake of simplicity, we chose its influence  
functions to be Zernike polynomials. For the same reasons, we assume  
that the wavefront sensor measurements directly give the coefficients  
of the Zernike expansion of the measured wavefront.

\subsection{Optimal reconstruction of the on-axis phase from off-axis  
measurements}
\label{sec:43}
We focus here on the reconstruction of the on-axis galaxy's phase in  
the pupil, from the off-axis wavefront sensors measurements.
The goal is therefore to use NGS for wavefront sensing, which in that  
case are very likely to be located far outside of the isoplanatic angle  
$\angleIso$, and to determine the best commands to apply to the DM from  
off-axis measurements, in order to correct the wavefront coming from  
the galaxy's direction.\newline

We recall in this section the main points of tomographic  
reconstruction, as developped by \cite{Fusco-a-2001}, himself inspired  
by \cite{Ragazzoni-a-1999b}. At a given altitude the phase perturbation  
is described by its expansion on a modal basis, defined over a disk  
wider than the pupil and covering any beam imprint of the field  
(\cite{Ragazzoni-a-1999b} call \textit{metapupil} this area). Assuming  
a finite number of turbulent layers, the phase in the whole volume is  
reduced to one vector, and the game is to express the phase expansion  
within the pupil, for a given direction, as a matrix product with this  
vector. Once this relation has been written, the inversion of this  
linear relation has various solutions, depending on the approach  
followed by the authors (least-square, maximum likelihood, etc).

Let us discretise the turbulence profile so that it can be modeled as a  
finite number $N_t$ of turbulent layers. Assuming near-field  
approximation, the resulting phase at a position $\V{r}$ in the pupil  
for a sky direction $\V{\alpha_i}$ can therefore be written as the sums  
of the phase
$\varphi_j$ produced by each turbulent $j^{th}$ layer at the altitude  
$h_j$:
\begin{equation}\label{eq:eq6}	 
\Phi(\grasR,\V{\alpha_i})=\sum_{j=1}^{N_t}\varphi_j(\grasR+h_j\V{\alpha_ i})
\end{equation}

If our system of coordinates is centered on the galaxy (on-axis,  
$\V{\alpha_i}=\V{0}$), then the phase perturbation from the galaxy will  
be written as:
\begin{equation}\label{eq:eq7}
	\Phi_G(\grasR,\V{0})=\sum_{j=1}^{N_t}\varphi_j(\grasR)
\end{equation}

Each phase $\Phi(\grasR,\V{\alpha_i})$ can be expanded as a sum of  
Zernike polynomials where we omit the piston term $Z_1(\grasR)$:
\begin{equation}\label{eq:eq8}
	\Phi(\grasR,\V{\alpha_i})=\sum _{k=2} ^{\infty} a_{\V{\alpha_i},k}  
Z_k(\grasR)
\end{equation}

We assume now that we sense the wavefronts coming from a number  
$N_{GS}$ of stars around the galaxy, i.e. we have $N_{GS}$ WFSs, and  
that we use the combined measurements of all those WFSs to compute the  
on-axis phase coming from the galaxy. Moreover we assume that the WFSs  
measurement are the coefficients 2 to $k_{max}$ of the Zernike  
expansion of the phase, plus some noise. We will now call the  
\textit{measured phase} the following quantity:

\begin{equation}\label{eq:eq9}
	\Phi^m(\grasR,\V{\alpha_i})=\sum _{k=2} ^{k_{max}} a_{\V{\alpha_i},k}  
Z_k(\grasR)+n_i(\grasR)
\end{equation}

where $n_i(\grasR)$ is the measurement noise for the WFS looking at the  
direction $\V{\alpha_i}$, i.e. the propagated noise on the Zernike  
polynomials in the wavefront reconstruction process. We can therefore  
write:
\begin{equation}\label{eq:eq10}
	n_i(\grasR)=\sum _{k=2} ^{k_{max}} \V{n_i}_{,k} Z_k(\grasR)
\end{equation}

The $\V{n_i}_{,k}$ coefficients are stored in a $\V{n_i}$ vector, which  
is a random gaussian vector whose statistics is given by its covariance  
matrix $\V{C_{n,i}}$, dependent of the type of WFS as well as GS  
magnitude and readout noise.\newline

We call now $\V{\Phi_{\alpha_i,k_{max}}}$ the vector storing the  
coordinates of $\Phi(\grasR,\V{\alpha_i})$ up to the Zernike polynomial  
$k_{max}$:
\begin{equation}\label{eq:eq11}
	\V{\Phi_{\alpha_i,k_{max}}}=\left(
		\begin{array}{c}
			a_{\V{\alpha_i},2}\\
			a_{\V{\alpha_i},3}\\
			\vdots\\
			a_{\V{\alpha_i},k_{max}}
		\end{array}
		\right)
\end{equation}

This allows us to define the vector $\V{\Phi^m_{\alpha_i}}$ storing the  
coefficients of the phase $\Phi^m(\grasR,\V{\alpha_i})$:

\begin{equation}\label{eq:eq12}
	\V{\Phi^m_{\alpha_i}}=\left(
	\begin{array}{c}
		a_{\V{\alpha_i},2}+\V{n_i}_{,2}\\
		a_{\V{\alpha_i},3}+\V{n_i}_{,3}\\
		\vdots\\
		a_{\V{\alpha_i},k_{max}}+\V{n_i}_{,k_{max}}
	\end{array}
	\right)=\V{\Phi_{\alpha_i,k_{max}}}+\V{n_i}
\end{equation}

Let us consider now the phase perturbation $\varphi_j(\V{\rho}_j)$ in  
the $j^{th}$ turbulent layer. This latter can be expressed also as a  
sum of Zernike polynomials. We call $\V{\varphi_j}$ the vector storing  
the coefficients of this phase up to the index $N_j$, and we have:
\begin{equation}\label{eq:eq13}
	\V{\varphi_j}=\left(
	\begin{array}{c}
		a_{j,2}\\
		a_{j,3}\\
		\vdots\\
		a_{j,N_j}
	\end{array}
	\right)
\end{equation}

If we concatenate now all the vectors  
$\V{\varphi_j},~j=1,2,\ldots,N_t$, we have the vector $\V{\varphi}$  
storing the coefficients of the phase in the volume:
\begin{equation}\label{eq:eq14}
	\V{\varphi}=\left(
		\begin{array}{c}
			\V{\varphi_1}\\
			\V{\varphi_2}\\
			\vdots\\
			\V{\varphi_{N_t}}\\
		\end{array}
	\right)
\end{equation}

\cite{Ragazzoni-a-1999b,Fusco-a-2001,Femenia-a-2003} have shown that  
there is a linear relation, that we will not detail here, between  
$\V{\Phi_{\alpha_i,k_{max}}}$ and $\V{\varphi}$. It takes the form:
\begin{equation}\label{eq:eq15}
	\V{\Phi_{\alpha_i,k_{max}}}=\matInterTomoi \V{\varphi}
\end{equation}

where the matrix $\matInterTomoi$ is called a "modal projection"  
matrix, performing the sum of the contributions of each wavefront  
$\varphi_j(\V{\rho}_j)$ on the telescope pupil for a given direction  
$\V{\alpha_i}$. Therefore, if we concatenate all the  
$\V{\Phi^m_{\alpha_i}}$ vectors, $i=1,2,\ldots,N_{GS}$ into the single  
vector $\V{\Phi^m}$, we can write \citep{Fusco-a-2001}:
\begin{equation}\label{eq:eq16}
	\V{\Phi^m}=\matInterTomo \V{\varphi}+\V{n}
\end{equation}

where $\V{n}$ is the concatenation of all the measurement noises  
$\V{n_i},~i=1,2,\ldots,N_{GS}$, and $\matInterTomo$ is the  
generalisation to several directions of the matrix  
$\matInterTomoi$.\newline

Let us consider now the on-axis galaxy's phase $\Phi_G(\grasR,\V{0})$.  
This latter can also be written as a sum of Zernike polynomials:
\begin{equation}\label{eq:eq17}
	\Phi_G(\grasR,\V{0})=\sum_{k=2}^\infty a_{G,k}Z_k(\grasR)
\end{equation}

Let $\V{\Phi_{G}}$ be the vector storing all the $a_{G,k}$ coefficients  
up to $\infty$, and $\V{\Phi_{G,k_{max}}}$ be the vector storing these  
coefficients up to the Zernike polynomial $k_{max}$. This last vector  
is also linked to the phase in the volume $\V{\varphi}$ by the  
following relation:
\begin{equation}\label{eq:eq18}
	\V{\Phi_{G,k_{max}}}=\matT \V{\varphi}
\end{equation}

where $\matT$ is the matrix summing the contributions of each wavefront  
$\varphi_j(\V{\rho}_j)$ on the telescope pupil for the on-axis  
direction $\V{\alpha_i=0}$. Our goal is to find the best estimation  
$\V{\hat{\Phi}_{G,k_{max}}}$, which will be the vector storing the  
commands to be applied to the DM, provided that its influences  
functions are also the Zernike polynomials 2 to $k_{max}$. Therefore,  
the correction phase $\hat{\Phi}_G(\grasR,\V{0})$ provided by the DM  
has the following expression:
\begin{equation}\label{eq:eq19}
	\hat{\Phi}_G(\grasR,\V{0})=\sum _{k=2} ^{k_{max}} \hat{a}_{G,k}  
Z_k(\grasR)
\end{equation}

and we seek a relation in the form:
\begin{equation}\label{eq:eq20}
	\V{\hat{\Phi}_{G,k_{max}}}=\V{W}~\V{\Phi^m}
\end{equation}

where $\V{W}$ is the reconstruction matrix linking the commands  
$\V{\hat{\Phi}_{G,k_{max}}}$ to the off-axis measurements $\V{\Phi^m}$.  
Let us define a minimum mean-square error (MMSE) for the  
reconstruction. In our case we want to minimize the residual variance  
of the wavefront $\sigma^2_{\varphi,res}$:

\begin{eqnarray}
\sigma^2_{\varphi,res}&=&\frac{1}{S}\int_{\mathcal{P}}\Moy{\Crochets{\Phi_G(\grasR)-\hat{\Phi}_G(\grasR)}^2 d\grasR}\\
&=&\Moy{\ModuleVect{\V{\Phi_G}-\V{W}~\V{\Phi^m}}^2}\\
&=&\sum_{k=2}^{k_{max}}\Moy{\moiPar{a_{G,k}- 
\hat{a}_{G,k}}^2}+\sum_{k=k_{max}+1}^\infty  
\Moy{a^2_{G,k}}\label{eq:eq21}
\end{eqnarray}

This residual variance is therefore the sum of two terms:
\begin{itemize}
	\item $\sum_{k=2}^{k_{max}}\Moy{\moiPar{a_{G,k}-\hat{a}_{G,k}}^2}$,  
the reconstruction error due to anisoplanatism and noise propagation
	\item $\sum_{k=k_{max}+1}^\infty \Moy{a^2_{G,k}}$, the uncorrected  
variance due to the finite numbers of Zernike polynomials used to  
correct the wavefront.
\end{itemize}

We therefore want to minimize  
$\sum_{k=2}^{k_{max}}\Moy{\moiPar{a_{G,k}-\hat{a}_{G,k}}^2}$. The  
optimal reconstruction matrix satisfying this condition can be written  
as \citep{Wallner-a-1983}:
\begin{equation}\label{eq:eq22}
\V{W}_{opt}=\Moy{\V{\Phi}_{G,k_{max}}~\moiPar{\V{\Phi^m}}^T}\Moy{\V{\Phi 
^m}~\moiPar{\V{\Phi^m}}^T}^{-1}
\end{equation}

\textit{i.e.} this matrix is the product of two matrices: the  
covariance matrix of the unknowns and of the measurements, and the  
inverse of the covariance matrix of the measurements. This expression  
is equivalent to the one obtained by \cite{Fusco-a-2001} with a  
{maximum a posteriori} approach.\newline

Using the equations \eqref{eq:eq16} and \eqref{eq:eq18}, and as the  
noise is uncorrelated from the turbulent phase, the equation  
\eqref{eq:eq22} can be rewritten as:
\begin{equation}\label{eq:eq23}
	\V{W_{opt}}=\matT~\V{C_{\varphi}}\moiPar{\V{M^{N_t}_{\alpha}}}^T  
\Crochets{\V{M^{N_t}_{\alpha}}\V{C_{\varphi}}\moiPar{\V{M^{N_t}_{\alpha} 
}}^T+\V{C_b}}^{-1}
\end{equation}

We recognize in this expression the product of two terms:
\begin{itemize}
	\item the matrix $\V{C_{\varphi}}\moiPar{\V{M^{N_t}_{\alpha}}}^T  
\Crochets{\V{M^{N_t}_{\alpha}}\V{C_{\varphi}}\moiPar{\V{M^{N_t}_{\alpha} 
}}^T+\V{C_b}}^{-1}$, which gives the best tomographic estimation of the  
phase in the volume $\V{\varphi}$ from off-axis measurements. We find  
here the same expression than in the equation (18) of  
\cite{Fusco-a-2001}
	\item the matrix $\matT$: as explained before, it sums the  
contributions of each wavefront $\varphi_j(\V{\rho}_j)$ on the  
telescope pupil for the on-axis direction $\V{\alpha_i=0}$.\newline
\end{itemize}

An important point here is the presence of the matrices  
$\V{C_{\varphi}}$ and $\V{C_b}$, introduced by \cite{Fusco-a-2001},  
which are the generalisation for several layers and for several GSs of  
the classical turbulence and noise covariance matrices. Thanks to the  
information contained in those matrices, it is possible to regularize  
the inversion and increase the field of view where off-axis NGS can be  
picked off to perform wavefront sensing, as well as to use fainter NGS  
than in classical least-square methods. However those matrices require   
some \textit{a-priori} knowledge on the noise measurement, and on the  
turbulence profile.

On a practical point of view, equation \eqref{eq:eq22} shows that  
"half" of the optimal matrix can be measured in-situ: the covariance  
matrix of the measurements is something that can be obtained from the  
experiment itself. Only the covariance matrix between real, actual  
phase and measurements has to be computed.\newline

It must be noticed that we used Zernike polynomials in our simulation  
because they are easy to use, useful mathematical tools. However the  
current study can be applied to any other modal basis. Further work  
about the optimal reconstruction matrix in the case of real AO  
components (Shack-Hartmann or pyramid WFS, segmented or continous  
facesheet DMs) has already been started \citep{Assemat-t-2004}.\newline

\subsection{Detailled presentation of the simulation code}
The elements of this code are the following:
\begin{itemize}
	\item a mono or multi-layer turbulent atmosphere. The number of  
layers, their altitude, the outer scale $\echExt$ as well as the  
strength of the turbulence in each layer (given by a local $\ro$) are  
adjustable. The atmospheric phase screens are simulated using Fourier  
filtering methods \citep{Shaklan-t-1989}, and take into accounts the  
effects of the outer scale $\echExt$ by introducing a proper Von-Karman  
spectrum
	\item any number of NGS, whose position in the field as well as  
magnitudes are adjustable
	\item any number of WFSs positioned on the off-axis NGSs. The WFS
	is supposed to give the Zernike expansion of 
	the phase. Some noise is added to the true Zernike coefficients. The  
noise is related to the guide star magnitude using the propagation  
properties in the reconstruction of Zernike polynomials. It is then  
possible to have $N_{GS}$ measurement vectors $\V{\Phi^m_{\alpha_i}}$,  
which are concatenated in the vector $\V{\Phi^m}$
	\item the computation of the reconstruction matrix $\matRecTomo$ from  
the GS geometry and magnitudes, the turbulent profile and the number  
$k_{max}$ of measured and corrected Zernike polynomials
	\item the computation of the vector  
$\V{\hat{\Phi}_{G,k_{max}}}=\matRecTomo~\V{\Phi^m}$ storing the  
coefficients of the reconstructed on-axis phase on the Zernike  
polynomials 2 to $k_{max}$
	\item a DM with $(k_{max}-1)$ actuators, whose influence functions are  
also the Zernike polynomials 2 to $k_{max}$, and whose commands are  
stored in the vector $\V{\hat{\Phi}_{G,k_{max}}}$, allowing to compute  
the corrected phase $\hat{\Phi}_G(\grasR,\V{0})=\sum _{k=2} ^{k_{max}}  
\hat{a}_{G,k} Z_k(\grasR)$
	\item the computation of the on-axis residual phase  
$\Phi_{res}(\grasR,\V{0})=\Phi_G(\grasR)-\hat{\Phi}_G(\grasR)$
	\item the computation of the short exposure AO corrected PSFs at  
different imaging wavelengths 
	\item the computation of the long exposure AO corrected PSFs by  
averaging the AO short exposure corrected PSFs.\newline
\end{itemize}

We considered in all those simulations a 8 meter telescope (VLT case),  
whose pupil was simulated on a discrete grid of $128 \times 128$  
pixels. Assuming Nyquist sampling in the focal plane ($\lambda/2D$),  
this leads to simulated PSFs covering respectively a field of $4.12  
\times 4.12~arcsec^2$ in J band and $5.44 \times 5.44~arcsec^2$ in H  
band. Each long exposure AO corrected PSF was the average of 100  
independent short exposure PSFs, as we assumed an open-loop system and  
that we did not consider any temporal error.\newline

\subsection{Simulation conditions}\label{sec:condSimul}
We are giving in this paragraph more details about the conditions of  
the simulations we performed.

\subsubsection{Atmospheric conditions}
Let us first deal with atmospheric  
conditions. We used data from ESO AO's department, who provided us with  
some statistics about the seeing $\theta_{1/2}$, the isoplanatic angle  
$\angleIso$ and the coherence time $\tau_0$. We therefore considered  
some median atmospheric conditions, leading us to the following  
quantities (all data given for zenith and for a wavelength of $0.5~\mu  
m$):
\begin{itemize}
	\item a median seeing $\theta_{1/2}=0.81~arcsec$. This latter is  
linked to the Fried parameter $\ro$ by the relation  
$\theta_{1/2}=0.976~\lambda/\ro$. We therefore find a median Fried  
parameter $\ro=12.42~cm$ at $0.5~\mu m$ for the whole turbulence  
profile
	\item a median isoplanatic angle $\angleIso=2.42~arcsec$
	\item a median coherence time $\tau_0=3.04~ms$
\end{itemize}

We therefore used the median seeing and isoplanatic angle to define our  
turbulence profile, and find that a profile made of 3 turbulent layers  
located at altitudes of 0 (ground layer), 1 and 10 km, and with  
respectively $20\%$, $65\%$ and $15\%$ of the whole turbulence allowed  
to reproduce these atmospheric conditions.\newline

Another important issue is the outer scale $\echExt$, which has a  
direct influence on the variance of low orders of the turbulence and  
the FWHM of uncorrected images \citep{Tokovinin-a-2002}.  
\cite{Martin-a-2000} give some statistics for this parameter at the  
Cerro Paranal, and we adopted their median value, i.e.  
$\echExt=24~m$.\newline

\subsubsection{AO system parameters}
We focus now on the components of the AO system. We considered for each  
simulation case a tomographic system, with 3 off-axis wavefront sensors  
and one deformable mirror. As said before, we assumed a correction degree
ranging from 0 to 120 Zernike polynomials.
We assumed we  
use wavefront slope sensors (Shack-Hartmann or pyramid), leading to a  
propagated noise variance on Zernike polynomials following a law in  
$(n+1)^{-2}$ \citep{Rigaut-a-1992}.

\subsubsection{AO system limiting magnitude}
The result of our study depends on the WFS noise level, and this latter is a function of the photon flux.
We assumed to be in a  
regime dominated by photon noise. In that case, the noise variance is  
inversely proportional to the number of photoelectrons per frame  
\citep{Rigaut-a-1992}, $N_{ph}^{-1}$. Hence, the noise behavior of the sensor will entierely be defined versus flux and for any Zernike mode, when the constant of proportionnality of the law is given.

The way we propose to fix this constant of proportionnality is to define the flux level that produces a particular noise variance. In order to make this flux level number physically meaningful, we will choose for our noise variance value, the one which corresponds to the limiting magnitude of the system. Doing this, setting a limiting magnitude implies a particular noise variance, and consequently defines the complete noise behavior for any flux level. The usual way of doing it is to derive the WFS performance (and the limiting magnitude) from the list of the WFS characteristics : detector noise, optical throughput, spectral bandpass, number of lenslets, etc. We think it is not reasonnable to undertake such a study : not only because this would always be possibly technically questionnable, out of the scope of the article, but also because a study based on current existing designs might be obsolete in a few years.\newline

Instead, our approach is to consider the WFS limiting magnitude as the input parameter to our study, not as an output number resulting from a complete modelisation of the wavefront sensor. Here, it should be seen as a specification number, when it's a matter of designing the wavefront sensor. The question is then to decide of a value for the noise variance that characterizes the limiting magnitude. The choice is just a matter of definition. In order to give meanigful numbers, we took a criterion that agrees with actual AO-users' experience. Practically speaking, 
the criteria is reached when the user might be discouraged trying higher magnitudes, considering that the correction only brings marginal improvement. 
From our experience with the NAOS Adaptive Optics system, behind this fuzzy rule of thumb is hidden a clear-cut number : we choose a criterion of a noise level value of $250~rd^2$ at 0.5 microns. This number really fits with what is actually called "limiting magnitude" by the AO community. It corresponds to a magnitude of $R = 17$ \citep{Fusco-a-2004} on the NAOS system (whose detector on the WFS has a read-out noise equal to 4 e- rms per pixel). At this magnitude, the NAOS system uses $7 \times 7$ subapertures and controls 42 mirror modes, only 10-15 of them being efficiently compensated.\newline

Now, the choice of the value for the limiting magnitude has to be done. We cannot arbitrarily increase it, and in particular we can set an upper limit, translating into limiting magnitude the classical laws of photon-noise limited wavefront sensing, that are widely used in the litterature. Following \citet{Rigaut-a-1992}, when the subapertures are larger than $r_0$, the noise variance can be written as:
\begin{equation}
\sigma^2 = \frac{2\pi}{n_0 r_0^2}
\end{equation}
with $n_0$ the number of photoelectrons per surface area. Using that formula, if we consider a spectral interval of $0.4~\mu m$, a sampling frequency of 100 Hz, and a global optical throughput of 0.5, then the value $\sigma^2 = 250~rd^2$ is attained when $n_0 = 1.63 phe^-/m^2/frame$, i.e. at a limiting magnitude of $R=20.0$. This number must be taken as the very upper limit, since this theoretical formula only assumes quantum-noise limitation, excluding any other instrumental effect.\newline

We know that in a near future, progress in the fast readout detectors may lead to much better results than NAOS. As an example, L3CCD \citep{Daigle-p-2004} are extremely good candidates, that could bring the noise level below 1 $e^-$ rms per pixel, i.e. increase the limiting magnitude of +1.5 compared to NAOS. However, we have decided to be rather conservative, and we will study in this article two cases of limiting magnitudes, $R_{max}=16$ and $R_{max}=17$, that we think to be realistic. In that case, one may then wonder how our results shown in section \ref{sec:sec57} can exhibit corrections up to 120 Zernike polynomials, when we noticed that a sensor working at its limiting magnitude -considering our definition- can only efficiently compensate for 10 to 15 Zernike modes (out of 42 effectively handled, and partially corrected). The answer is that the magnitudes of the 3 guide stars are not simultaneously equal to $R_{max}$: as shown in the figure \ref{fig:fig13}, the star surface density increases as a function of the $R$ magnitude by a factor 2 every 1.5 magnitudes for magnitudes $R>13$. It means that the median magnitude of a field of stars brighter than $R=R_{max}$ is equal to $R_{max}-1.50$. Then as the signal of the 3 wavefront sensors are combined together through an optimal reconstructor where both statistics of measurement noise and anisplanatism are taken into account, thus increasing the final SNR compared to the one that would be given by a single WFS, this still allows us to partially compensate for high-order modes. And as it was explained in section \ref{sec:sec32}, such a correction is particularly effective at increasing the ensquared energy of the AO corrected PSF. \newline

\subsubsection{Star density}
Let us consider then the three off-axis NGS used to perform off-axis  
wavefront sensing and on-axis wavefront reconstruction. The angular  
distance of the first, second and third closest NGS to the central  
object, is directly linked to the galactic latitude, as the surface  
density of stars decrease with galactic latitude  
\citep{Bahcall-a-1980}, and has a direct impact on the performance of  
the tomographic AO system. In order to have some estimation of the sky  
coverage of the FALCON's AO system, we therefore decided to perform our  
simulations on three stellar fields, located at the following galactic  
coordinates:
\begin{enumerate}
	\item $l=280.2\Moideg,~b=-88.5\Moideg$
	\item $l=0.3\Moideg,~b=-61.4\Moideg$
	\item $l=44.5\Moideg,~b=-30.9\Moideg$ \newline
\end{enumerate}

\begin{table*}
	\centering
		\begin{tabular}{|c|ccc|}
			\hline
			$b$&$N(R_{max}=16)$&$N(R_{max}=17)$&$N(R_{max}=18)$\\
			\hline
			$-30\Moideg$&1271&2048&3096\\
			$-60\Moideg$&400&639&946\\
			$-90\Moideg$&246&361&522\\
			\hline
		\end{tabular}
	\caption{Number of stars per square degree as a function of the  
galactic latitude $b$ and the magnitude $R_{max}$}
\label{tab:table2}
\end{table*}

\begin{figure*}
\centering
\includegraphics[width=0.33\linewidth]{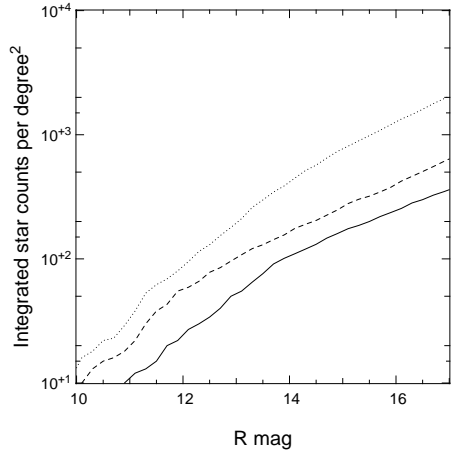}
\caption{Integrated star counts per square degree in R band as a  
function of the R magnitude given by Mod\`ele de Besan\c con for three  
galactic latitudes: $b\approx -90 \Moideg$ (full line), $b\approx -60  
\Moideg$ (dashed line) and $b\approx -30 \Moideg$ (dotted line).}
\label{fig:fig13}
\end{figure*}

We used statistics given by the \textit{Mod\`ele de Besan\c con}  
\citep{Robin-a-2003} to simulate star fields for these three galactic  
latitudes. This requires to know the surface density of stars, which  
are shown on the figure \ref{fig:fig13}, and the table \ref{tab:table2}  
shows the integrated number of stars per square degree for the limiting  
magnitudes $R_{max}=16$, $R_{max}=17$ and $R_{max}=18$. Then, for each  
field, we choose 100 random positions. For each position, we chose the  
three closest NGS with a magnitude $R\leq R_{max}$ to perform wavefront  
sensing, and computed the reconstruction matrix given by equation  
\eqref{eq:eq23} for each maximal degree of correction. We were then  
able to reconstruct the on-axis phase from off-axis measurements,  
compute the on-axis residual phase, and the long exposure AO corrected  
PSF for different correction orders in J and H band.\newline

The quality of the on-axis wavefront reconstruction is directly linked  
to the geometry of the closest NGS used to perform wavefront sensing:  
anisoplanatism will be increased when the NGS will be far away of the  
scientific target. Moreover, as we choose the three closest NGS, it is  
very likely that those latest will be faint, meaning an important  
measurement noise. We are therefore giving in the next paragraph some  
analytical formulae for the distances of the first, second and third  
closest NGS.

\subsection{Distances of the closest NGS}
Let us call $\sigma_*$ the surface density of stars, i.e. the number of  
stars per angular surface on the sky. As an exemple, the previous  
paragraph has shown that we have $\sigma_*=246$ stars per square degree  
for $R_{max}=16$ and $b \approx -90 \Moideg$, thus $\sigma_*=0.07$  
stars per square arcmin. Therefore, if we call $\mathcal{N}(r)$ the  
average number of stars in a circle of radius $r$ (in arcmin), we'll  
have $\mathcal{N}(r)=\pi \sigma_* r^2$. As the spatial distribution of  
stars on the sky follows an uniform probablity law, the number of stars  
in an area of $\pi \sigma_* r^2~arcmin^2$  follows a Poisson law. So,  
the probability to find $k$ stars in an area of $\pi r^2~arcmin^2$ is  
equal to:
\begin{equation}
	P(\xi=k)=\frac{\mathcal{N}^k(r)}{k!}\exp \moiPar{-\mathcal{N}(r)}
\end{equation}

The probability that there is then \textbf{one} star at a distance  
greater than $r$ is also equal to the probability that there is  
\textbf{no} star in an area equal to $\pi D r^2~arcmin^2$, i.e.  
$P(\xi=0)=\exp{-\mathcal{N}(r)}=\exp\moiPar{-\pi \sigma_* r^2}$. Then,  
the probability that there is \textbf{one} star at a distance $D_1 \leq  
r$ is equal to
\begin{eqnarray}
	P_1(r)&=&P(D_1 \leq r)=1-P(\xi=0)\nonumber \\
	&=&1-\exp\moiPar{-\mathcal{N}(r)}\nonumber \\
	&=&1-\exp \moiPar{-\pi \sigma_* r^2}\label{eq:eq29}
\end{eqnarray}

and we recognize the cumulative distribution function (cdf) $P_1(r)$ of  
the random variable $D_1$, this latter being the distance of the first  
closest NGS. Then, the probability of having \textbf{one} star at a  
distance between $r$ and $r+\Delta r$ will be equal to the derivative  
of the cdf of $D_1$, i.e. its probability density function (pdf)  
$p_1(r)$, which is equal to:
\begin{eqnarray}
	p_1(r)&=&\frac{d}{dr}P_1(r)\nonumber \\
	&=&2\pi \sigma_* r \exp \moiPar{-\mathcal{N}(r)}\nonumber \\
	&=&2\pi \sigma_* r \exp \moiPar{-\pi \sigma_* r^2}
\end{eqnarray}

\begin{figure*}
	\centering
	\begin{tabular}{cc}
		\includegraphics[width=0.4\textwidth]{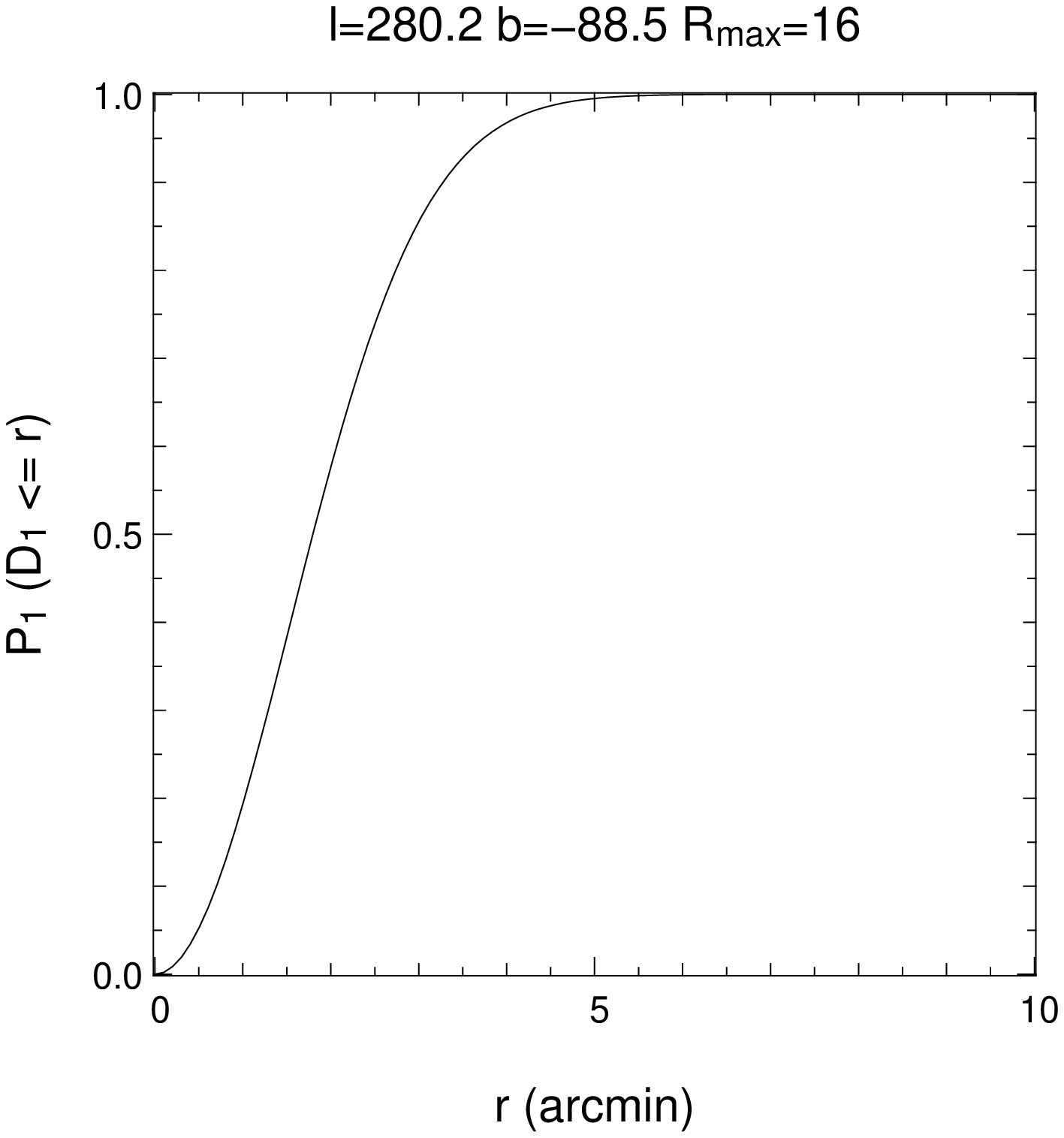}&
		\includegraphics[width=0.4\textwidth]{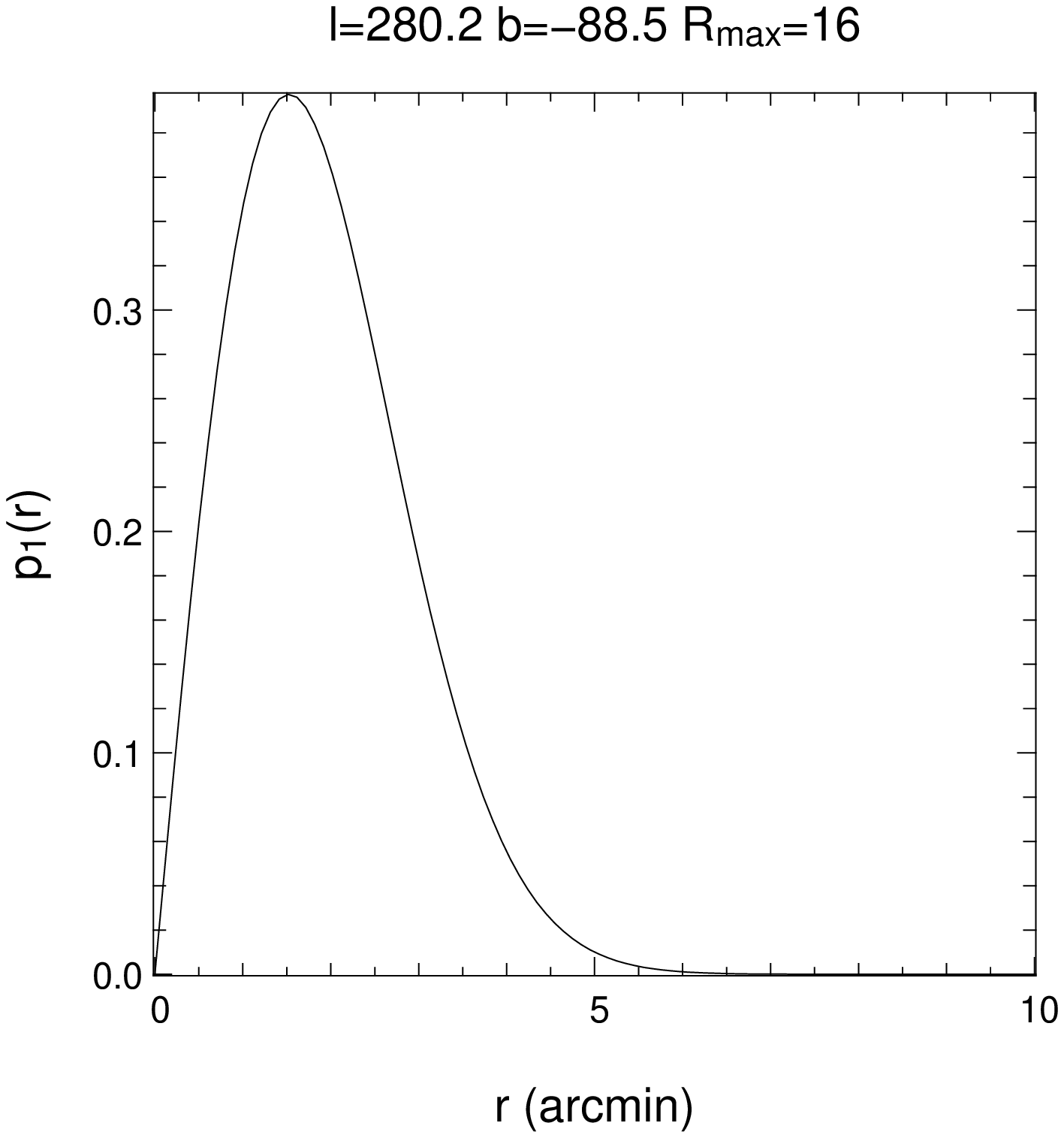}\\
	\end{tabular}
\caption{Cumulative distribution function (left) and probability  
density function (right) of the distance of the first closest NGS  
computed from the star density at the galactic pole ($b\approx -90  
\Moideg$, $R_{max}=16$) provided by \textit{Mod\`ele de Besan\c con}.}
\label{fig:fig14}
\end{figure*}

As an example, the figure \ref{fig:fig14} shows the cdf and the pdf of  
$D_1$ for a galactic latitude $b \approx -90 \Moideg$ and a limiting  
magnitude $R_{max}=16$. Knowing it, it is possible to know two  
interesting quantites. First the average value of $D_1$:
\begin{eqnarray}
	\Moy{D_1}&=&\int_{0}^{\infty}r~p_1(r)dr \nonumber \\
	&=&2\pi \sigma_* \int_{0}^{\infty}r^2\exp \moiPar{-\pi \sigma_* r^2}dr  
\nonumber \\
	&=&\frac{1}{2\sqrt{\sigma_*}}\label{eq:eq31}
\end{eqnarray}

and the most likely value of $D_1$, that we'll note $D_{1,ml}$, given  
by the maximum of the pdf $p_1(r)$. The determination of this maximum  
requires to compute the derivative of $p_1(r)$, whose expression is:
\begin{equation}
	\frac{d}{dr}p_1(r)=2 \pi \sigma_* \exp\moiPar{-\pi \sigma_*  
r^2}\moiPar{1-2\pi \sigma_* r^2}
\end{equation}

and which is equal to 0 for
\begin{equation}\label{eq:eq32}
	D_{1,ml}=\frac{1}{\sqrt{2 \pi \sigma_*}}
\end{equation}

Let us focus now on $D_2$, the distance of the second closest NGS.  
Doing the same assumptions than before, the probability to find two  
stars at a distance $D_2 \leq r$ is equal to:
\begin{eqnarray}
	P_2(r)&=&P(D_2 \leq r)\nonumber \\
	&=&1-\Crochets{P(\xi=0)+P(\xi=1)} \nonumber \\
	&=&1-\exp\moiPar{-\mathcal{N}(r)}(1+\mathcal{N}(r))\nonumber \\
	&=&1-\exp\moiPar{-\pi \sigma_* r^2}(1+\pi \sigma_* r^2)
\end{eqnarray}

The pdf of $D_2$ has then the following expression:
\begin{eqnarray}
	p_2(r)&=&2 \pi \sigma_* r \mathcal{N}(r) \exp  
\moiPar{-\mathcal{N}(r)}\nonumber \\
	&=& 2\pi^2 \sigma_*^2 r^3 \exp \moiPar{-\pi \sigma_* r^2}
\end{eqnarray}

and the average value of $\Moy{D_2}$ can be computed:
\begin{eqnarray}
	\Moy{D_2}&=&\int_0^\infty r~p_2(r)dr \nonumber \\
	&=&2 \pi \sigma_* \int_0^\infty r^2 \mathcal{N}(r) \exp  
\moiPar{-\mathcal{N}(r)}\\
	&=&\frac{3}{4\sqrt{\sigma_*}}\label{eq:eq34}
\end{eqnarray}

The derivative of $p_2(r)$ is equal to:
\begin{eqnarray}
	\frac{d}{dr}p_2(r)&=&2 \pi \sigma_* \mathcal{N}(r)  
\exp{-\mathcal{N}(r)}\Crochets{3-2\mathcal{N}(r)} \nonumber \\
	&=&2 \pi^2 \sigma_*^2 r^2 \exp{- \pi \sigma_* r^2}\Crochets{3-2\pi  
\sigma_* r^2}
\end{eqnarray}

and is equal to 0 for
\begin{equation}\label{eq:eq35}
	D_{2,ml}=\sqrt{\frac{3}{2 \pi \sigma_*}}
\end{equation}

We can now focus on $D_3$, the distance of the third closest NGS. The  
probability that there are three stars at a distance $D_3\leq r$ is  
equal to:
\begin{eqnarray}
	P_3(r)&=&P(D_3 \leq r)=1-\Crochets{P(\xi=0)+P(\xi=1)+P(\xi=2}  
\nonumber \\
	&=&1-\exp\moiPar{-\mathcal{N}(r)}\moiPar{1+\mathcal{N}(r)+\frac{\mathcal{N}^2(r)}{2}}\nonumber \\
&=&1-\exp \moiPar{-\pi \sigma_* r^2}\moiPar{1+\pi \sigma_*r^2+\frac{\pi^2 \sigma_*^2 r^4}{2}}
\end{eqnarray}

The pdf has then the following expression:
\begin{eqnarray}
	p_3(r)&=&\pi \sigma_* r \mathcal{N}^2(r) \exp  
\moiPar{-\mathcal{N}(r)}\nonumber \\
	&=& \pi^3 \sigma_*^3 r^5 \exp \moiPar{-\pi \sigma_* r^2}
\end{eqnarray}

and can be used to compute the average distance $\Moy{D_3}$:
\begin{eqnarray}
	\Moy{D_3}&=&\int_0^\infty r~p_3(r)dr \nonumber \\
	&=&\pi \sigma_* \int_0^\infty r^2 \mathcal{N}^2(r) \exp  
\moiPar{-\mathcal{N}(r)} \nonumber \\
	&=&\frac{15}{16\sqrt{\sigma_*}}
\end{eqnarray}

The first derivative of $p_3(r)$ is equal to
\begin{eqnarray}
	\frac{d}{dr}p_3(r)&=&\pi \sigma_* \mathcal{N}^2(r)  
\exp{-\mathcal{N}(r)}\Crochets{3-2\mathcal{N}(r)} \nonumber \\
	&=&2 \pi^2 \sigma_*^2 r^2 \exp{- \pi \sigma_* r^2}\Crochets{5-2\pi  
\sigma_* r^2}
\end{eqnarray}

and is equal to 0 for
\begin{equation}\label{eq:eq36}
	D_{3,ml}=\sqrt{\frac{5}{2 \pi \sigma_*}}
\end{equation}

\begin{table*}
	\centering
	\begin{tabular}{|c|c|c|c|}
	\hline
	&Expression&$\Moy{D}$&$D_{ml}$\\
	\hline
	$p_1$&$2 \pi \sigma_* r \exp \moiPar{- \pi \sigma_* r^2}$&$\frac{1}{2  
\sqrt{\sigma_*}}$&$\sqrt{\frac{1}{2 \pi \sigma_*}}$\\
	\hline
	$p_2$&$2 \pi^2 \sigma_*^2 r^3 \exp \moiPar{- \pi \sigma_*  
r^2}$&$\frac{3}{4 \sqrt{\sigma_*}}$&$\sqrt{\frac{3}{2 \pi \sigma_*}}$\\
	\hline
	$p_3$&$\pi^3 \sigma_*^3 r^5 \exp \moiPar{- \pi \sigma_*  
r^2}$&$\frac{15}{16 \sqrt{\sigma_*}}$&$\sqrt{\frac{5}{2 \pi  
\sigma_*}}$\\
	\hline
	\end{tabular}
	\caption{Summary table for the probability density function, the  
average and the most likely value of the distance of the closest first,  
second and third NGS used for tomography as a function of the star  
surface density $\sigma_*$}
\label{tab:table3}
\end{table*}

\begin{table*}
	\centering
		\begin{tabular}{|c|cc|cc|cc|cc|}
			\hline
$b$&$R_{max}$&$\sigma_*$&$\Moy{D_1}$&$D_{1,ml}$&$\Moy{D_2}$&$D_{2,ml}$&$ 
\Moy{D_3}$&$D_{3,ml}$\\
	\hline
	$-30\Moideg$&16&0.35&40&50&70&76&90&94\\
	$-30\Moideg$&17&0.57&32&40&55&60&71&75\\
	\hline
	$-60\Moideg$&16&0.11&72&90&124&135&160&169\\
	$-60\Moideg$&17&0.17&57&71&98&107&127&133\\
	\hline
	$-90\Moideg$&16&0.07&92&115&158&172&205&215\\
	$-90\Moideg$&17&0.10&76&95&130&142&169&177\\
	\hline
	\end{tabular}
	\caption{Average and most likely distance of the closest first, second  
and third NGS as a function of the galactic latitude and the limiting  
magnitude. From left to right: the galactic latitude $b$, the limiting  
magnitude $R_{max}$, the number of stars per square arcmin $\sigma_*$,  
and the average and most likely value of the distances of the off-axis  
NGS (in arcsec). From star densities given by Mod\`ele de Besan\c con.}
\label{tab:table4}
\end{table*}

The table \ref{tab:table3} summarises the above results, whereas the  
table \ref{tab:table4} gives some values of distances for the three  
galactic latitudes we studied and for the two cases of limiting  
magnitude. We can notice there that the distances of the closest NGS  
fastly increase with the galactic latitude $b$. As an example, for a  
limiting magnitude $R_{max}=16$, the distance $D_{1,ml}$ goes from  
$50~arcsec$ for $b=-30 \Moideg$ to $115~arcsec$ at the galactic pole.  
The difference is more important for the third NGS: we have  
$D_{1,ml}=94~arcsec$ at $b=-30 \Moideg$ (which remains usable thank to  
tomography), but at the galactic pole, the third NGS is very likely to  
be at a distance of more than 3 arcmin ! The gain brought by a higher  
sensitivity becomes then obvious: if fainter NGS (i.e. $R_{max}=17$)  
are used, these distances dramatically decrease. At a galactic latitude  
$b=-30 \Moideg$, the distances of the closest first and third NGS are  
respectively 10 and $20~arcsec$ smaller than for $R_{max}=16$, and for  
greater galactic latitudes ($\Module{b} \geq 60 \Moideg$), these  
distances are 20 and $35~arcsec$ smaller. Such differences are not  
negligible for AO and particularly for anisoplanatism's correction, and  
show that the improvement of WFS's sensitivities is really mandatory  
for future systems.\newline

\subsection{Sky coverage: definition of a criterium}
One of the key issues of AO systems is their sky coverage, i.e. the  
probability to reach a certain performance thanks to the correction  
provided by AO. For the studies of new AO systems, it is very useful to  
be able to know the variability law of the sky coverage as a function  
of observing conditions, like the galactic latitude or the limiting  
magnitude of off-axis guide stars.\newline

This implies therefore to set the performance we want to reach. As we  
have seen in the section \ref{sec:sec3}, ensquared energies of 30\% in  
J band and 40\% in H band could allow to reach a sufficient SNR to  
perform the dynamical studies of galaxies located at $z\geq 0.9$ with  
an angular resolution of $0.25~arcsec$ and a spectral resolution  
$R=10000$, and a minimum exposure time of 3 hours. Such values of  
\textbf{absolute} ensquared energy can therefore be used as a basis to  
design the instrument, and we show in the next sections the results we  
obtained for this specification. Definitive values of the absolute  
ensquared energy would however require to made the same simulations for  
a broader sample of galaxies, which currently are not possible because  
of the current lack of 3D spectroscopic data for distant galaxies.

\subsection{Results}
\label{sec:sec57}
\begin{figure*}
	\centering
	\begin{tabular}{cc}
		\includegraphics[width=0.45\textwidth]{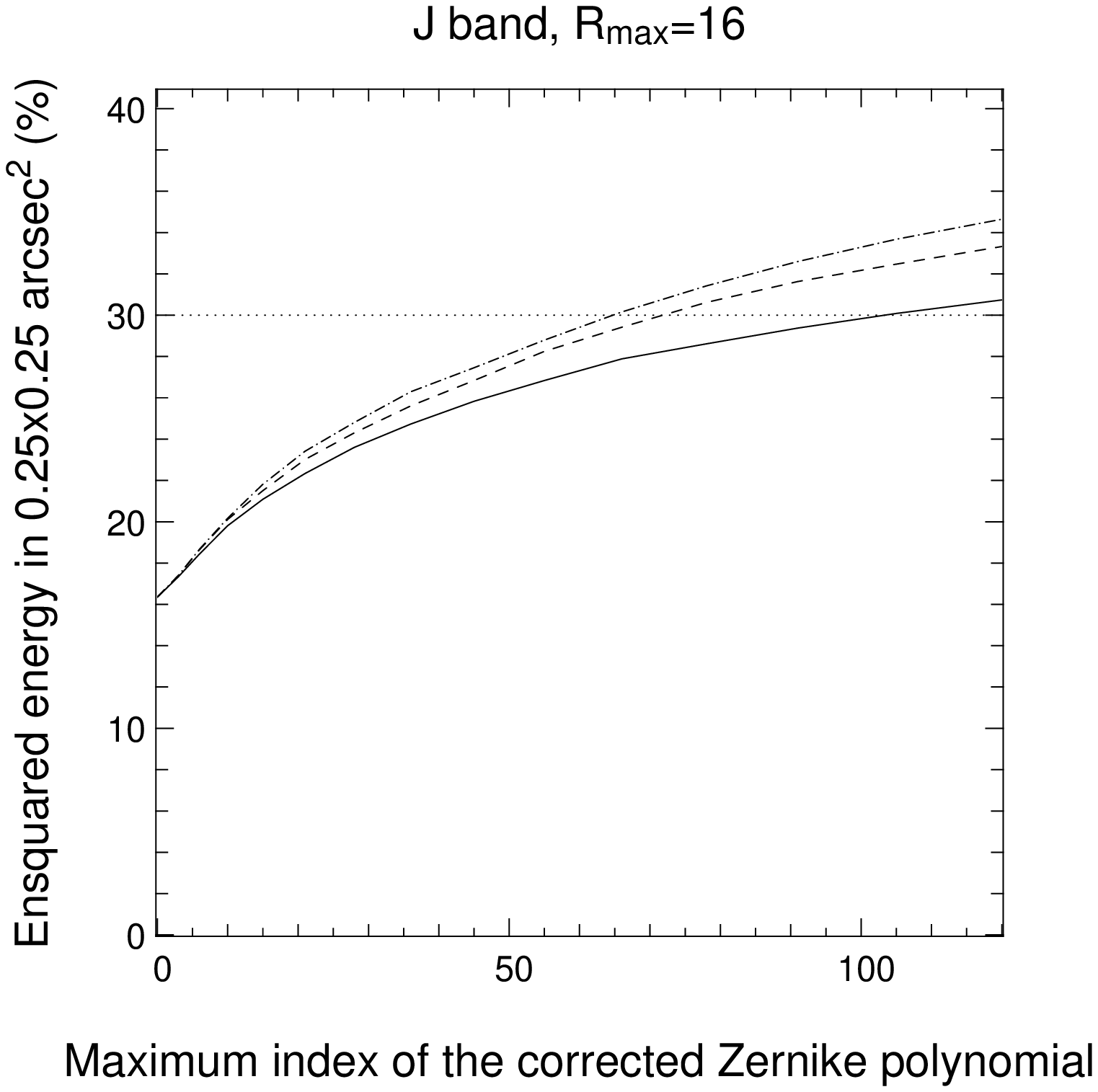}&
		\includegraphics[width=0.45\textwidth]{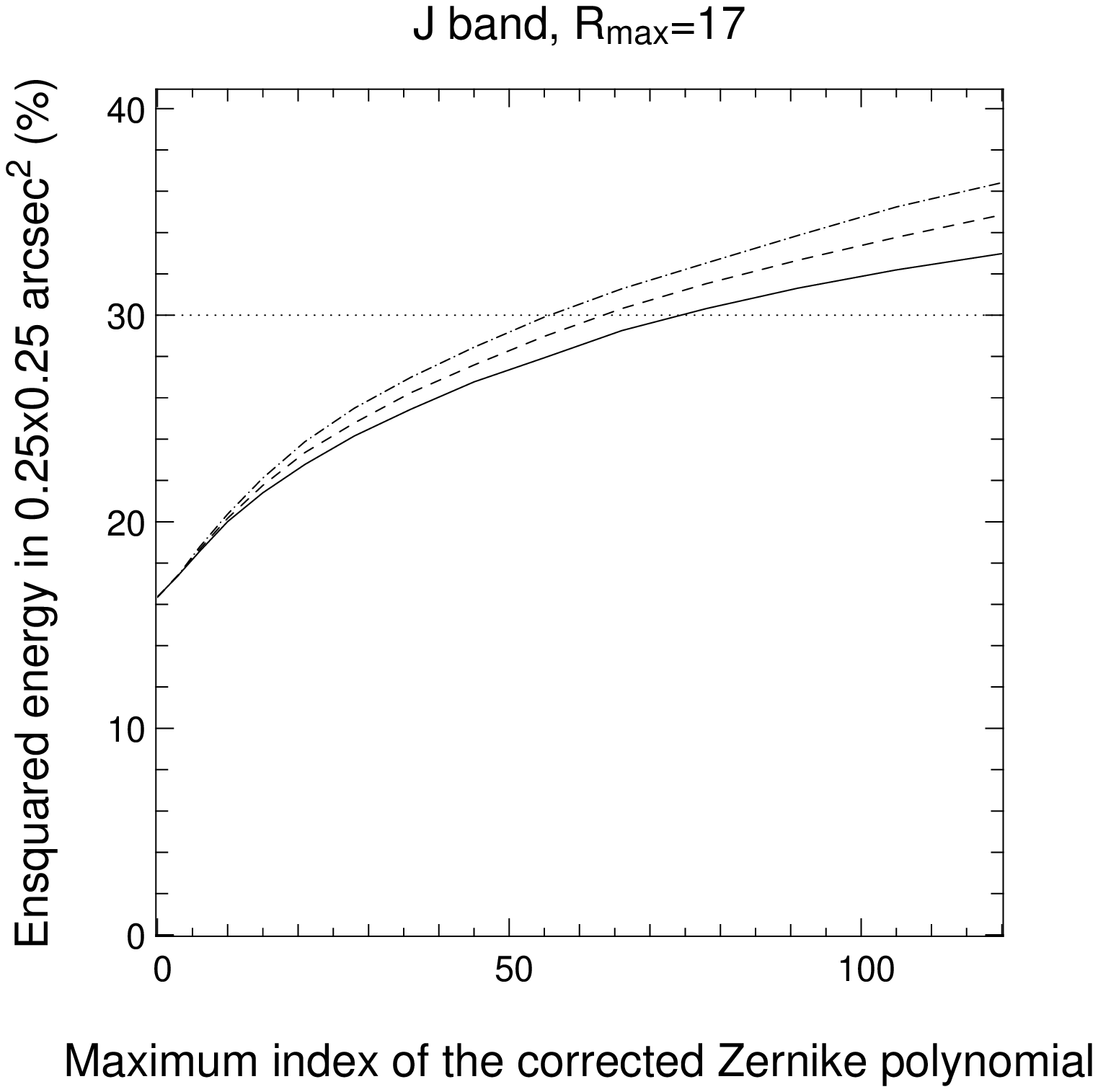}\\\\
		\includegraphics[width=0.45\textwidth]{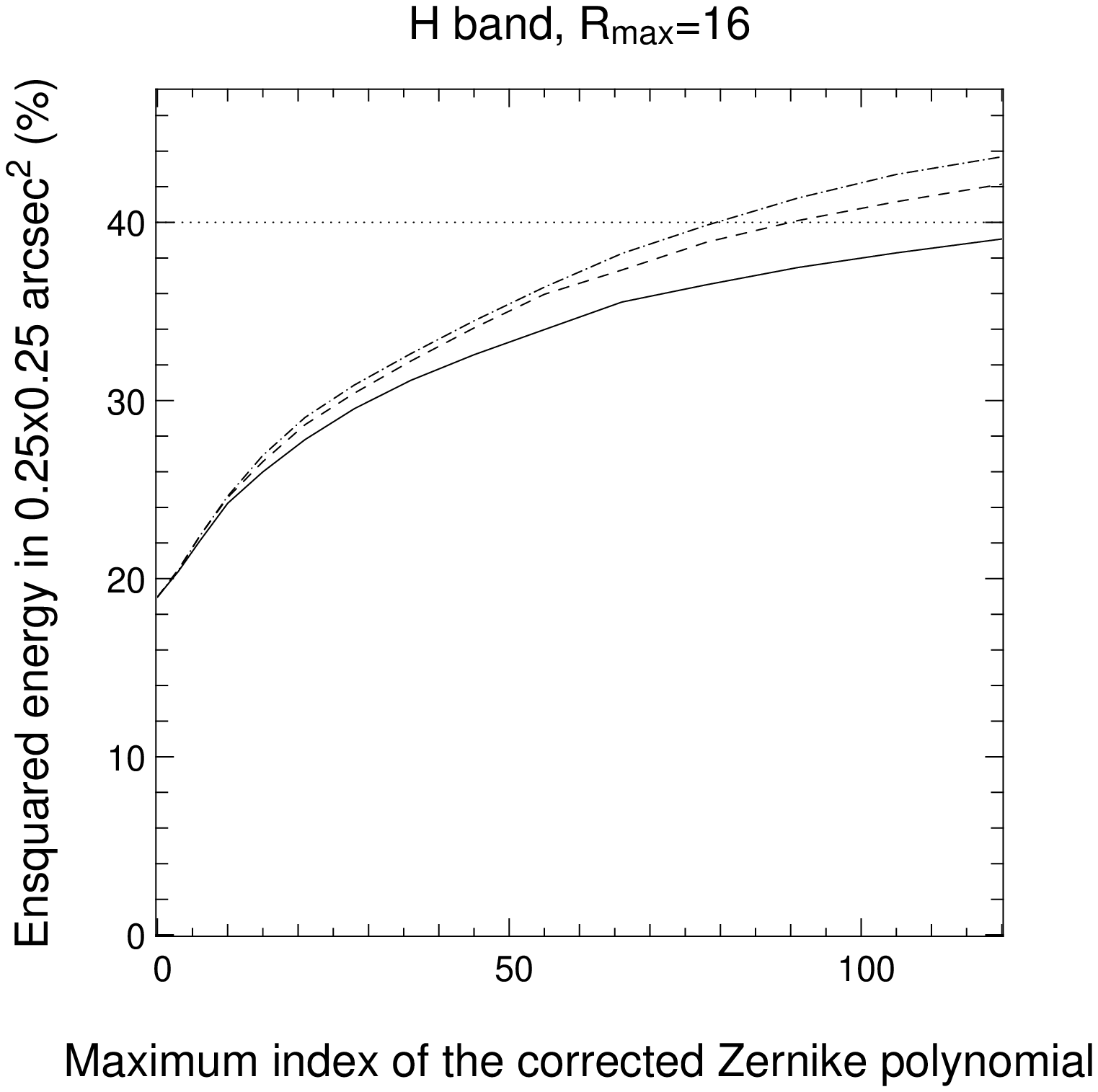}&
		\includegraphics[width=0.45\textwidth]{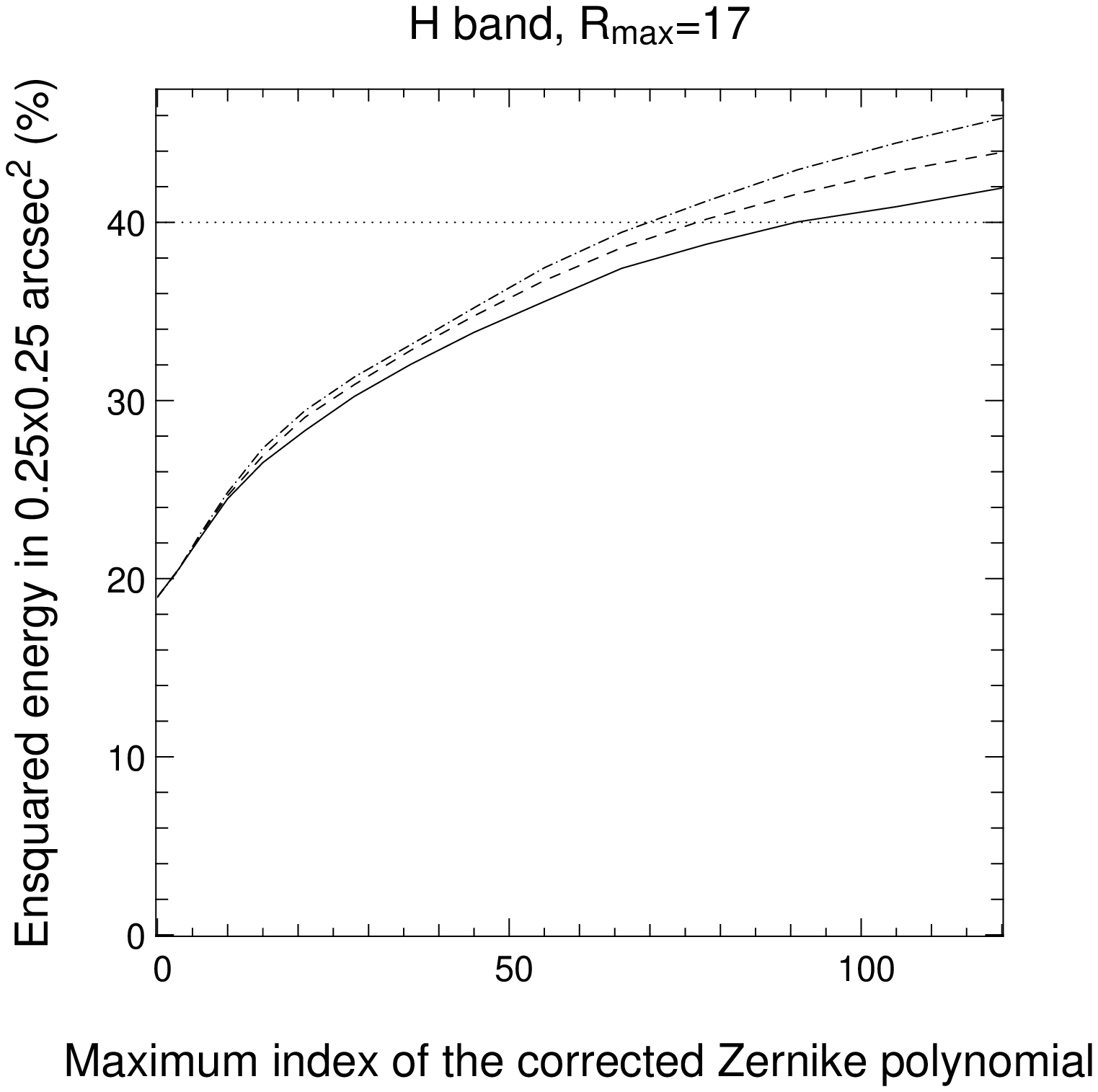}\\\\
	\end{tabular}
\caption{Evolution of the median ensquared energy in a $0.25 \times  
0.25~arcsec^2$ square aperture for each galactic latitude $b$ and each  
limiting magnitude $R_{max}$ in the J and H band, as a function of the  
number of corrected Zernike polynomials. The full line corresponds to  
the median performance at the galactic pole, the dashed one to the  
median performance at $b\approx -60 \Moideg$, and the dash-dotted one  
at $b\approx -30 \Moideg$. The dashed horizontal line shows the  
specification, i.e. 30\% in J band and 40\% in H band.}
\label{fig:fig15}
\end{figure*}

\begin{figure*}
	\centering
	\begin{tabular}{cc}
		\includegraphics[width=0.45\textwidth]{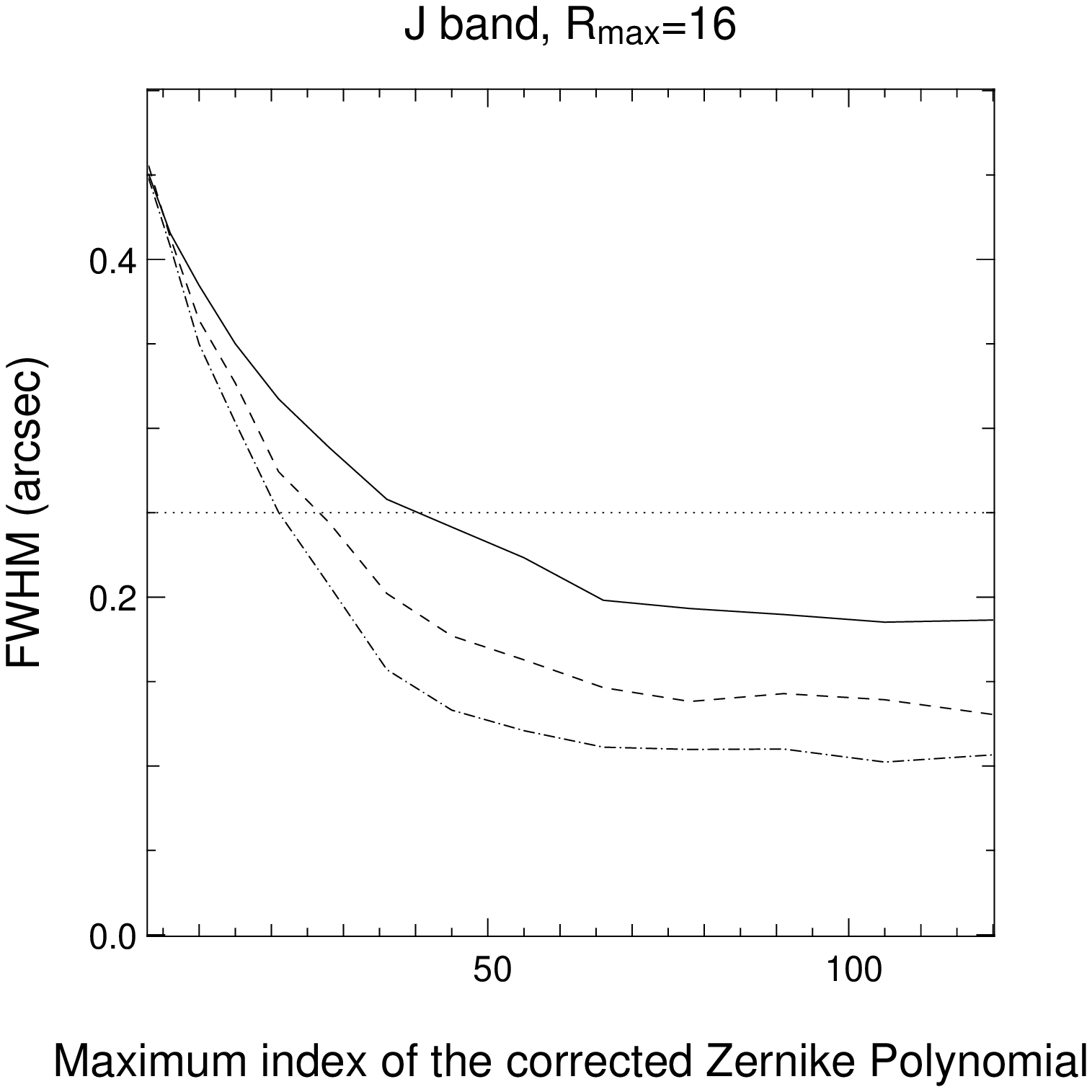}&
		\includegraphics[width=0.45\textwidth]{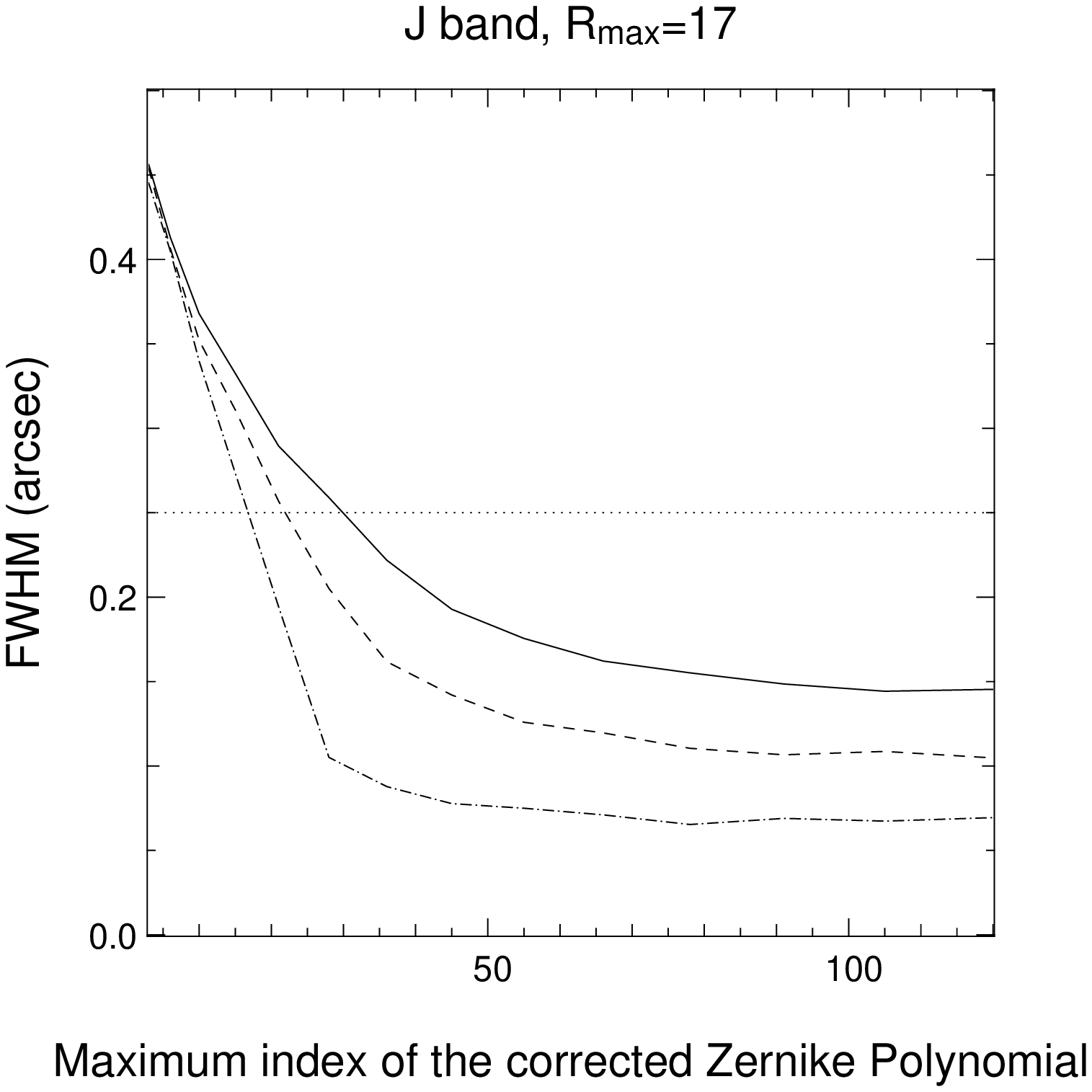}\\\\
		\includegraphics[width=0.45\textwidth]{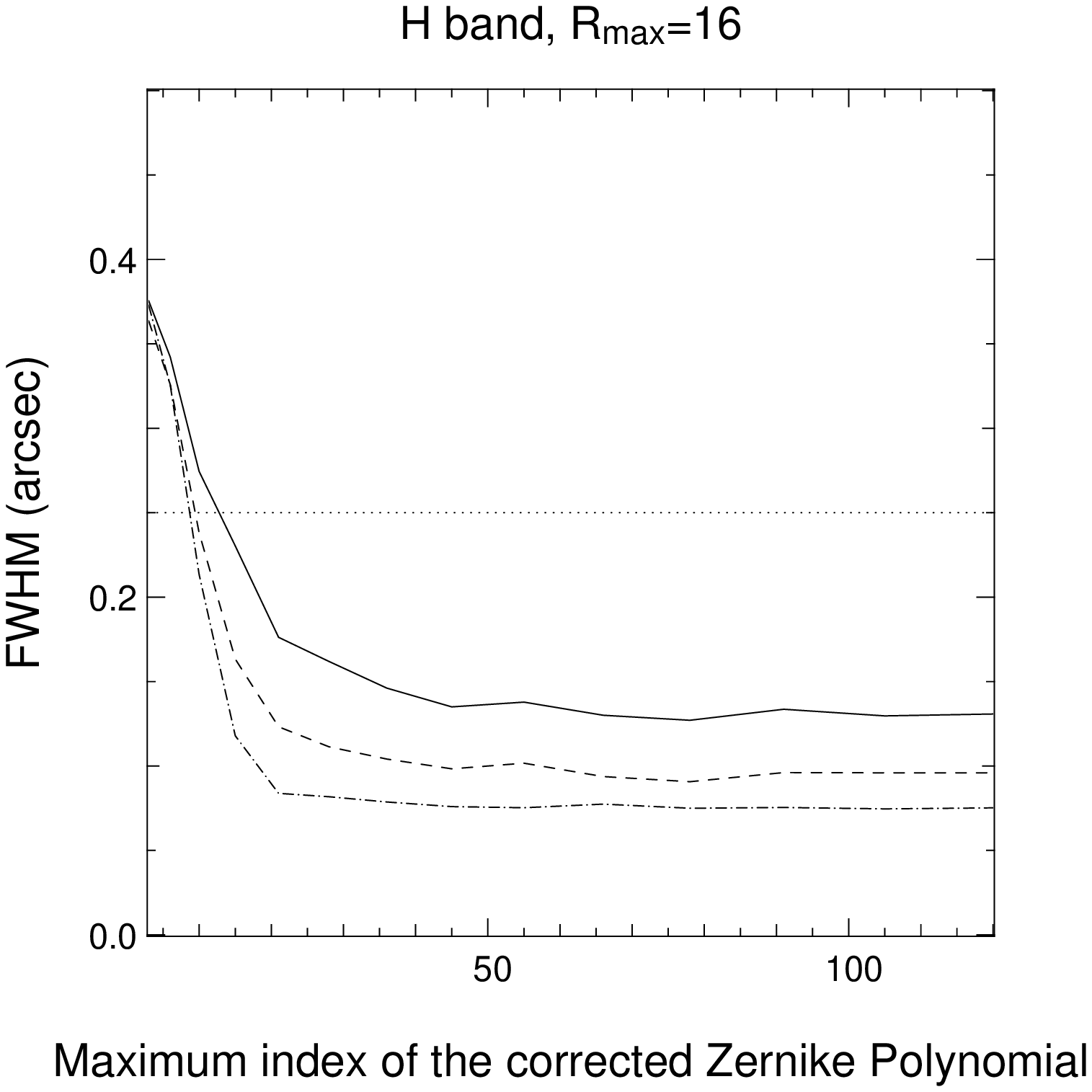}&
		\includegraphics[width=0.45\textwidth]{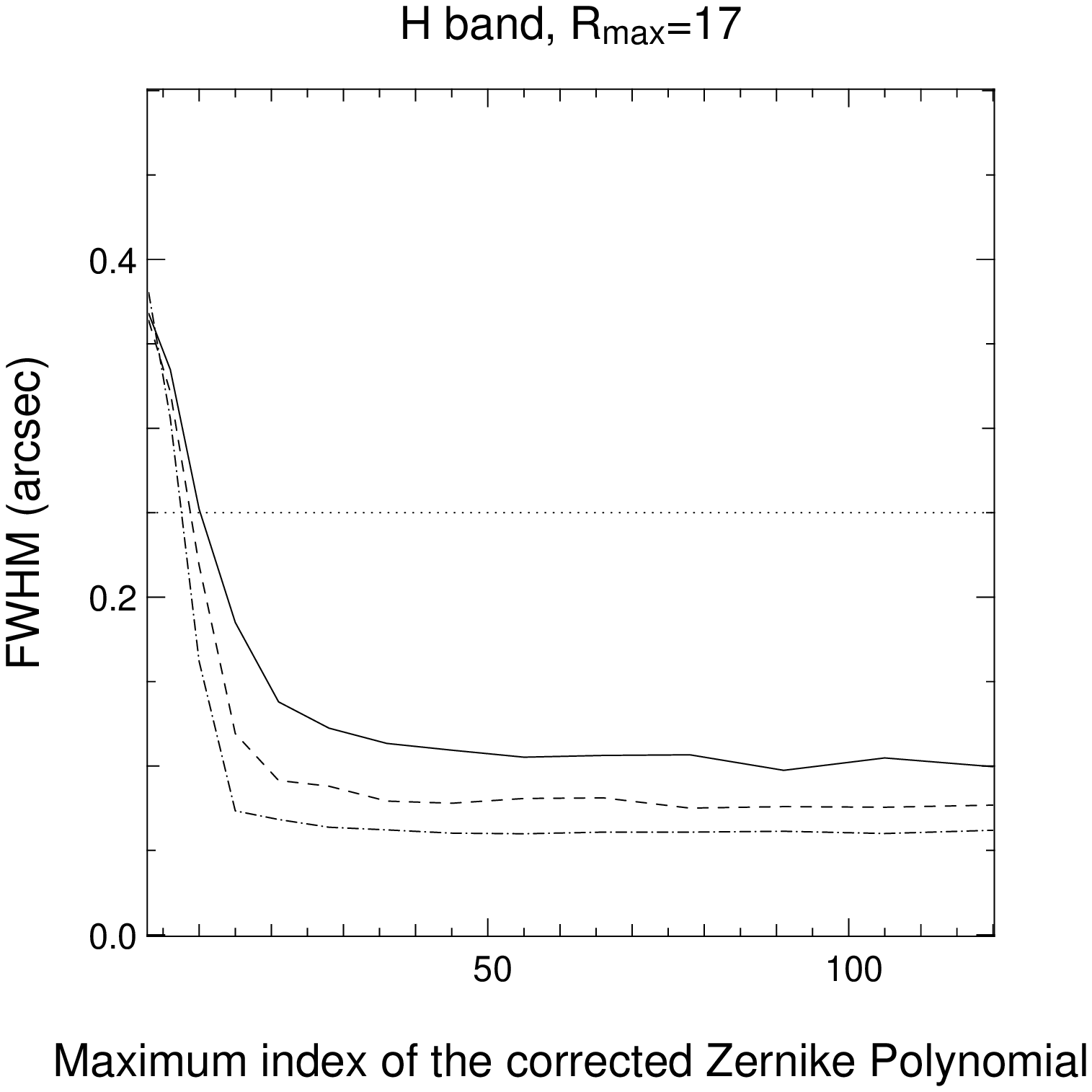}\\\\
	\end{tabular}
\caption{Evolution of the median full width at half maximum (in arcsec)  
for each galactic galactic latitude $b$ and each limiting magnitude  
$R_{max}$ in the J and H band, as a function of the number of corrected  
Zernike polynomials. The full line corresponds to the median  
performance at the galactic pole, the dashed one to the median  
performance at $b\approx -60 \Moideg$, and the dash-dotted one at  
$b\approx -30 \Moideg$. The dashed horizontal line shows the  
specification, i.e. $FWHM=0.25~arcsec$.}
\label{fig:fig16}
\end{figure*}

The figures \ref{fig:fig15} and \ref{fig:fig16} show the median  
performance (over 100 NGS triplets) of the FALCON's AO system, i.e. the  
evolution of the ensquared energy in an $0.25 \times 0.25~arcsec^2$  
square aperture and the evolution of the FWHM as a function of the  
number of corrected Zernike polynomials. The figures on the left assume  
a WFS with a limiting magnitude $R_{max}=16$, the ones on the right a  
WFS with the same noise variance than the NAOS visible WFS  
($R_{max}=17$). On each figure, 3 plots have been reported: they  
correspond to galactic latitudes of 30, 60 and 90 degrees, and  
therefore show the performance for 50\% of the simulation cases,  
corresponding to a \textit{sky coverage} of 50\%. The specification,  
i.e. an absolute ensquared energy of 30\% or 40\%, or a FWHM of  
$0.25~arcsec$ is also drawn on these figures.\newline

One should notice the global shape of these curves. For the ensquared  
energy curves, we can see that they still significantly increase, even  
at a high number of corrected modes. However, the FWHM curves quickly  
saturate to values which are always less than our specification  
($FWHM=0.25~arcsec$). In fact, as shown in the section \ref{sec:sec32},  
high order modes need to be corrected in order to gather the light in  
the inner regions of the PSF and to improve the ensquared energy.  
Therefore, to reach our required specifications for the ensquared  
energy, a non negligible number of modes must be corrected. For a WFS  
with a moderate quality, at least 100 (for J band) and 120 (in H band)  
Zernike polynomials must be corrected so that a sky coverage of 50\%  
can be reached for any galactic latitude $b$. With a more sensitive WFS  
($R_{max}=17$), allowing to use fainter and closer NGS, the correction  
order should be only 65 to 85 Zernike polynomials to reach the same sky  
coverage value: the better the WFS, the lower should be the correction  
order, as there is a transfer of the phase error between noise and  
anisoplanatism. The consequence is therefore that a rather large number  
of actuators would be required, even with the moderate angular  
resolution required.\newline

These curves show the median values over 100 NGS triplets of the  
expected performance expected thanks to the FALCON's AO system. This  
means that the performance will be better than shown for 50\% cases,  
and worse for the other 50\%. In other words, it means that the  
performance shown here stand for a sky coverage of 50\%.

\begin{figure*}
	\centering
	\includegraphics[width=0.45\textwidth]{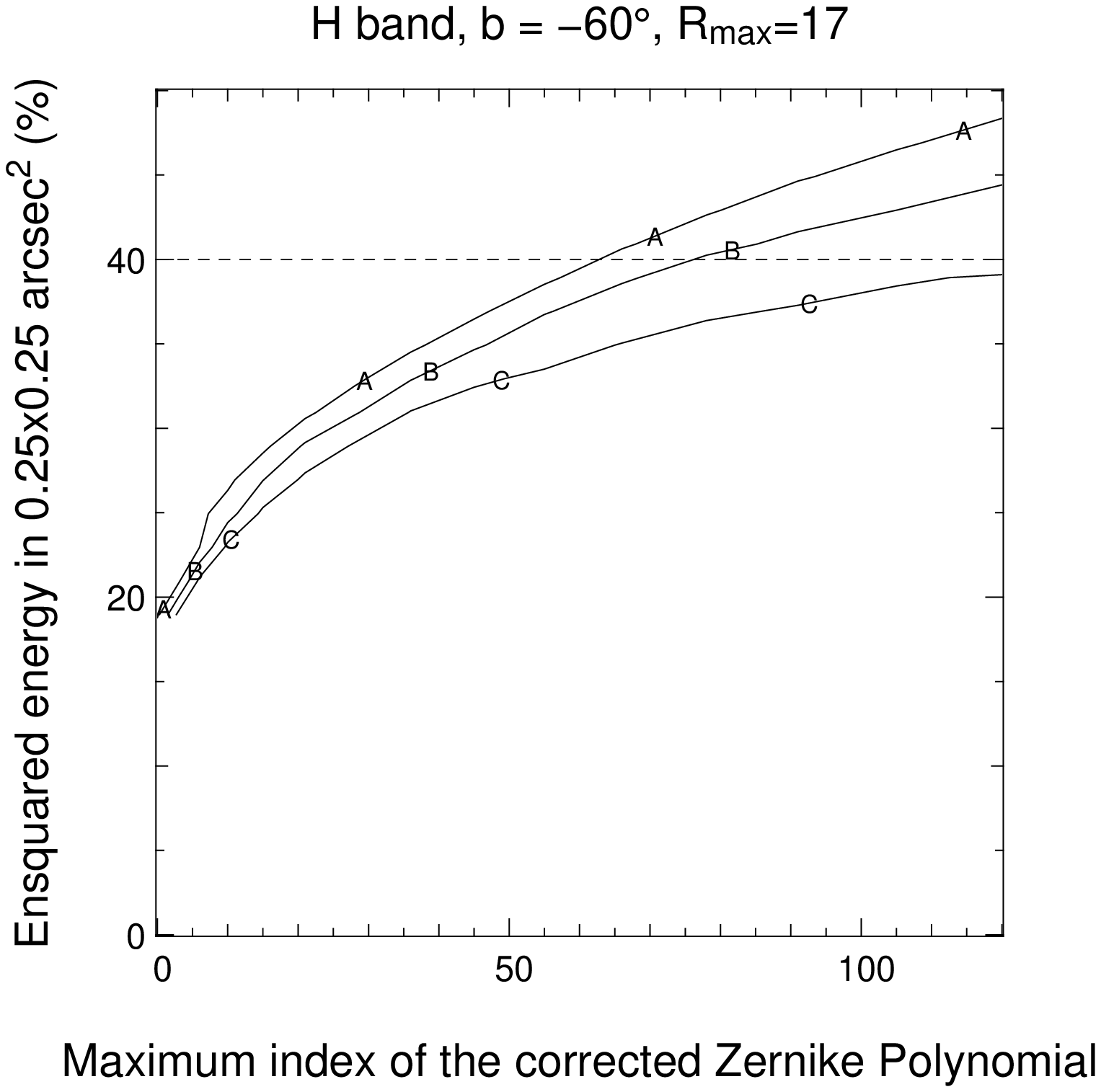}
	\caption{Evolution of the ensquared energy  in H band for the field  
located at $b\approx -60 \Moideg$ as a function of the corrected number  
of Zernike polynomials, and for a limiting magnitude $R_{max}=17$. The  
"A" line shows the performance reached for 10\% of the simulation  
cases, the "B" one the performance reached for 50\% of the cases, and  
the "C" one the performance reached for 90\% of the cases. The dashed  
horizontal line shows the specification, i.e. an ensquared energy  of  
40\%.}
\label{fig:fig17}
\end{figure*}

Immediately, a question arises, that is to know whether 50\% sky  
coverage is a sufficient number or not. This seems to be a critical  
issue, as demonstrated on the figure \ref{fig:fig17}, where the  
ensquared energy has been plotted as a function of the number of  
corrected Zernike polynomials and for sky coverages equal to 10\%, 50\%  
and 90\%. We see that for a specification of 40\%, the number of  
corrected modes should be respectively equal to 60, 75, or more than  
120 ! This way of thinking leads therefore to the conclusion that the  
design of the system is strongly dependent of the sky coverage.  
However, this is only partially true, because a sky coverage of 50\%  
does not mean that only half of the objects can be observed whereas the  
others cannot: it means that 50\% of the objects are observable (some  
of them with an even better quality than expected), and most of the  
other 50\% are observable too, but with a moderate image quality. This  
latter could however be sufficient, as it will be in any case better  
than the uncompensated seeing-limited image.

As an example, when assuming a median corrected ensquared energy  equal  
to 40\%, 75 Zernike polynomials have to be corrected. With such a  
correction, 95\% of the galaxies can be observed with a performance  
ranging between 36\% and 43\%. In other words, the dispersion of the  
ensquared energy  provided by AO correction is very low, making the  
50\% sky coverage a very acceptable performance.

\subsection{Conclusion}
Thanks to numerical simulations, we have been able to show in this  
section the expected performance of a MOAO system using atmospheric  
tomography methods to reconstruct the on-axis wavefront from off-axis  
measurements performed on NGS, in order to provide simulaneously the 3D  
spectroscopy of distant galaxies in a $10 \times 10~arcmin^2$ field of view. For  
any galactic latitude, our results show that an AO system using three  
very sensitive WFS (allowing to measure the wavefront on NGS with a  
magnitude $R\leq 17$) and one deformable mirror conjugated to the pupil  
per scientific target would allow to reach a spatial resolution better  
than $0.25~arcsec$ in J and H bands for 95\% of the objects, and to  
achieve an ensquared energy better than 30\% in J band and 40\% in H  
band for 50\% of the objects. Greater sky coverage (up to 95\%) seem to  
be achievable for slighty less values of ensquared energy. As explained  
before, this latter has not been definitely fixed, especially for  
galaxies with $z \geq 1.5$, as further dynamical and photometric data  
are required to constrain it. However this requires to correct a non  
negligible number of modes: deformable mirrors and wavefront sensors  
with at least $10 \times 10$ actuators and subapertures are required  
to achieve such performance.\newline

From a technological point of view, such a system is very demanding. In  
particular, the miniaturisation of AO components in order to avoid  
field obstruction is critical, and requires further developments. Some  
of them have already started: microdeformable mirrors  
\citep{Zamkotsian-p-2003} would be very good candidates to be used as  
wavefront correctors in each integral field unit. But the miniaturised  
and high-dynamic WFS architecture used in the FALCON's AO system  
remains to be defined.\newline

Therefore, if such an instrumental concept is implemented on  
the VLT, it will then provide a huge gain in terms of observing time  
efficiency for the dynamical study of high redshift galaxies. As an  
example, recent observations made by the AO-assisted spectrograph  
SINFONI \citep{Forster-a-2006} have allowed to retrieve the rotation  
curves of 14 $z\approx 2$ galaxies, with exposure times between 2 and 8  
hours. However, as SINFONI was used in its NGS mode, only two galaxies  
were observed with a spatial resolution better than atmospheric seeing,  
the other ones being observed with a spatial resolution of  
$0.5~arcsec$. In other words we can say that SINFONI in NGS mode as a  
sky coverage of $\approx 14\%$. Moreover the spectral resolution for  
these observations was between $R\approx 1900$ and $R\approx 4500$.  
These numbers (spatial and spectral resolution, sky coverage) have to  
be compared with the ones shown in the sections \ref{sec:sec34} and  
\ref{sec:sec57}, where we can see that the performance of an instrument  
like FALCON are far beyond the capabilities of SINFONI in its NGS mode.  
The performances of SINFONI in its LGS mode will probably be better  
in terms of spatial resolution and ensquared energy. However, the fact that
very long exposures (see section \ref{sec:sec342})are mandatory to probe the kinematics
of distant sources with an 8 meter telescope provides an enormeous advantage to a multiplex instrument such as FALCON.

\section[Discussion]{Discussion}
\subsection{Comparison with GLAO}
The previous section has shown that thanks to atmospheric tomography,  
the correction provided by a MOAO system like FALCON allows a huge gain  
in terms of ensquared energy improvement as well as FWHM reduction,  
thus allowing to perform the 3D spectroscopy of distant galaxies.  
However the use of tomography is quite complex, in particular because  
it requires to have a real-time knowledge of the turbulent profile for  
the computation of the turbulent profile.\newline

If we look at the turbulence profile definined at the section  
\ref{sec:condSimul}, we can see that 85\% of the turbulence is located  
between altitudes of 0 and 1 kilometers, so very close to the ground.  
Ground-Layer Adaptive Optics (GLAO) could therefore be used to improve  
the ensquared energy as well as the angular resolution  
\citep{Rigaut-p-2002,Tokovinin-a-2004}. GLAO consists in averaging the  
measurements from several off-axis guide stars, and is normally very  
efficient when most of the turbulence is located at or close to the  
telescope pupil.\newline

The key issue here is to see if GLAO provides a correction as effective  
as the one provided by tomography, as this would decrease the AO  
system's complexity. To answer to this question, we did again the same  
simulations than the one described in the section \ref{sec:sec5}. But  
this time, instead of using the optimal tomographic reconstruction  
matrix $\matRecTomo$ to compute the on-axis correction phase, we simply  
did an average of the off-axis measurements coming from the three  
closest GS of the science target.

\begin{figure*}
	\centering
	\begin{tabular}{cc}
		\includegraphics[width=0.45\textwidth]{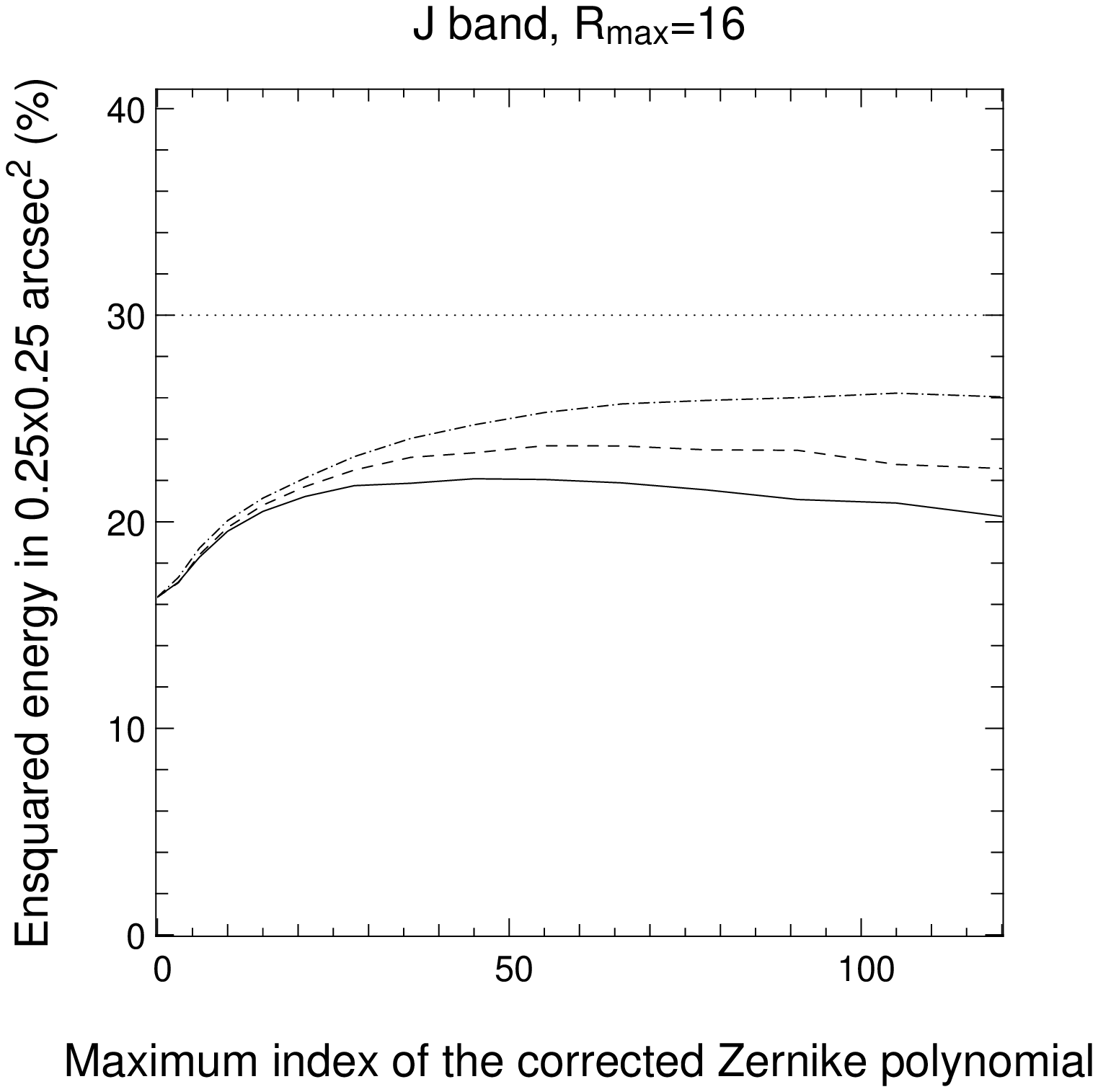}&
		\includegraphics[width=0.45\textwidth]{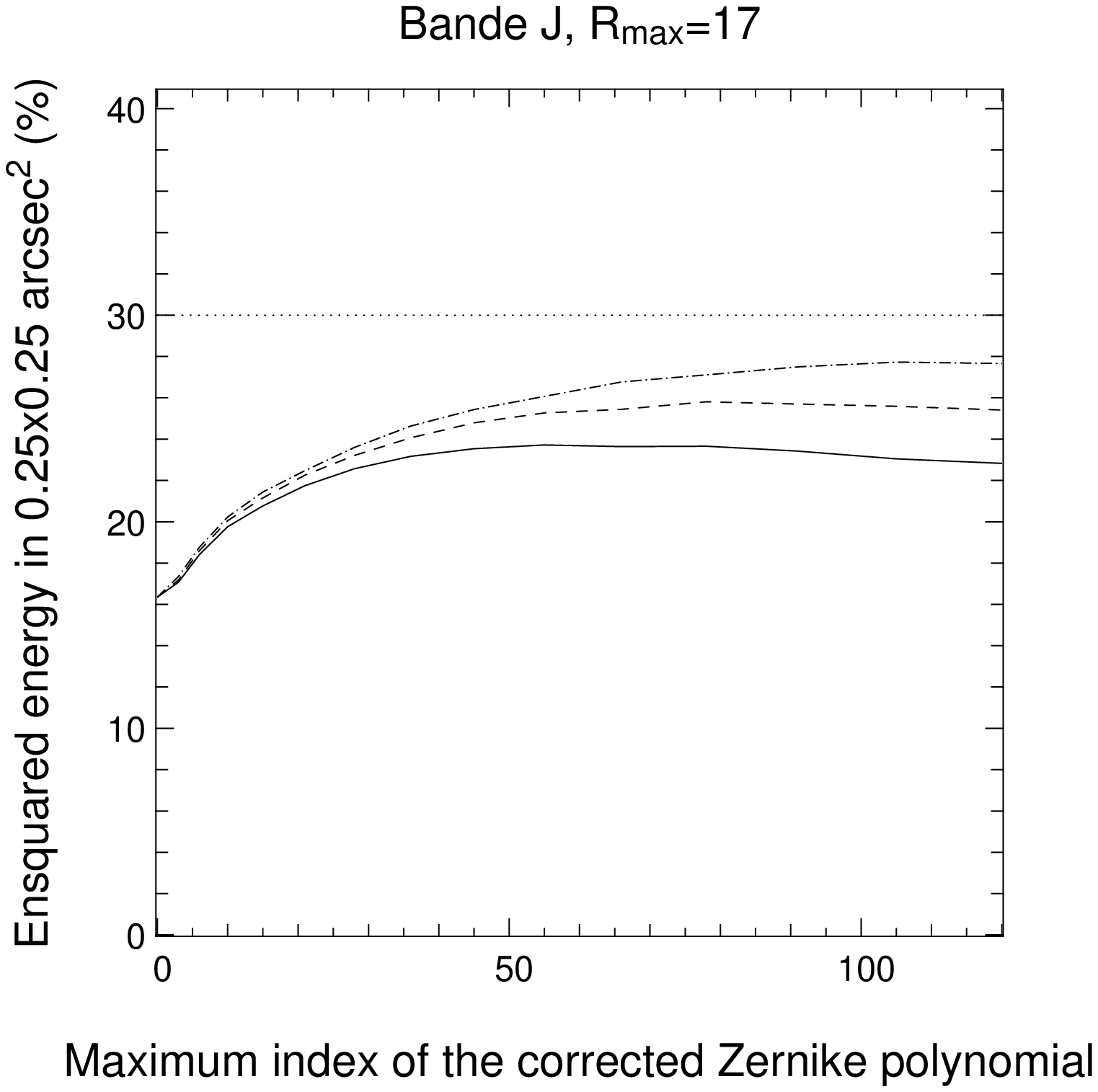}\\\\
		\includegraphics[width=0.45\textwidth]{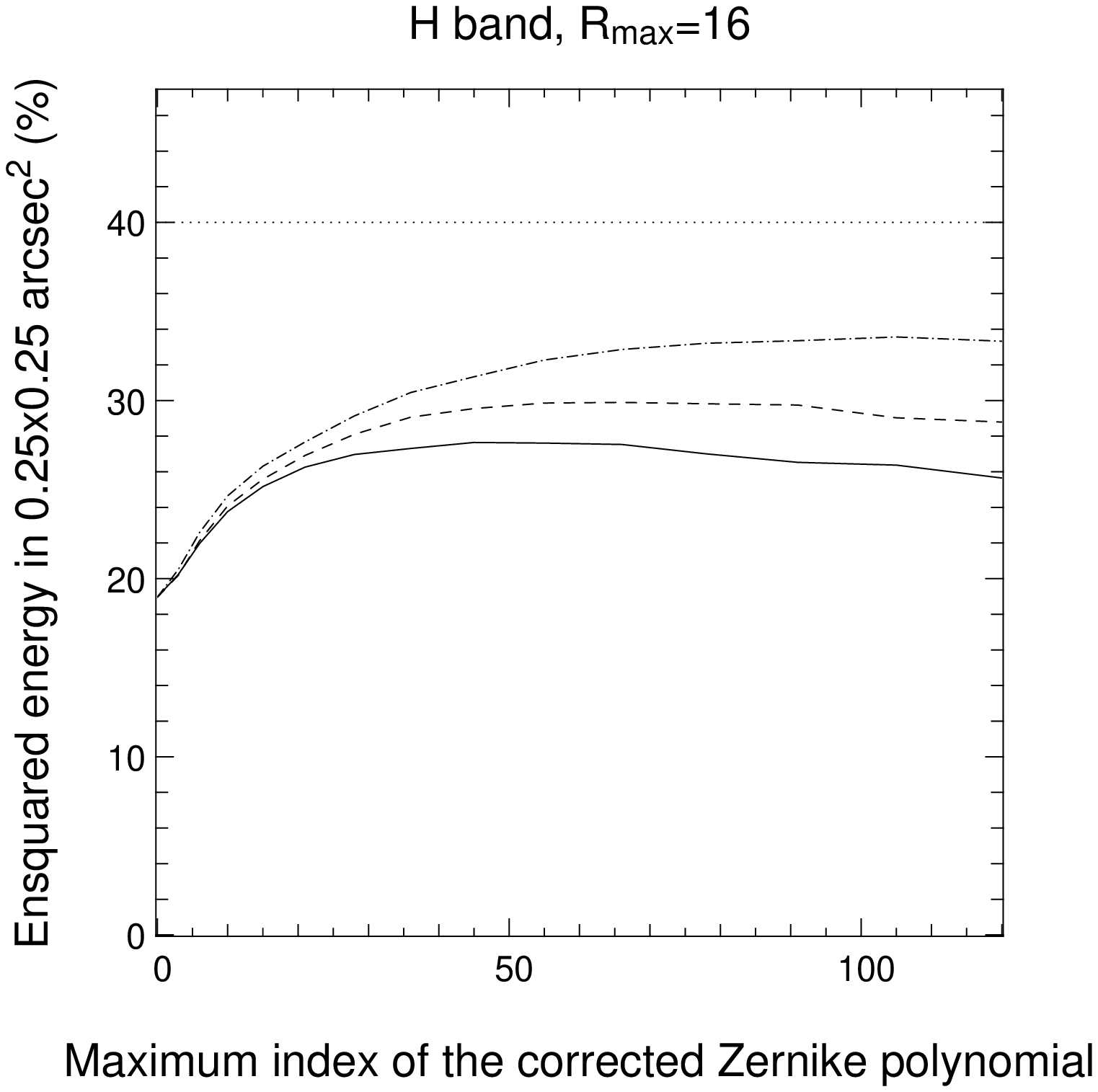}&
		\includegraphics[width=0.45\textwidth]{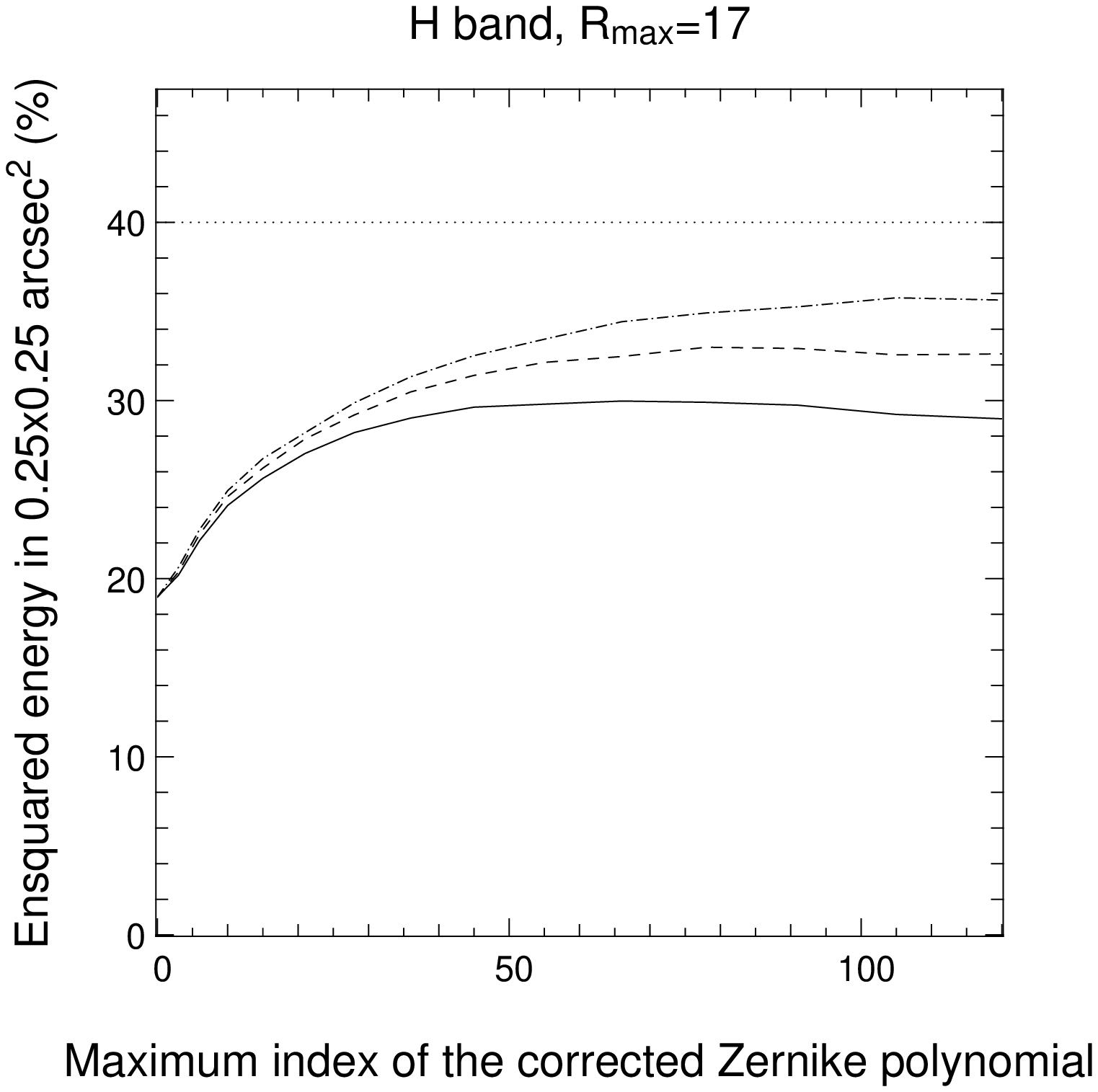}\\\\
	\end{tabular}
\caption{Evolution of the median ensquared energy in a $0.25 \times  
0.25~arcsec^2$ square aperture for each galactic latitude $b$ and each  
limiting magnitude $R_{max}$ in the J and H band, as a function of the  
number of corrected Zernike polynomials, in the case of GLAO correction  
(averaging of off-axis measurements). The full line corresponds to the  
median performance at the galactic pole, the dashed one to the median  
performance at $b\approx -60 \Moideg$, and the dash-dotted one at  
$b\approx -30 \Moideg$. The dashed horizontal line shows the  
specification, i.e. 30\% in J band and 40\% in H band.}
\label{fig:fig18}
\end{figure*}

\begin{figure*}
	\centering
	\begin{tabular}{cc}
		\includegraphics[width=0.45\textwidth]{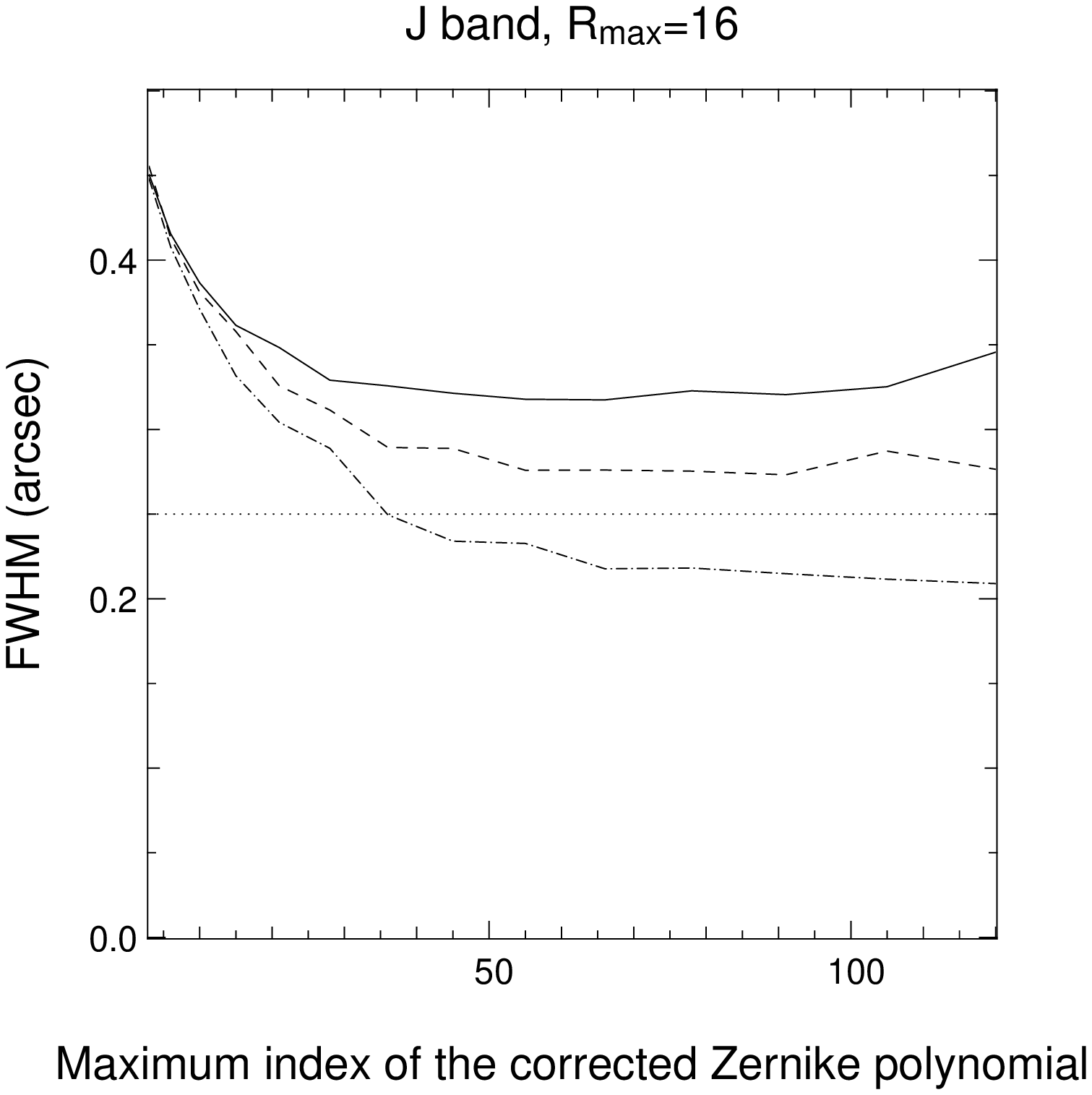}&
		\includegraphics[width=0.45\textwidth]{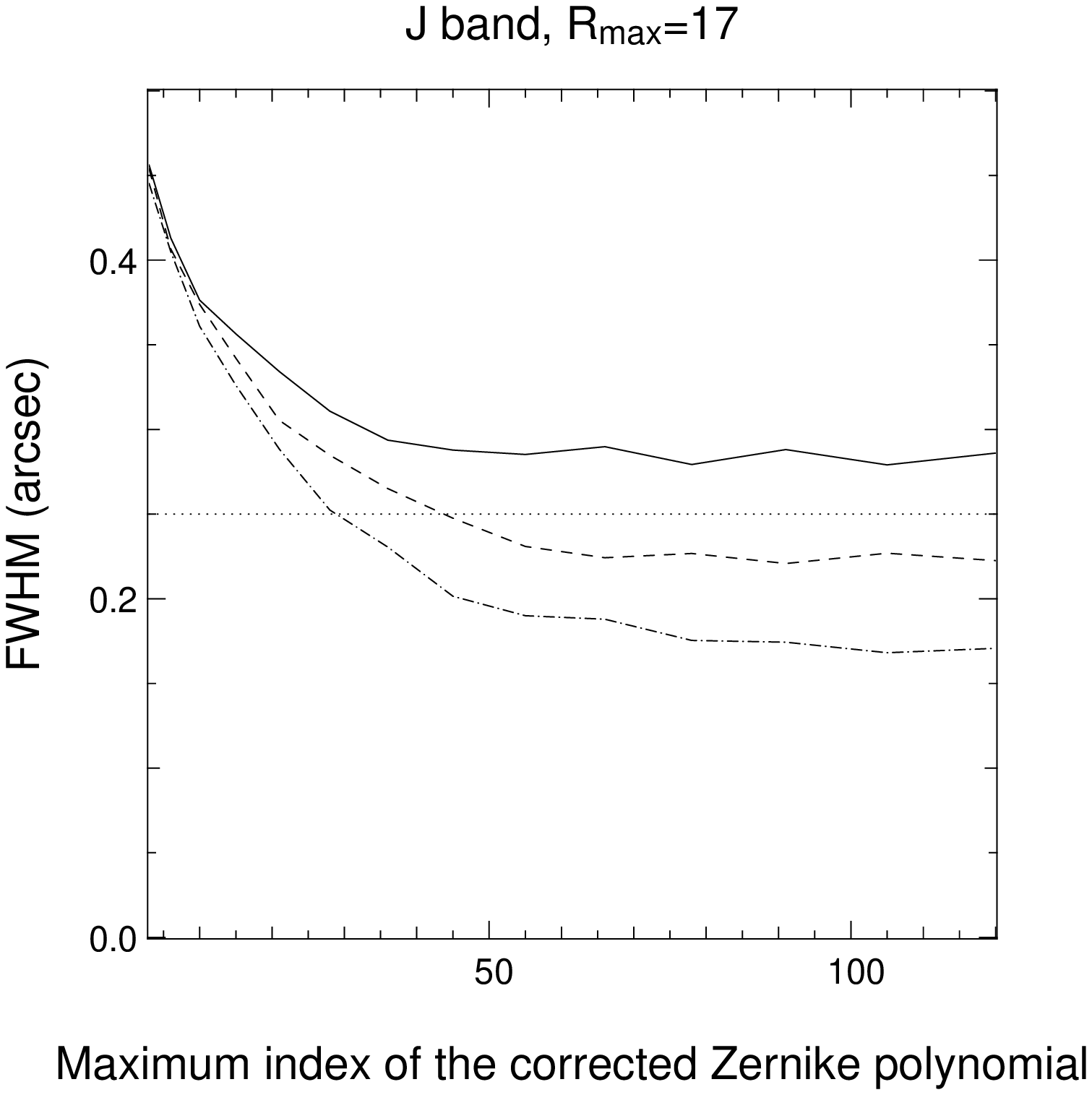}\\\\
		\includegraphics[width=0.45\textwidth]{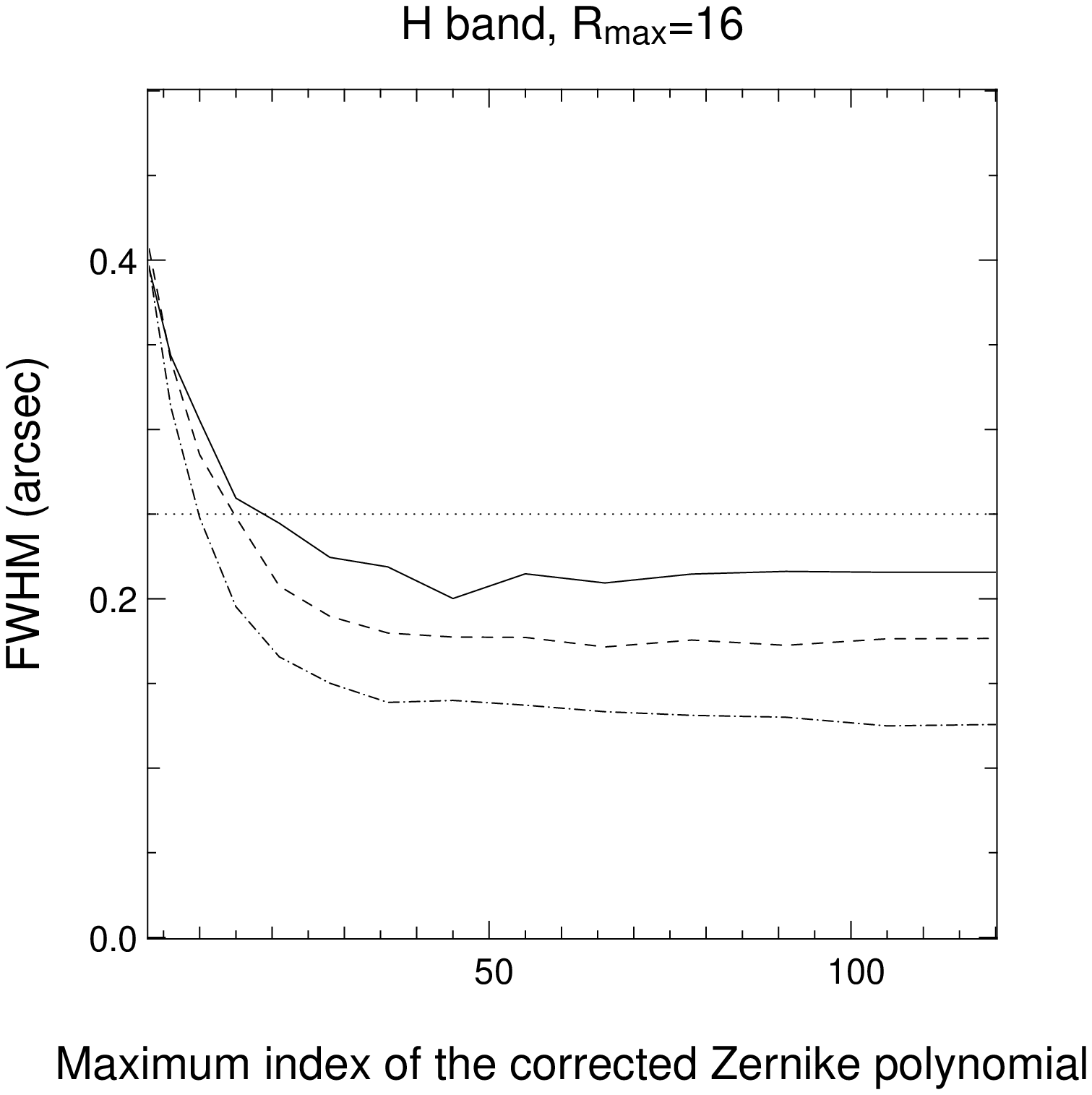}&
		\includegraphics[width=0.45\textwidth]{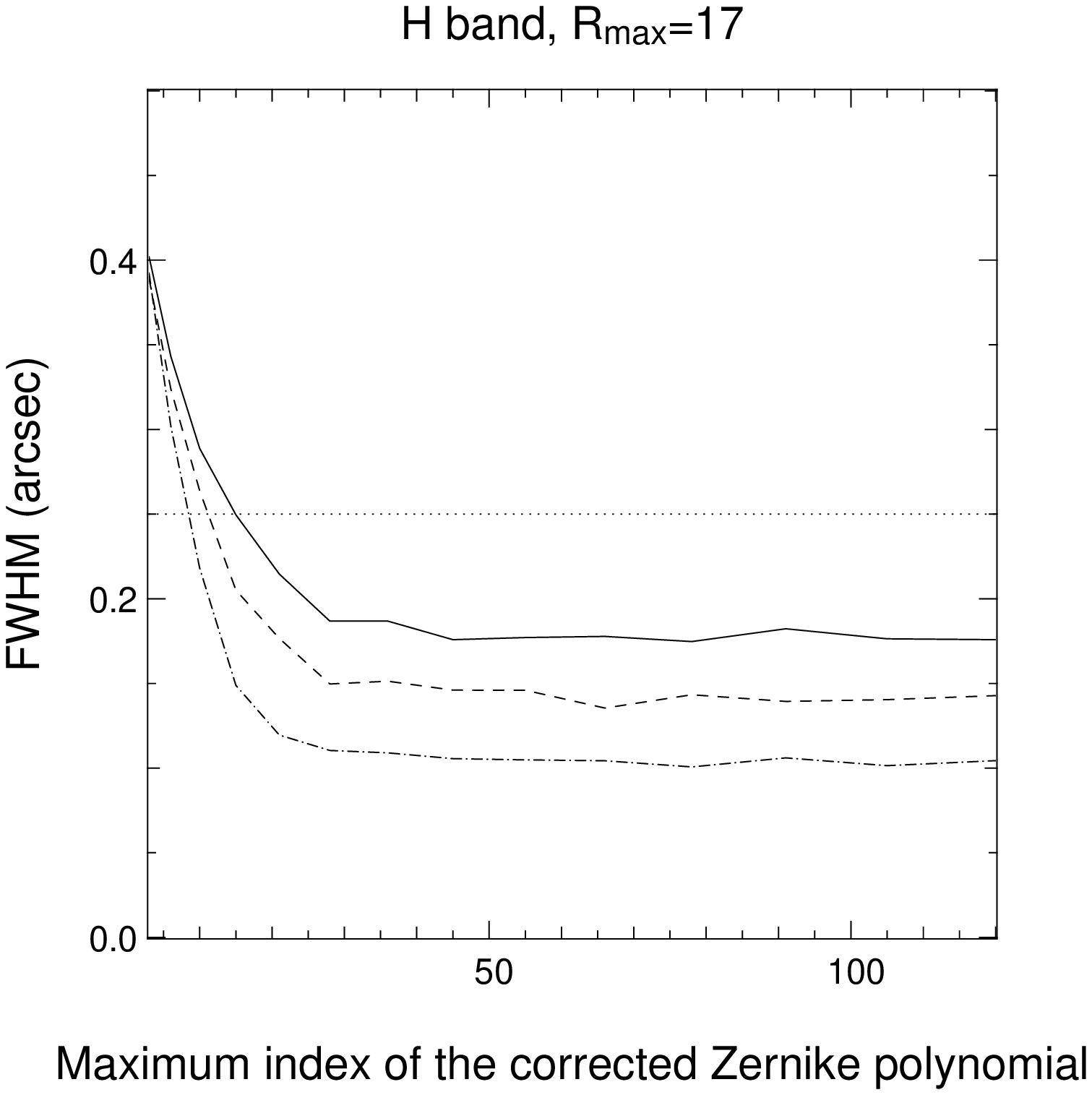}\\\\
	\end{tabular}
\caption{Evolution of the median full width at half maximum (in arcsec)  
for each galactic galactic latitude $b$ and each limiting magnitude  
$R_{max}$ in the J and H band, as a function of the number of corrected  
Zernike polynomials, in the case of GLAO correction (averaging of  
off-axis measurements). The full line corresponds to the median  
performance at the galactic pole, the dashed one to the median  
performance at $b\approx -60 \Moideg$, and the dash-dotted one at  
$b\approx -30 \Moideg$. The dashed horizontal line shows the  
specification, i.e. $FWHM=0.25~arcsec$.}
\label{fig:fig19}
\end{figure*}

The figures \ref{fig:fig18} and \ref{fig:fig19} show therefore the  
median performance (over 100 NGS triplets) of the evolution of the  
ensquared energy and the FWHM for the three cases of galactic latitudes  
and the two cases of limiting magnitudes already studied, but this time  
for GLAO correction. For the FWHM, we observe also a saturation of the  
FWHM, but we can see that even the use of stars with magnitudes $R \leq  
17$ does not allow to reach the required angular resolution for all the  
galactic latitudes, as the median angular resolution reached at the  
galactic pole is equal to $0.30~arcsec$. The results are worst for the  
ensquared energy curves, which show that it is never possible to reach  
the required ensquared energy  values: a GLAO system has a \textbf{0\%  
sky coverage} for the FALCON specification.

Moreover, a careful examination of the ensquared energy curves show  
that their behaviour is quite different from the ones shown on the  
figures \ref{fig:fig15} : contrary to these latter, the curves start to  
increase, reach an upper limit, and then systematically decrease. This  
result has to be related with the study of the AO corrected PSF made at  
the section \ref{sec:sec32}. Indeed, we have shown there that high  
order modes must be corrected to bring the energy into the inner parts  
of the PSF. In GLAO, as an averaging of off-axis measurements is done,  
there is an important residual anisoplanatism which degrades high-order  
correction, thus explaining the decrease of the ensquared energy for an  
important number of corrected modes. This effect is moreover going to  
be degraded by the noise propagation because of the faint stars used  
for wavefront sensing.

However we do not observe such effects when the optimal tomographic  
reconstruction matrix $\matRecTomo$ is used. Indeed, the analytical  
expression of this matrix at equation \eqref{eq:eq23} shows that it  
uses some \textit{a-priori} about the turbulence profile as well as the  
noise, allowing to compensate for anisoplanatism \textbf{and}  
noise.\newline

One of the main justification of GLAO is to have \textit{reduced-seeing  
limited} images. Although providing better angular resolution, the  
corrected PSF is still seeing-limited, meaning that it is not sharpened  
enough to gather the light into its inner regions and increase the  
spectroscopic SNR. As a conclusion, we have shown here that even if  
GLAO seems to be very promising, it is however not sufficient to allow  
to perform the 3D spectroscopy of distant galaxies.  
\citet{LeLouarn-a-2004} have suggested to combine GLAO with LGS, and  
found that using 4 LGS allow to improve the ensquared energy over a  
field of view equal to 1 arcmin. However, this compensated field of  
view is still one order of magnitude lower that the $10 \times 10~arcmin^2$ required.  
This means that using GLAO with LGS over the Nasmyth FoV would again  
require at least one LGS per scientific target, and we have seen that  
such an architecture dramatically increase the cost of such an  
instrument. As a conclusion, GLAO is not usable in a system like  
FALCON, and the combination of MOAO with atmospheric tomography is the  
best architecture to reach our scientific goals.

\subsection{Influence of other parameters}
The goal of the results shown in the section \ref{sec:sec5} was to  
allow us to give some first order of specifications for the required  
FALCON AO system, in terms of sky coverage, limiting magnitude and  
number of actuators. We therefore found that using stars with  
magnitudes $R \leq 17$ would allow us to satisfy our specifications  
with a sky coverage of at least 50\% for any galactic latitude.  
However, as we focused only on the spatial aspect of tomographic  
reconstruction, we did several simplifications before running our  
simulations, whose consequences are far from being negligible on the AO  
system's definitive design.\newline

First we assumed that we did not have any focal plane obstruction  
problem, meaning that the physical size of the AO components (the WFS  
or the IFUs) would not be a problem. For a real system, it is possible  
that the size of the optomechanical devices will be too large, implying  
that the first closest NGS to the scientific target will not be usable  
to perform wavefront sensing, and that it will be required to use more  
distant NGS. In fact, the equation \eqref{eq:eq29} gives us the  
expression of the cdf $P_1(D_1 \leq r)$ of $D_1$, the distance of the  
first closest NGS, and its expression is $P_1(D_1 \leq r)=1-\exp{-\pi  
\sigma_* r^2}$. If we assume that all the devices have the same  
diameter $D_B$, and that they can work even when they are tangent, then  
the distance $D_1$ will indeed be the device's diameter $D_B$, and the  
function $P_1(D_1 \leq r)$ will give us the probability $P_N$ of not  
being able to use the first closest NGS. We therefore found that $P_N$  
and $D_B$ are linked by the following relation:
\begin{equation}\label{eq:eq37}
	D_B=\sqrt{\frac{1}{\pi \sigma_*}\log\moiPar{\frac{1}{1-P_N}}}
\end{equation}

$\sigma_*$ being a number of stars per solid angle, the above  
expression has to be multiplied by the focal plane scale  
($35~mm~arcmin^{-1}$ for the VLT Nasmyth focus) to give a physical  
size. As an example, for $b\approx -30 \Moideg$, we had $\sigma_*=0.35$  
stars per square arcmin ($R_{max}=16$) or $\sigma_*=0.57$ stars per  
square arcmin ($R_{max}=17$). Then, if we accept to loose 50\% of the  
first closest NGS, we found respective diameters $D_B$ of 28 mm or 22  
mm. These numbers will of course be lower if we want to loose less  
stars.\newline

Another critical issue for AO is the temporal error due to time delay,  
which we did not take into account for our simulations, and which is  
going to degrade the performance of the AO system, in particular for  
high wind velocities. Additional simulations taking into account time  
delay are required to fix the temporal bandwidth of the system. For  
MCAO systems, \citet{LeRoux-a-2004} have shown that the use of a Kalman  
filter allows to reach very good performance. Such a system therefore  
needs to be simulated to see if it could be incorporated into the  
FALCON's AO system.\newline

We have also to emphasize the fact that we simulated an open-loop  
system, where the off-axis wavefront sensors do not get any feedback  
from the on-axis DM. Moreover we assumed in our simulations perfect  
components with no errors and with instantaneous response. As we work  
in open-loop, the presence of errors like WFS aliasing or misalignents,  
DM hysterisis, and more generally non linearity, is going to have some  
non-negligible consequences on the performance of the AO system. A  
tolerance analysis is therefore required to quantify the maximum errors  
we can accept for each AO subsystem.\newline

At last, the simulations we performed assumed static turbulence  
conditions, these latter being the median ones at the Cerro Paranal.  
However the turbulence is going to significantly change during the long  
exposures times required by our science objectives. Additional  
simulations with non-static turbulence conditions are therefore  
required in order to have some definitive realistic instrument  
performance.

\section{Conclusions}
We have proposed in this paper FALCON, a new concept of multi-object  
spectrograph for the ESO Very Large Telescope. Thanks to the  
combination of adaptive optics and atmospheric tomography methods,  
FALCON will allow to perform the 3D spectroscopy of several galaxies  
located up to $z=1.5$ with a spatial resolution of $0.25~arcsec$ and a  
spectral resolution $R=10000$, in a wide FoV. Such a performance  
implies to use Multi-Objects Adaptive Optics systems, and we have shown  
that for median Cerro Paranal atmospheric conditions, such systems will  
allow to reach sky coverages of at least 50\% \textbf{for any galactic  
latitude}.\newline

In terms of AO components, our simulations have shown that very  
sensitive wavefront sensors measuring the wavefront from guide stars  
with magnitudes $R \leq 17$ will be required, with $10 \times 10$  
subapertures or actuators. The same number of actuators will be  
required for the deformable mirror in each integral field unit.  
Moreover each deformable mirror will require measurements from three  
off-axis wavefront sensors, as atmospheric tomography methods are  
needed to improve the sky coverage.\newline

Such an instrumental concept is going to be a real technological  
challenge. Firstly because of the required miniaturisation of the AO  
components in order to be able to use in parallel several AO systems in  
the VLT focal plane and to avoid focal plane obstruction: the maximum  
size of the wavefront sensors and integral field units should not  
exceed 22 mm. Moreover, as atmospheric tomography is required in order  
to reach high sky coverages for any galactic latitude, each independent  
AO system will work in open-loop, meaning to use non-classical AO  
architectures. We have indeed shown that atmospheric tomography is  
definitely required for the required science objectives, as Ground  
Layer Adaptive Optics will not provide sufficient performance.\newline

Additional studies are required for the final design of the instrument,  
in particular for the temporal bandwidth of the AO system and the  
sensitivity to the variation of atmospheric parameters. But the  
extrapolation of a concept like FALCON to the next generation of  
Extremely Large Telescopes is very promising. As an example, for a 42  
meter telescope with a $F/15$ focal ratio, a $10\times 10~arcmin^2$  
field will cover a physical size of $\approx 2 \times 2$ square meters in  
the focal plane. A whole field instrumentation will therefore be  
impossible to manufacture because of the sizes of the optical  
components, and a MOAO system will be the best solution for wide field  
observations, which will be needed for the studies of the very early  
universe. Then, the use of an instrument like FALCON, but scaled to an  
ELT, will allow to detail galaxy physics with scales of $400~pc$ up to  
$z=7$, and to understand how galaxies formed since the reionization  
epoch.

\section*{Acknowledgments}
Fran\c cois Ass\'emat is grateful to V\'eronique Cayatte at  
Observatoire de Paris-Meudon, and Simon Morris, Richard Myers and Ray  
Sharples at Durham University for helpful discussions. The authors  
acknowledge the Institut National des Sciences de l'Univers (INSU) and  
the ESO Adaptive Optics department for financial support.
\bibliographystyle{mn2e}

%

\label{lastpage}

\end{document}